\documentclass{fundam}

\pdfoutput=1

\usepackage{url} 

\usepackage{graphicx}

\usepackage{amsmath}
\usepackage{amsfonts}
\usepackage{amssymb}
\usepackage{graphicx}%
\usepackage{url}

\begin{document}

\setcounter{page}{1}
\publyear{22}
\papernumber{2150}
\volume{189}
\issue{1}

\finalVersionForARXIV

\title{Unfoldings and Coverings of  Weighted  Graphs}

\author{Bruno Courcelle\thanks{Address for correspondence: Bordeaux University, LaBRI,
                              F-33400 Talence, France. \newline \newline
          \vspace*{-6mm}{\scriptsize{Received December 2022; \ accepted May 2023.}}}
\\
LaBRI, CNRS (UMR 5800), University of Bordeaux, France\\
courcell@labri.fr}

\maketitle

\runninghead{B. Courcelle}{Unfoldings and Coverings}

\begin{abstract}
\emph{Coverings} of undirected graphs are used in distributed computing, and
\emph{unfoldings} of directed graphs in semantics of programs. We study these
two notions from a graph theoretical point of view so as to highlight their
similarities, as they are both defined in terms of surjective graph
homomorphisms.\ In particular, \emph{universal coverings} and \emph{complete
unfoldings} are infinite trees that are \emph{regular} if the initial graphs
are finite.\ \emph{Regularity} means that a tree has finitely many subtrees up
to isomorphism.\ Two important theorems have been established by Leighton and
Norris for coverings of finite graphs. We prove similar results for unfoldings
of finite directed graphs.\
Moreover, we generalize coverings and similarly, unfoldings to graphs and
digraphs equipped with finite or infinite \emph{weights} attached to edges of
the covered or unfolded graphs.\ This generalization yields a canonical
"factorization" of the universal covering of any finite graph, that (provably)
does not exist without using weights.\ Introducing $\omega$\ as an infinite
weight provides us with finite descriptions of regular trees having nodes of
\emph{countably infinite degree}. Regular trees (trees having finitely many
subtrees up to isomorphism) play an important role in the extension of Formal
Language Theory to infinite structures described in finitary ways.\ Our
weighted graphs offer effective descriptions of the above mentioned regular
trees and yield decidability results. We also generalize to weighted graphs
and their coverings a classical factorization theorem of their
\emph{characteristic polynomials}.
\end{abstract}


\section{Introduction}

We first review\ informally some basic notions and results. The notion of
\emph{covering of an undirected graph} has been introduced by Reidemeister
\cite{Red} as a discrete analogue of coverings of surfaces.\ It has proved to
be useful in the theory of \emph{distributed computing} where a network is
considered as an undirected graph $N$\ whose edges represent communication
channels.\ The questions are whether certain problems such as the
\emph{election problem} (consisting in distinguishing a unique node of the
network) can be solved by a distributed algorithm (of a certain type).\ This
is possible if the graph $N$\ is minimal for the \emph{covering relation},
equivalently if the \emph{universal coverings} of $N$\ defined from any two
different nodes are not isomorphic rooted trees. The universal covering of an
undirected graph is an infinite tree.\ It has a characterization in the sense
of Category Theory and can be constructed as the infinite tree of the walks in
the graph originated from a node and that do not take the same edge twice in a
row (in opposite directions).\ Starting from any two nodes yields isomorphic
trees (without roots).\ Detailed definitions will be given in Section 4. The
universal covering of a finite graph is a \emph{regular tree}, \emph{i.e.}, a
tree that has finitely\ many subtrees \emph{up to isomorphism} (\emph{i.e.},
finitely many isomorphism classes of subtrees). The application of coverings
to distributed computing was initiated by Angluin in \cite{Ang}.

\medskip
\emph{Unfoldings} \emph{of directed graphs} are used in the study of abstract
programs called \emph{transition systems} in order to represent their
semantics \cite{Arn,Cou9,CouWal}. In particular, the \emph{complete
unfolding\footnote{It is simply called \emph{unfolding} in
\cite{Arn,Cou9,CouWal}. }} of a directed graph equipped with a distinguished
vertex (representing the "begin" instruction) is a rooted tree that is
infinite if the graph has directed cycles. The complete unfolding of the graph
representing a transition system $S$\ encodes all computations of the program
abstracted into $S$. If the graph is finite, its complete unfolding is a
regular tree.\ Precise definitions will be given in Section 3.\

We are interested in unfoldings and coverings from a \emph{graph theoretical
point of view}.\ Both notions are defined in terms of \emph{surjective graph
homomorphisms} that are bijective on the neighbourhoods of vertices related by
the considered homomorphisms. The notion of neighbourhood is thus a parameter
that gives rise to different but related notions: unfoldings, coverings and
even others \cite{CouWal}. For unfoldings of directed graphs, the
neighbourhood of a vertex $x$ is the set of edges outgoing from $x$. For
coverings of undirected graphs, it is the set of edges incident to $x$. We
study unfoldings and coverings by means of \emph{graph homomorphisms},
\emph{quotient graphs}, \emph{infinite trees} and, in particular,
\emph{regular} ones. One of our objectives is to highlight the similarities
between the two notions, regarding the definitions and also some results
without using any cumbersome categorical framework.

In the theory of coverings, a theorem by Norris \cite{Nor} states that two
regular rooted trees $T_{x}$ and $T_{y}$, that define the universal covering
$T$ of a finite undirected graph with $p$ vertices\ by starting the walks from
$x$ and $y$ are isomorphic if their truncations at depth $p-1$ are isomorphic.
Another important theorem by Leighton \cite{Lei} states that, if two finite
undirected graphs have isomorphic universal coverings, then they have a
\emph{common finite} covering.\ Its proof is quite difficult.\ We prove a
special case that subsumes the known case of regular graphs \cite{AngGar}.\

\medskip
\emph{Weighted graphs.}

Moreover, we extend the definitions of unfoldings and coverings in the
following ways. A directed graphs is \emph{weighted} if each edge has a
\emph{weight}, a positive integer or the infinite cardinal $\omega$. An edge
of weight 3 (resp. $\omega$) \emph{unfolds into} 3 directed edges (resp.
countably many) with the same origin. We define complete unfoldings
accordingly, and we obtain \emph{regular trees} from finite graphs.\ These
trees have nodes of infinite degree\footnote{Here, we extend the notion of
regular tree that arises from the theory of recursive program schemes
\cite{Cou83}.} in the case where some edges have weight $\omega$, which
generalizes the usual definitions. We call \emph{complete unfolding} what is
usually called \emph{the} unfolding (this tree is unique up to isomorphism),
and we define as \emph{unfolding} of a weighted directed graph $H$\ a weighted
directed graph that lies inbetween $H$ and its complete
unfolding.\ "Inbetween" is formally defined in terms of surjective
homomorphisms that are locally bijective as explained above.\ Each regular
rooted tree $T$\ is the complete unfolding of a finite unique \emph{canonical}
weighted directed graph, that can be used as a finite description of $T$. We
extend to weighted directed graphs the theorems by Leighton and Norris
described above\footnote{In the forthecoming article \cite{CouNext}, we will
establish the \emph{first-order\ definability} of regular trees, among all
trees, and also of the universal coverings of finite weighted graphs, as
described below. These proofs use our extensions of Norris's Theorem.}.

We also extend the notion of covering to \emph{weighted undirected
graphs}.\ In this case, weights in $\mathbb{N}_{+}\cup\{\omega\}$ are attached
to \emph{half-edges}: an edge that is not a loop has two half-edges and thus
two weights. A loop is a half-edge (without any matching opposite half-edge)
and has a single weight. Each such graph $H$\ has a unique \emph{universal
covering} (unicity is up to isomorphism) that is an infinite tree $T$\ without
root denoted by $\boldsymbol{UC}(H)$.\ It is formally defined from the
unfolding of a directed graph, where $p$ parallel directed edges from a vertex
$x$ to $y$ replace a weight $p$ attached to an half-edge incident with $x$
whose matching half-edge is incident with $y$.\ It is not from walks as easily
as in the case of unweighted graphs.\

We call \emph{strongly regular} a tree $T$\ of the form $\boldsymbol{UC}(H)$
for some finite weighted graph $H$.\ This means that $T$ yields finitely many
regular rooted trees $T_{x}$, up to isomorphism, by taking its different nodes
$x$ as roots. This is a new notion. Each strongly regular tree is the
universal covering of a \emph{canonical} (it is unique up to isomorphism)
finite weighted graph of minimal size, and thus, has a finitary description.
It can be seen as a kind of minimal factorization.\ The infinite rooted binary
tree is regular, but it is not strongly regular after forgetting its root.

\medskip
\emph{Our new definitions and main results}

1) We define and study coverings and unfoldings in close connection by
considering them as two instances of the same notion of a locally bijective
homomorphism, based on different types of neighbourhood. In both cases we
introduce weights on edges.\ Infinite weights yield trees of infinite degree
having finite descriptions.

2) Our first main result states that two finite graphs have isomorphic
universal coverings if and only if they are coverings of a unique minimal
weighted graph.\ Using weighted graphs is here necessary.

3) Our second main theorem extends that by Norris to universal coverings and
to complete unfoldings of finite, weighted, graphs and directed graphs.

4) Our third main theorem extends that by Leighton to complete unfoldings of
weighted directed graphs. We give an easy proof of it for coverings of graphs
in a special case that subsumes the previously known cases and yields new cases.

5) Finite weighted undirected graphs are defined by matrices in a natural way.
Our fourth main theorem extends to them a factorization of the characteristic
polynomials of their coverings that is known in the case of finite graphs
without weights. Hence, our approach fits nicely in Algebraic Graph Theory.

6) We identify as \emph{strongly regular} the universal coverings of the
finite weighted graphs.\ They form a proper subclass of regular trees that we
study more in \cite{CouNext}.

\medskip
\emph{Summary of the article}: Basic definitions are in Section 2.\ Unfoldings
of weighted directed graphs are defined and studied in Section 3.\ Coverings
of weighted undirected graphs are defined and studied in Section 4.\ We study
universal coverings of weighted graphs in Section 5 and we discuss Leighton's
Theorem for graphs in Section 6.

\section{Basic definitions}

This section reviews notation and some easy lemmas. Definitions for graphs and
trees are standard, but we make precise some possibly ambiguous terminological points.

\subsection{Sets, multisets and weighted sets.}

All sets, graphs and trees are finite or countably infinite (of cardinality
$\omega$).

The cardinality of a set $X$ is denoted by $\left\vert X\right\vert
\in\mathbb{N}\cup\{\omega\}$. This latter set is equipped with an addition +
that is the standard one on $\mathbb{N}$ together with the rule $\omega
+x=x+\omega=\omega$ for all $x$ in $\mathbb{N}\cup\{\omega\}$.

We denote by $[p]$ the set $\{1,\dots,p\}$ and by $\mathbb{N}_{+}$\ the set of
positive integers.

A \emph{weighted set} is a pair $(X,\lambda)$ where $X$ is a set and $\lambda$
is a mapping $X\rightarrow\mathbb{N}_{+}\cup\{\omega\}$.\ We call $\lambda(x)$
the \emph{weight of} $x$, and, for $Y\subseteq X$, we define\footnote{For
typographical reasons, we use the notation $\Sigma\{\lambda(x)\mid x\in Y\}$
\ rather than $\sum_{x\in Y}\lambda(x)$ and we will do the same below in
Sections 3.1 and 4.2.} $\lambda(Y):=\Sigma\{\lambda(x)\mid x\in Y\}$. A
weighted set can be seen as a \emph{multiset}, where $\lambda(x)$\ is the
number of occurrences of $x$. From a set $X,$ we get the weighted set denoted
by $(X,\boldsymbol{1})$ where all weights are 1. We define $Set(X,\lambda
):=\{(x,i)\mid x\in X,i\in\mathbb{N}_{+},1\leq i\leq\lambda(x)\}$ so that
$\lambda(X)=\left\vert Set(X,\lambda)\right\vert .$

We denote by $\uplus$\ the union of multisets, equivalently of weighted sets:
$(X,\lambda)\uplus(Y,\lambda^{\prime}):=(X\cup Y,\lambda^{\prime\prime})$
where $\lambda^{\prime\prime}(x)$ is $\lambda(x)+\lambda^{\prime}(x)$ if $x\in
X\cap Y$\ and $\lambda(x)$\ or $\lambda^{\prime}(x)$ otherwise.

Let $(X,\lambda)$ and $(Y,\lambda^{\prime})$ be weighted sets.\ A surjective
mapping $\kappa:X\rightarrow Y$ is a \emph{weighted surjection} or a
\emph{surjection of multisets}: $(X,\lambda)\rightarrow(Y,\lambda^{\prime})$
if, for every $y\in Y$, we have $\lambda^{\prime}(y)=\lambda(\kappa^{-1}(y))$,
hence is the sum of weights of the $x$'s such that $\kappa(x)=y$.\ If $X$ is a
set, hence, if $\lambda$\ has value 1 for all $x$ $\in X$, a weighted
surjection $\kappa:X\rightarrow Y$ satisfies $\lambda^{\prime}(y)=\left\vert
\kappa^{-1}(y)\right\vert $ for every $y\in Y$. Figure 1 illustrates this
notion, see Example 2.2(1).%

\bigskip
\noindent\textbf{Lemma 2.1:} Let $(X,\lambda)$ and $(Y,\lambda^{\prime})$ be
weighted sets. \smallskip

1) A mapping $\kappa:X\rightarrow Y$ is a weighted surjection if and only if
there exists a bijection $\kappa^{\prime}:Set(X,\lambda)\rightarrow
Set(Y,\lambda^{\prime})$ such that\footnote{To simplify notation, we write
$\kappa^{\prime}(x,i)$ instead of $\kappa^{\prime}((x,i))$ and we will do the
same in many similar cases.} $\kappa^{\prime}(x,i)=(y,j)$ implies
$\kappa(x)=y.$

2) If there are weighted surjections $\kappa:(X,\lambda)\rightarrow
(Y,\lambda^{\prime})$ and $\alpha:Z=(Z,\boldsymbol{1})\rightarrow
(Y,\lambda^{\prime})$, there exists a weighted surjection $\beta
:Z=(Z,\boldsymbol{1})\rightarrow(X,\lambda)$ such that $\alpha^{\prime}%
=\kappa^{\prime}\circ\beta^{\prime}$, where $\alpha^{\prime},\kappa^{\prime
},\beta^{\prime}$ are related to $\alpha,\kappa,\beta$ as in 1). For each
triple $p,q,r$ such that $\kappa(p)=\alpha(r)=q,$ we can define $\beta$ such
that $\beta(r)=p$.

3) We have $\lambda(X)=\lambda^{\prime}(Y)$ if and only if there exists a set
$S\subseteq X\times Y$ and a weight function $\mu$ on $S$ such that
$\mu(S)=\lambda(X)=\lambda^{\prime}(Y)$ and for every $x\in X$, $\lambda
(x)=\mu(S\cap\{(x,y)\mid y\in Y\})$ and similarly, for every $y\in Y$,
$\lambda^{\prime}(y)=\mu(S\cap\{(x,y)\mid x\in X\})$.

\begin{proof}
Let $(X,\lambda)$ and $(Y,\lambda^{\prime})$ be weighted sets. \smallskip

1) Assume that we have $\kappa:X\rightarrow Y$ and a bijection
$\kappa^{\prime}:Set(X,\lambda)\rightarrow Set(Y,\lambda^{\prime})$ as in the
statement.\ Then $\kappa$ is surjective.\ For each $y\in Y$, the mapping
$\kappa^{\prime}$ induces a bijection $Set(\kappa^{-1}(y),\lambda)\rightarrow
Set(\{y\},\lambda^{\prime})$, hence $\lambda^{\prime}(y)=\lambda(\kappa
^{-1}(y)).$ Hence, $\kappa$ is a weighted surjection.\

Conversely, let $\kappa:X\rightarrow Y$ be a weighted surjection.\ For each
$y$ in $Y$, since $\lambda^{\prime}(y)=\lambda(\kappa^{-1}(y))$, we can define
a bijection: $Set(\kappa^{-1}(y),\lambda)\rightarrow Set(\{y\},\lambda
^{\prime})$.\ The union of all these bijections defines $\kappa^{\prime}$ as desired.

2) Let $\kappa$ and $\kappa^{\prime}$\ be as in 1).\ We have a bijection
$\alpha^{\prime}:Z=Set(Z,\boldsymbol{1})\rightarrow Set(Y,\lambda^{\prime}%
)$.\ We define $\beta^{\prime}:Z\!=\!Set(Z,\boldsymbol{1})\rightarrow
Set(X,\lambda)$ by $\beta^{\prime}:=\kappa^{\prime-1}\circ\alpha^{\prime}$,
from which we get the desired weighted~surjection $\beta:(Z,\boldsymbol{1}%
)\rightarrow(X,\lambda)$ such that $\alpha^{\prime}=\kappa^{\prime}\circ
\beta^{\prime}$. The condition on $p,q,r$ is straightforward to satisfy.

3) Assume we have $\lambda(X)=\lambda^{\prime}(Y)$.\ Consider any bijection
$\mu^{\prime}:Set(X,\lambda)\rightarrow Set(Y,\lambda^{\prime}).$ Then, we
define $\mu(x,y)$ as the cardinality of the set $\{((x,i),(y,j))\mid
\mu^{\prime}(x,i)=(y,j)\}$ if it is not empty. We let $S\subseteq X\times Y$
be the set of all pairs $(x,y)$ such that $\mu^{\prime}(x,i)=(y,j)$ for some
$i,j$.\ We obtain the desired weight function on $S$. The converse is clear.
\end{proof}

In Assertion 3), we call $S$ a \emph{witness of the equality} of weights
$\lambda(X)=\lambda^{\prime}(Y)$. If $X$ and $Y$ are disjoint, we can consider
it as a bipartite graph whose edges are between $X$ and $Y$, and are weighted
by $\mu$. The weight $\lambda(x)$ of vertex $x$ is the sum of the weights of
its incident edges.\ See Example 2.2(3). %

\begin{figure}[b]
\begin{center}
\includegraphics[height=1.3in, width=1.9597in]{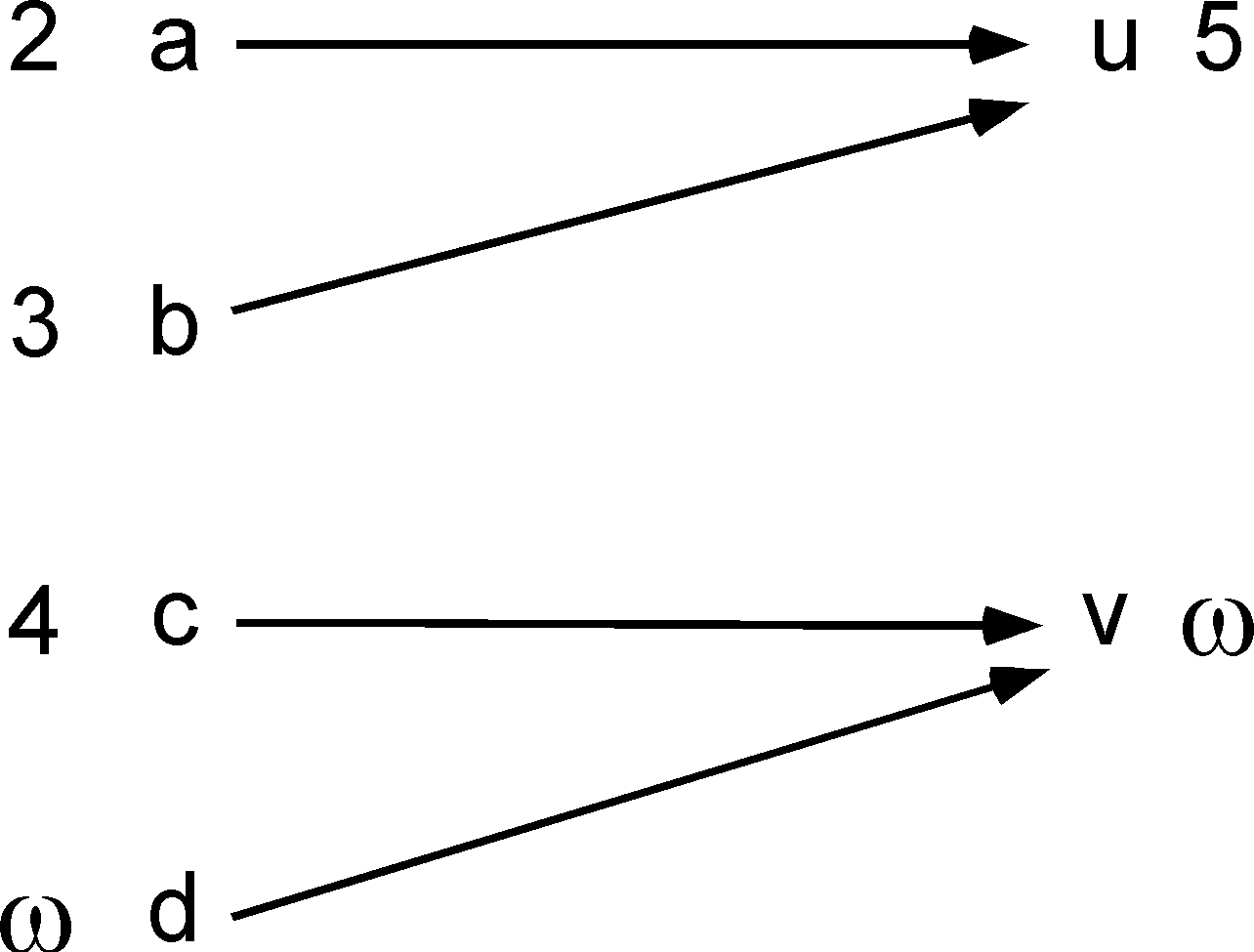}%
\caption{A weighted surjection, see Example 2.2(1).} \vspace*{-1mm}%
\end{center}
\end{figure}

\bigskip
\noindent\textbf{Examples 2.2:} \emph{Weighted relations between weighted  sets.}\smallskip

(1) Let $X$ consist of $a,b,c,d$ of respective weights $2,3,4$ and $\omega$
\ and $Y$ consist of $u$ and $v$ of respective weights 5 and $\omega.$ The
mapping $\kappa$: $a\longmapsto u,b\longmapsto u,c\longmapsto v,d\longmapsto
v$ is a weighted surjection, illustrated in Figure 1. One possible bijection
$\kappa^{\prime}$ satisfying Assertion (1) of Lemma 2.1 is: $(a,i)\longmapsto
(u,i)$ for $i=1,2$, $(b,i)\longmapsto(u,i+2)$ for $i=1,2,3$, $(c,i)\longmapsto
(v,i)$ for $i=1,\dots,4$, $(d,i)\longmapsto(v,i+4)$ for $i\geq1.$

(2) We examplify Assertion (2).\ Let $X,Y,\kappa,\kappa^{\prime}$ be as above
and $Z:=\mathbb{N}_{+}$.\ Let $\alpha:Z\rightarrow Y$ that maps $i\mapsto u$
for $i=1,\dots,5$ and $i\mapsto v$ for $i>5$.\ We obtain $\beta^{\prime}$\ that
maps $i\mapsto(a,i)$ for $i=1,2$, $i\mapsto(b,i-2)$ for $i=3,4,5$,
$i\mapsto(c,i-5)$ for $i=6,\dots,9$, and $i\mapsto(d,i-9)$ for $i>9.$ We deduce
the weighted surjection $\ \beta:Z\ \rightarrow Y.$ This construction works if
we are given $p:=c,q:=v$ and $r:=7$ (cf.\ the last point of Assertion
(2)).\ If $p:=d,$ we can modify accordingly the definition of $\beta^{\prime}$.

(3) To illustrate Assertion (3), we use $X$ consisting of $a,b,c,d$ of
respective weights $\omega,4,2$ and $\omega$ and $Y$ consisting of $u,v,w,x,y$
of respective weights $\omega,4,3,5,1.$ We can take $S$ to consist of $(a,u)$
and $(d,u)$\ of weight $\omega,$ $(a,v),(c,v),$ $(c,w)$\ \ and $(d,y)$ of
weight 1, $(b,v)$ and $(b,w)$ of weight 2 and $(d,x)$ of weight 5. See
Figure~2.\ This is clearly not the unique way to define $S$.

\begin{figure}[h]
\vspace{2mm}
\begin{center}
\includegraphics[height=1.8109in, width=1.7564in]{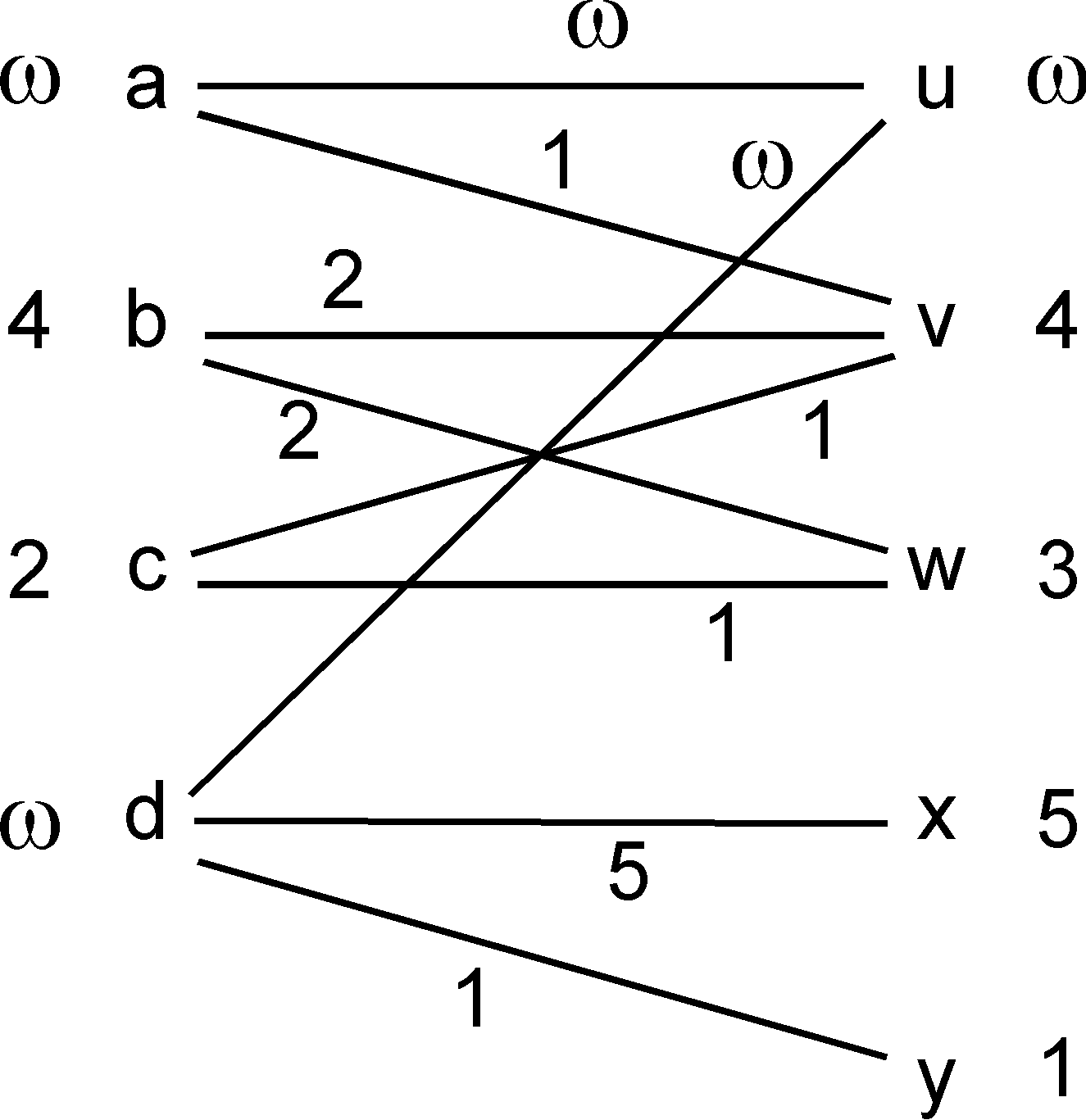}%
\caption{The weighted set S of Example 2.2(3).}%
\end{center}
\end{figure}

If, with the same weighted set $X,$ we take $Y$ consisting of $y_{1}%
,\dots,y_{n},\dots$ \ all of weight\ $\omega$, then we can take $S$ to consist of
$(b,y_{1})$ of weight 4, $(c,y_{1})$ of weight 2 and $(a,y_{i})$ and
$(d,y_{i})$ of weight $\omega$ for all $i$. $\square$

\subsection{Graphs}

By a \emph{graph} we mean an undirected graph, and we call \emph{digraph} a
directed graph, for shortness sake.

A graph is defined as a triple $G=(V,E,Inc)$ where $V$ is the set of vertices,
$E$ is the set of edges, and $Inc$ is the \emph{incidence relation}.\ The
notation $e:x-y$ indicates that edge $e$ links vertices $x$ and $y$, called
its \emph{ends}, equivalently, that $(e,x)$ and $(e,y)$ belong to the set
$Inc\subseteq E\times V$. A triple $(V,E,Inc)$ defines a graph if and only if
$V$ and $E$ are disjoint, $Inc\subseteq E\times V$, and for each $e\in E$,
there are one or two vertices $x\in V$ such that $(e,x)\in Inc$.

A pair in $Inc$ is called a \emph{half-edge}. We write $e:x-x$ if $e$ is a
\emph{loop} \emph{at} $x$, i.e., \emph{incident with} $x$. It is equivalent to
a single half-edge. We denote by $E(x)$ the set of edges incident with $x$,
and by $N(x)$ the set $\{y\in V\mid x-y\}$.\ We have $x\in N(x)$ if there is a
\emph{loop} at $x$. A graph is \emph{simple} if no two edges have the same set
of ends. Hence, it has no two parallel edges.\ It may have loops, where at
most one loop is incident with any vertex.

A \emph{walk} starting at a vertex $x$ is a possibly infinite sequence
$x_{0},e_{1},x_{1},\dots,e_{n},$\ \ $x_{n},\dots$\ such that $x=x_{0}%
,x_{1},\dots,x_{n},\dots$\ are vertices and each $e_{i}$ is an edge whose ends are
$x_{i-1}$ and $x_{i}$. It is a \emph{path} if the vertices $x_{0}%
,\dots,x_{n},\dots$\ are pairwise distinct. In both cases, we say that each
$x_{i}$ \emph{is accessible from} $x_{0}$. Its \emph{length} is the number of
edges. A path $x_{0},\dots,x_{n}$ defines a \emph{cycle} if $n\geq2$ and there
an edge between $x_{0}$ and $x_{n}$. Its length is $n+1$.

A\ \emph{directed graph} (a \emph{digraph}) is defined similarly as a triple
$G=(V,E,Inc)$.\ Its edges are called \emph{arcs}.\ An arc $e$ is directed from
its \emph{tail }$x$ to its \emph{head} $y$, and we denote this by
$e:x\rightarrow y$. Its two \emph{half-arcs} are $(x,e)$ and $(e,y),$ which
encodes the direction of $e$. Hence $Inc\subseteq(V\times E)\cup(E\times V)$.
A triple $(V,E,Inc)$ defines a digraph if and only if $V$ and $E$ are
disjoint, $Inc\subseteq(V\times E)\cup(E\times V)$, and for each $e\in E$,
there are vertices $x,y\in V$ such that $(x,e)$ and $(e,y)$ belong to $Inc$.\

A \emph{loop} $e$ at $x$ has two half-arcs $(x,e)$ and $(e,x)$. A digraph is
\emph{simple} if, for any $x,y$, it has no two arcs from $x$ to $y$. In that
case, $G$ can be defined as a pair $(V,E)$ where $E\subseteq V\times V$. To
simplify notation, we will also define such $G$ as a pair $(V,E)$ where an arc
in $E$ is defined the pair of a tail and a head.

We denote by $E^{+}(x)$ the set of arcs outgoing from $x$, and by $N^{+}(x)$
the set of heads of the arcs in $E^{+}(x).$ We have $x\in N^{+}(x)$ if there
is a loop at $x$.\

A \emph{directed walk} starting at a vertex $x$ is a possibly infinite
sequence $x_{0},e_{1},$\ \ $x_{1},\dots,e_{n},x_{n},\dots$\ \ as above such that
$x=x_{0}$ and $e_{i}:x_{i-1}\rightarrow x_{i}$ for each $i$. Without ambiguity
unless it is reduced to the single vertex $x_{0}$, it can be specified as the
sequence of arcs $e_{1},\dots,e_{n},\dots$. Its \emph{length} is its number of
arcs.\ It is a \emph{directed path} if the vertices $x_{0},\dots,x_{n},\dots$\ are
pairwise distinct. We say that each $x_{i}$ \emph{is accessible from} $x_{0}$.
A digraph is \emph{strongly connected} if any two vertices are accessible from
each other. A directed path $x_{0},\dots,x_{n}$ defines a \emph{directed cycle}
if $n\geq1$ and there is an arc $x_{n}\rightarrow x_{0}$.

A \emph{rooted digraph} $G$\ has a distinguished vertex called the
\emph{root}, denoted by $rt_{G}$, from which all vertices are accessible by a
directed path. We denote by $G/x$ the induced subgraph of $G$\ whose vertices
are those accessible from $x$ by a directed path. (The study of rooted trees
uses this notion with same notation). We define $x$ as its root.

We denote by $Und(G)$ the graph underlying a digraph $G$: each arc
$e:x\rightarrow y$ of $G$ is made into\ an edge $e:x-y$ of $Und(G)$. Hence, it
need not be simple if $G$\ is.

We write $V_{G},E_{G},E_{G}(x),E_{G}^{+}(x),N_{G}^{+}(x),Inc_{G}$\ etc. to
specify, if necessary, the relevant graph or digraph $G$.

For graphs and digraphs, \emph{inclusion }is denoted by $\subseteq$,
\emph{i.e.} $G=(V,E,Inc)\subseteq H=(V^{\prime},E^{\prime},Inc^{\prime})$\ if
and only if $V\subseteq V^{\prime},E\subseteq E^{\prime}$\ and $Inc\subseteq
Inc^{\prime}$. \emph{Induced inclusion} denoted by $\subseteq_{i}$ holds if,
furthermore, $E$ is the set of edges or arcs of $E^{\prime}$\ whose ends,
tails and heads are in $V$. We write then $G=H[V]$.

A \emph{homomorphism} $\eta:G\rightarrow H$ of graphs or of digraphs maps
$V_{G}$\ to $V_{H}$, $E_{G}$ to $E_{H}$, $Inc_{G}$ to $Inc_{H}$ and preserves
incidences in the obvious way. It maps loops to loops but can map a nonloop
edge or arc to a loop.\ If $G$ and $H$ are rooted, it maps the root of $G$ to
that of $H$.\ Isomorphism is denoted by $\simeq$ and the isomorphism class of
$G$ by $[G]_{\simeq}.$

If $\eta:G\rightarrow H$ is a homomorphism of graphs or of digraphs, we make
$G$ into a labelled graph or digraph $G_{\eta}$ by equipping each vertex, edge
or arc $x$ by the \emph{label} $\eta(x)$.\ Formally, $G_{\eta}=(V,E,Inc,\eta
).$ Hence, this labelled graph encodes $G$ and $\eta$.\ We will use this
notion when $H$ is finite. Other graph labellings will be defined at the
relevant places.

We extend the notion of a homomorphism by allowing "forgetful" operations. A
homomorphism $Und(G)\rightarrow H$\ where $G$ is directed and $H$ is not is
also considered as a homomorphism $G\rightarrow H$.\ Similar conventions
concern labelled graphs.

\bigskip
\noindent\textbf{Definition 2.3:} \emph{Quotient graphs and digraphs} \smallskip

(a) An equivalence relation $\sim$ on a graph $G=(V,E,Inc)$ is an equivalence
relation on $V\cup E$ such that each equivalence class is either a set of
vertices or a set of edges, and, if $e$ and $e^{\prime}$ are equivalent
edges\footnote{An edge $e:x-y$ and a loop $f:z-z$ are equivalent if $x\sim
y\sim z$.\ }, then each end of $e$ is equivalent to an end of $e^{\prime}$.

(b) The \emph{quotient graph} $G/\sim$\ is then defined as $(V/\sim
,E/\sim,Inc_{G/\sim})$ such that $([e]_{\sim},[v]_{\sim})\in$ $Inc_{G/\sim}$
if and only if $(e^{\prime},v^{\prime})\in Inc$\ for some $e^{\prime}\sim e$
and $v^{\prime}\sim v$.

(c) The definition is similar for a digraph $G$: we require that if $e$ and
$f$ are equivalent arcs, then the tail (resp.\ the head) of $e$ is equivalent
to that of $f$. The quotient digraph is defined as for graphs.

(d) In both cases, we have a surjective homomorphism $\eta_{\sim}:G\rightarrow
G/\sim$ that maps a vertex, an edge or an arc to its equivalence class. An
edge $e:x-y$ is mapped to a loop in $G/\sim$ if $x\sim y$. The same holds for arcs.

\bigskip

\noindent\textbf{Remark 2.4: }An equivalence relation $\sim$\ on the vertex
set $V$\ of $G=(V,E,Inc)$ can be extended to edges or arcs as follows: two edges are equivalent if and only if each end of one is equivalent to some
end of the other;  two arcs are equivalent if and only if their tails are equivalent and so are
their heads.\

A notion of  quotient graph of a digraph follows then by Definition  2.3.\ $\square$

\subsection{Trees}

A \emph{tree} is a nonempty simple connected graph without loops or cycles. We
call \emph{nodes} its vertices.\ This convention is useful in the frequent
case where we discuss simultaneously a graph and a tree constructed from it.\

The set of nodes of a tree $T$ is denoted by $N_{T}$.\ A \emph{subtree} of a
tree $T$ is a connected subgraph, hence, it is a tree.\ A tree has
\emph{(locally) finite degree} if each node has finite degree.\ It has
\emph{bounded degree} if the degrees of its nodes are bounded by a same integer.\

A \emph{rooted tree} is a tree $T$\ equipped with a distinguished node $r$
called its \emph{root}.\ We denote it sometimes by $T_{r}$ to specify
simultaneously the root and the underlying undirected tree $T.$\ In a way
depending on $r$, we direct its edges so that every node is accessible from
$r$ by a directed path. If $x\rightarrow y$ in $T_{r}$, then $y$ is called a
\emph{son} of $x$, and $x$ is the (unique) \emph{father} of $y$.\ The
\emph{depth} of a node is its distance to the root (the root has depth 0). The
\emph{height} of a rooted tree is the least upper-bound of the depths of its
nodes. A \emph{star} is a rooted tree of height 1.

Let $R$ be a rooted tree; its root is $rt_{R}$.\ By \emph{forgetting} its root
and making its arcs undirected, we get a tree $T:=Unr(R)$. Hence,
$R=T_{rt_{R}}.$ If $x$ is a node of $R$, then the digraph $R/x$ is a\ rooted
tree with root $x$, called the \emph{subtree of R issued from} $x$.\ It is
induced on the set of nodes accessible from $x$ by a directed path.\ If
$i\in\mathbb{N}$, the \emph{truncation at depth} $i$ \emph{of} $R,$ denoted by
$R\upharpoonright i$, is the induced subgraph of $R$\ whose nodes are at
\emph{distance} at most $i$ \ from the root, that is, are accessible from it
by a (unique) directed path of length at most $i$. It is a rooted tree with
the same root as $R$ and $R\upharpoonright0$ is the tree reduced to the root
$rt_{R}$.

A \emph{homomorphism of rooted trees}: $R\rightarrow R^{\prime}$ is a
homomorphism of directed graphs that maps $rt_{R}$ to $rt_{R^{\prime}}.$ A
homomorphism from\ a rooted tree $R$ to a tree $T$\ is defined as a
homomorphism of trees: $Unr(R)\rightarrow T$.

\medskip \smallskip
\noindent\textbf{Lemma 2.5}: An isomorphism of rooted trees $\eta:R\rightarrow
R^{\prime}$ induces, for each $u\in N_{R}$, an isomorphism: $R/u\rightarrow
R^{\prime}/\eta(u)$ and, in particular, a bijection $N_{R}^{+}(u)\rightarrow
N_{R^{\prime}}^{+}(\eta(u))$ such that $R/v\simeq R^{\prime}/\eta(v)$ if $v\in
N_{R}^{+}(u).$

\section{Unfoldings of directed graphs}

Certain abstract programs can be formalized as \emph{transition systems} that
are finite directed graphs with information attached to vertices and arcs. A
vertex of the graph is a \emph{state} of the corresponding transition system.
An \emph{initial state} $r$ is specified.\ The tree of directed walks starting
at $r$ collects all possible computations of the corresponding transition
system.\ It is called its \emph{unfolding} \cite{Arn,Cou9,CouWal}.

We will consider unfoldings from a graph theoretical point of view, without
offering any new application to semantics. We will generalize them and define
unfoldings of digraphs whose arcs have \emph{weights}.\ In particular, an arc
of weight $\omega$\ with head $y$ unfolds into countably many arcs whose heads
yield $y$ by the unfolding homomorphism. We will obtain a notion of regular
tree that generalizes the classical one in that the nodes can have infinite
outdegrees.\ These trees are the unfoldings of finite, weighted and rooted digraphs.

In this section, all trees are rooted and thus directed in a canonical
way.\ In \cite{Arn,Cou9,CouWal}\ the \emph{unfolding} of a rooted digraph
$G$\ is what we will call its \emph{complete unfolding}.\ We will call
\emph{unfolding} of such a digraph $G$\ a rooted digraph that lies inbetween,
via surjective homomorphisms, the digraph $G$ and its complete unfolding,
denoted by $\mathit{Unf}(G)$. This terminology is thus similar to that
concerning \emph{coverings} and \emph{universal coverings}.

The main contributions of this section are the use of possibly infinite
weights, the decidablility of isomorphism in Theorem 3.14, and two theorems
similar to those by Norris and Leighton for universal coverings of finite
undirected graphs, see Theorems 3.20 and 3.22.

Equality of trees and digraphs will be understood in the strict sense: same
nodes or vertices, and same arcs.\ Equality via an isomorphism is specified
explicitely in statements and proofs, and denoted by~$\simeq$.

\subsection{Weighted directed graphs and their unfoldings}

We will equip digraphs with weights in $\mathbb{N}_{+}\cup\{\omega\}.$ We
recall from Section 2.2 that a digraph can be defined as a pair $(V,E)$ where
each arc $e$ is an ordered pair of vertices.

\bigskip
\noindent\textbf{Definition 3.1:} \emph{Weighted digraphs.} \smallskip

A \emph{weighted digraph} is a triple $G=(V,E,\lambda)$ such that $(V,E)$ is a
digraph whose set of arcs $E$ is \emph{weighted}, that is, equipped with a
\emph{weight function} $\lambda:E\rightarrow\mathbb{N}_{+}\cup\{\omega\}.$ We
denote by $\overline{E^{+}}(u)$\ the weighted set ($E^{+}(u),\lambda)$ and by
$\overline{N^{+}}(u)$ the weighted set ($N^{+}(u),\lambda^{\prime})$ such that
$\lambda^{\prime}(v)=\Sigma\{\lambda(e)\mid e:u\rightarrow v\}$.

A digraph\footnote{\emph{Digraph} will mean "without weights" and possibly
with parallel arcs.} is a weighted digraph whose arcs have all the weight 1.

A weighted digraph is \emph{simple} or \emph{rooted} if the underlying digraph
is. If $x$ is a vertex of a weighted digraph $G,$ then $G/x$ (cf.\ Section
2.2) is a rooted and weighted digraph with root $x$.\ If $G$ is strongly
connected, the digraphs $G/x$ have all the same vertices and arcs as $G$.

In the special case where $G$ is simple digraph, then $\overline{E^{+}%
}(u)=(E^{+}(u),\boldsymbol{1})$, $\overline{N^{+}}(u)=(N^{+}(u),\boldsymbol{1}%
)$ and the head mapping is a bijection $E^{+}(u)\rightarrow N^{+}(u)$.
$\square$

We can handle parallel arcs by means of weights.\ That is, an arc $(x$,$y)$ of
weight $\lambda(x,y)>1$ encodes $\lambda(x,y)$ parallel arcs from $x$ to $y$.

\bigskip
\noindent\textbf{Definition 3.2:} \emph{Unfolding} \smallskip

Let $H$ and $G$ be rooted and weighted digraphs.

(a) A surjective homomorphism $\eta:G\rightarrow H$ is an \emph{unfolding }of
$H$ if it induces a weighted surjection $E_{G}\rightarrow E_{H}.$ In
particular, if $u\in V_{G}$\ and $\eta(u)=x$, then $\eta$ induces a weighted
surjection $\overline{E_{G}^{+}}(u)\rightarrow\overline{E_{H}^{+}}(x).$ If $G$
and $H$ are simple digraphs, then $\eta$ induces a bijection $E_{G}%
^{+}(u)\rightarrow E_{H}^{+}(x)$\ and a bijection $N_{G}^{+}(u)\rightarrow
N_{H}^{+}(x)$.

We will also say that $G$ is an \emph{unfolding }of $H$ or that $H$
\emph{unfolds into} $G$. From the accessibility condition in the definition of
a rooted digraph, unfoldings only concern connected graphs. They are called
\emph{op-fibrations} by Boldi and Vigna \cite{BolVig}.

(b) An unfolding $G\rightarrow H$ is \emph{complete} if $G$ is a rooted tree
without weights (equivalently, all weights are 1). We will also say that $G$
is \emph{a complete unfolding} of $H$ or that $H$ \emph{unfolds completely
into}~$G$.

\bigskip
\noindent\textbf{Examples 3.3:} (1) A loop of weight 1 (resp.\ 2) unfolds
completely into an infinite directed path (resp. into the infinite binary
rooted tree).

(2) An arc $x\rightarrow y$ of weight $\omega$ such that $x$\ is taken as root
unfolds (not completely) into any finite star, where at least one arc has
weight $\omega$. It unfolds completely into a star $S_{\omega}$,
\emph{i.e.,}\ any tree whose root has $\omega$\ sons that are
leaves\footnote{Any two such trees are isomorphic.\ By thinking of trees up to
isomorphism, which is adequate since any two complete unfoldings of a rooted
digraph are isomorphic, we can also write \emph{the} star $S_{\omega}$.}.\ If
in addition, there is a loop $y\rightarrow y$ of weight 1, this rooted and
weighted digraph unfolds completely into the union of $\omega$\ infinite
directed paths with the same origin, that are otherwise disjoint.

\bigskip

\noindent\textbf{Proposition 3.4:} (1) If $\eta:G\rightarrow H$ and
$\kappa:H\rightarrow K$ are unfoldings, then $\kappa\circ\eta$ is an unfolding
$G\rightarrow K$.

(2) If $\eta:G\rightarrow H$ is an unfolding, $u\in V_{G}$ and $x=\eta(u)$,
then $\eta$ is an unfolding\footnote{This is a short expression for "the
restriction of $\eta$ to $G/u$ is an unfolding $G/u\rightarrow H/x$".\ Similar
shortenings will be used at other places.}$G/u\rightarrow H/x$.

\begin{proof}
(1) The composition $\kappa\circ\eta$ induces a weighted surjection
$E_{G}\rightarrow E_{K}$ as $\kappa$ and $\eta$ do the same $E_{H}\rightarrow
E_{K}$ and $E_{G}\rightarrow E_{H}$ respectively. This observation proves the assertion.

(2) Clear from Definition 3.2.
\end{proof}

The following theorem implies that every rooted and weighted digraph $H$\ has,
up to isomorphism, a unique complete unfolding.

\bigskip
\noindent\textbf{Theorem 3.5:} Let $H$ be a rooted and weighted digraph. \smallskip

\noindent 1) $H$ has a complete unfolding.

\noindent 2) If $\beta:T\rightarrow H$\ is a complete unfolding, then:
\begin{quote}
(U) For every unfolding $\kappa:G\rightarrow H$, there is a complete unfolding
$\eta:T\rightarrow G$ such that $\beta=\kappa\circ\eta$.
\end{quote}
3) Any two complete unfoldings of $H$ are isomorphic.

\noindent 4) If $\beta:T\rightarrow H$ is an unfolding such that Condition (U) holds,
then $T$ is a rooted tree, hence a complete unfolding of $H$.

\medskip
Properties 2) and 4) show that the complete unfoldings of $H$ are
characterized by a universal property in the sense of Category Theory. One can
speak of \emph{the} complete unfolding of \emph{H}, well-defined up to
isomorphism. The following notion helps to approximate, level by level, a
complete unfolding. The \emph{height} of a rooted tree is the least
upper-bound of the distances of its nodes to the root.

\bigskip
\noindent\textbf{Definition 3.6}: \emph{Depth-limited unfoldings}. \smallskip

Let $A$ be a rooted tree of \emph{height} at most $i$ (cf.\ Section 2.3) and
$H$ be a rooted and weighted digraph. An $i$-\emph{unfolding} $\eta
:A\rightarrow H$ is a homomorphism (it is not necessarily surjective)
satisfying the following condition:
\begin{quote}
For every node $u$ of $A$ at distance at most $i-1$ from the root, if
$\eta(u)=x$ and $e$ is an arc of $H$ with tail $x$, then $\left\vert \eta
^{-1}(e)\cap E_{A}^{+}(u)\right\vert =\lambda_{H}(e).\ \ \ \ \square$
\end{quote}
The complete unfolding of a rooted unweighted digraph $H$\ can be constructed
as the tree of finite walks starting from the root.\ As weights in digraphs
represent parallel arcs, this construction must be adapted.\ This is the
purpose of the following definition, that replaces parallel arcs with sets of
parallel ones.

\bigskip
\noindent\textbf{Definition 3.7:}\emph{ The expansion of a weighted digraph.} \smallskip

Let $H=(V,E,\lambda)$ be a weighted digraph.\ Its \emph{expansion} is the
digraph $Exp(H)=(V,Set(E,\lambda))$ having the arc $(e,i):x\rightarrow y$ if
$e:x\rightarrow y$ in $H$ and $(e,i)\in Set(E,\lambda)$.\ (The mapping $Set$
is defined in Section 2.1) The digraph $Exp(H)$\ is infinite if some arc has
weight $\omega$, and/or, of course, if $V$ is infinite. If $H$ has a root,
then $Exp(H)$ has the same root.

\medskip
We now prove Theorem 3.5.

\begin{proof}
Let $H$ be a rooted and weighted digraph. \smallskip

1) The rooted digraph $Exp(H)$ is an unfolding of $H$, and we denote by
$\varepsilon$ the corresponding homomorphism $Exp(H)\rightarrow H$.\ By
Proposition 3.4, we need only construct a complete unfolding $T$ of $Exp(H)$.
We define it as the tree of directed walks in $Exp(H)$ that start from
$rt_{H}$, the common root of $Exp(H)$ and $H$. The father of a node
$(e_{1},\dots,e_{p})$ is $(e_{1},\dots,e_{p-1})$.

Let $\alpha:T\rightarrow Exp(H)$ map $(e_{1},\dots,e_{p})$ to the head of
$e_{p}$; if $p=0$, then $(e_{1},\dots,e_{p})$ is the empty walk, mapped to
$rt_{H}$; the arc from $(e_{1},\dots,e_{p-1})$ to $(e_{1},\dots,e_{p})$ is mapped
to $e_{p}$. We say that this arc of $T$ is \emph{of type} $e_{p}$.

Then $\alpha$ is a complete unfolding $T\rightarrow Exp(H)$ and $\beta
:=\varepsilon\circ\alpha$ yields a complete unfolding $T\rightarrow H$.\ We
will denote $T$\ by $\mathit{Unf}(H)$. Note that $\mathit{Unf}(H)$ is a
concrete tree made of walks in $Exp(H)$.

If $H$ is a rooted tree, then $Exp(H)\simeq H$ and $\alpha$ and $\beta$ are
isomorphisms as one checks easily. \smallskip

2) We let $\beta:\mathit{Unf}(H)\rightarrow H$\ be the particular complete
unfolding constructed in 1) and $\kappa:G\rightarrow H$ be any unfolding.\

By induction on $i$, we construct for each $i,$ an $i$-unfolding $\eta
_{i}:\mathit{Unf}(H)\upharpoonright i\rightarrow G$ such that $\kappa\circ
\eta_{i}$\ is the restriction of $\beta$ to $\mathit{Unf}(H)\upharpoonright i$
(the restriction of $\mathit{Unf}(H)$ to nodes at distance at most $i$ from
the root) in such a way that $\eta_{i+1}$ extends $\eta_{i}$.\ The union of
the mappings $\eta_{i}$ will be a complete unfolding $\eta:\mathit{Unf}%
(H)\rightarrow G$ such that $\beta=\kappa\circ\eta$.

We construct $\eta_{i+1}$ from $\eta_{i}$ as follows.\ Let $u=(e_{1}%
,\dots,e_{i})\in N_{T\upharpoonright i}$ be mapped to $w\in V_{G}$ by $\eta_{i}%
$.\ There is a weighted surjection $\mu_{u}:N_{\mathit{Unf}(H)}^{+}%
(u)\rightarrow\overline{N_{G}^{+}}(w)$\ such that $\kappa\circ\mu_{u}$ is the
restriction of $\beta$ to $N_{\mathit{Unf}(H)}^{+}(u)$.\ Its existence follows
from Lemma 1.1(2), as $N_{\mathit{Unf}(H)}^{+}(u)$\ is a set, equivalently,
the weighted set $\overline{N_{\mathit{Unf}(H)}^{+}}(u)=(N_{\mathit{Unf}%
(H)}^{+}(u),\boldsymbol{1})$.\ Then, we let $\eta_{i+1}$ be the union of
$\eta_{i}$ and all such mappings $\mu_{u}$ for all nodes $u$ of $\mathit{Unf}%
(H)$ at depth $i$. \smallskip

To prove 3) and to complete the proof of 2), we let $\kappa:G\rightarrow H$ be
a complete unfolding, hence, $G$ is a tree.\ Then, the complete unfolding
$\eta:\mathit{Unf}(H)\rightarrow G$ is an isomorphism.\ Hence, any two
complete unfoldings of $H$ are isomorphic and 2) holds for any complete
unfolding $\beta$ of $H.$ \smallskip

4) Let $\beta:T\rightarrow H$ be an unfolding such that Condition (U)
holds.\ Let $G$ be a complete unfolding of $H$. There is an unfolding
$\gamma:T\rightarrow G$.$\ $Since $G$ is a tree, $T$ is also a tree, hence a
complete unfolding of $H$.
\end{proof}

We will reserve the notation $\mathit{Unf}(H)$ to the complete unfolding
defined as a tree of walks in $H$.\ It is defined in \cite{Arn,Cou9,CouWal},
but not characterized by a universality property.

\subsection{Complete unfoldings and regular trees}

The notion of an infinite regular tree is important in applications to
semantics, in particular because the complete unfolding of a finite transition
system is regular \cite{Arn,Cou83,CouHdBk}, and more generally for the monadic
second-order logic of infinite structures, see \cite{Cou9,CouWal}. We will
consider regular trees that are complete unfoldings of finite digraphs.

A graph, a digraph or a tree can have labels attached to its vertices, nodes,
edges or arcs.

\bigskip
\noindent\textbf{Definition 3.8:} \emph{Regular trees}. \smallskip

A rooted, possibly labelled, tree $T$ is \emph{regular\footnote{Slightly
different notions of regular trees are studied in \cite{Cou83,Cou9,CouWal}.
However, they have in common the finiteness of the set of subtrees $T/x$ up to
isomorphism.}} if it has finitely many subtrees $T/x$ (inheriting the possible
labels of $T$), \emph{up to isomorphism}, which we will denote by
\emph{u.t.i.}, that is, if the set of isomorphism classes $\{[T/x]_{\simeq
}\mid x\in N_{T}\ \}$ is finite.\ In the latter case, its cardinality is the
\emph{regularity index} of $T,$ denoted by $Ind(T)$. If $T$ is regular, each
subtree $T/x$ is regular of no larger index because $(T/x)/y=T/y$ for
$y\leq_{T}x$ (which means that $x$ is on the directed path from the root to
$y)$. $\square$

\medskip
Every finite tree is regular.\ A rooted tree of height 1 (a \emph{star}) is
regular of index 2. We will prove that the complete unfolding of a finite,
rooted and weighted digraph $H$, that may have infinite weights, is regular of
index at most\ $\left\vert V_{H}\right\vert $ and has a canonical
"factorization" in terms of a finite weighted digraph analoguous to the
minimal automaton of a regular language.

\medskip
Let $G$\ be a weighted digraph. Let $\approx$ be the equivalence
relation\footnote{If $G$ is rooted so that $\mathit{Unf}(G)$ is defined, an
equivalent expression of $x\approx y$ is $\mathit{Unf}(G)/u\simeq
\mathit{Unf}(G)/v$ where $u,v$\ are nodes of $\mathit{Unf}(G)$ such that
$\alpha(u)=x$, $\alpha(v)=y$ and $\alpha:\mathit{Unf}(G)\rightarrow G$\ is the
complete unfolding.\ } on $V_{G}$\ such that $x\approx y$ if and only if
$\mathit{Unf}(G/x)\simeq\mathit{Unf}(G/y).$ According to the definitions of
Section 1.2, the quotient $H:=G/\approx$ is the simple digraph defined as
follows: $V_{H}:=\{[x]_{\approx}\mid x\in V_{G}\}$ and $E_{H}:=\{([x]_{\approx
},[y]_{\approx})\mid G$ has an arc $x\rightarrow y$\}.\ If $G$ is rooted, we
take $rt_{H}:=[rt_{G}]_{\approx}.$ We have a surjective homomorphism
$\eta:G\rightarrow H.$ We now define weights on the arcs of $H$.\ If $e$ is an
arc $x\rightarrow y$ of $G$, we define $\lambda_{G}^{\prime}(e):=\Sigma
\{\lambda_{G}(f)\mid f:x\rightarrow z$ is an arc of $G$ for some $z\approx
y\}$.

\bigskip
\noindent\textbf{Lemma 3.9:} Let $x,x^{\prime}$ be vertices of $G$ such that
$x\approx x^{\prime}$. \smallskip

(1) If there is an arc $x\rightarrow y$ for some $y$, then there is one
$x^{\prime}\rightarrow y^{\prime}$ such that $y^{\prime}\approx y$.

(2) If $e$ is an arc $x\rightarrow y,$ $e^{\prime}$ is an arc $x^{\prime
}\rightarrow y^{\prime}$ such that $y^{\prime}\approx y$, then $\lambda
_{G}^{\prime}(e)=\lambda_{G}^{\prime}(e^{\prime}).$

\begin{proof}
(1) Assume $x\approx x^{\prime}$.\ \ Let $\alpha:\mathit{Unf}(G/x)\rightarrow
G/x$ by the unfolding homomorphism, mapping the root $\overline{x}$ of
$\mathit{Unf}(G/x)$ to $x$, and similarly $\alpha^{\prime}:\mathit{Unf}%
(G/x^{\prime})\rightarrow G/x^{\prime}$ mapping the root $\overline{x^{\prime
}}$ of $\mathit{Unf}(G/x^{\prime})$ to $x^{\prime}$. Let $\mu$\ be an
isomorphism $\mathit{Unf}(G/x)\simeq\mathit{Unf}(G/x^{\prime}).$ It maps
$\overline{x}$ to $\overline{x^{\prime}}$.\

Let $e$ be an arc $x\rightarrow y$ of $G$.\ There is $u$ in $\mathit{Unf}%
(G/x)$ such that $\alpha(u)=y$ and $\overline{x}\rightarrow u$ in
$\mathit{Unf}(G/x).$\ Let $y^{\prime}:=\alpha^{\prime}(\mu(u)).$ We have
$\overline{x^{\prime}}\rightarrow\mu(u)$, hence an arc $x^{\prime}\rightarrow
y^{\prime}$\ in $G$.

We have $\mathit{Unf}(G/y)\simeq\mathit{Unf}(G/x)/u\simeq\mathit{Unf}%
(G/x^{\prime})/\mu(u)\simeq\mathit{Unf}(G/y^{\prime}).\ $Hence, $y^{\prime
}\approx y$.

(2) If $e$ is an arc $x\rightarrow y$ of $G$\ and with the same notation as in
(1), we observe that, since $\alpha$ is an unfolding, $\lambda_{G}^{\prime
}(e)$\ is the number of sons $u$ of the root of the tree $\mathit{Unf}(G/x)$
such that $\mathit{Unf}(G/x)/u\simeq\mathit{Unf}(G/y).\ $Then, if $e^{\prime}$
is an arc $x^{\prime}\rightarrow y^{\prime}$, we have similarly that
$\lambda_{G}^{\prime}(e^{\prime})$\ is the number of sons $u^{\prime}$ of the
root of the tree $\mathit{Unf}(G/x^{\prime})$ such that $\mathit{Unf}%
(G/x^{\prime})/u^{\prime}\simeq\mathit{Unf}(G/y^{\prime}).$ Since $y^{\prime
}\approx y$, we have $\lambda_{G}^{\prime}(e)=\lambda_{G}^{\prime}(e^{\prime
}).$
\end{proof}

 \smallskip
 \noindent\textbf{Definition 3.10:} \emph{The canonical quotient of a rooted
and weighted digraph}. \smallskip

Let $G$ be a rooted and weighted digraph and $H:=G/\approx$ as above. The
mapping $\eta$\ such that $\eta(x):=[x]_{\approx}$ if $x\in V_{G}$ and
$\eta(e):=([x]_{\approx},[y]_{\approx})$\ if $e$ is an arc $x\rightarrow y$ of
$G$ is a homomorphism $G\rightarrow H$ \ that is surjective by Lemma 3.9(1).

We define a weight function on $H$ by $\lambda_{H}([x]_{\approx},[y]_{\approx
}):=\lambda_{G}^{\prime}(e)$ for any arc $e:x\rightarrow z$ of $G$ such that
$z\approx y$. It is well-defined by Lemma 3.9(2).

Furthermore, if $G$ is vertex-labelled, then $x\approx y$ implies that $x$ and
$y$ have same label.\ The quotient digraph $H:=G/\approx$ is vertex-labelled
and the homomorphism $\eta:G\rightarrow H$\ preserves labels. $\ $

We define the \emph{size} $\left\vert G\right\vert $\ of a digraph $G$\ as
$\left\vert V_{G}\right\vert \ +\left\vert E_{G}\right\vert $. $\square$

\bigskip
\noindent\textbf{Proposition 3.11:} (1) The homomorphism $\eta:G\rightarrow
G/\approx$ is an unfolding. \smallskip

(2) If $G$ is finite, then $G/\approx$ is, up to isomorphism, the unique
rooted and weighted digraph\ of minimal size of which $G$ is an unfolding.

\begin{proof}
(1) As observed in Definition 3.10, the homomorphism $\eta:G\rightarrow
G/\approx$ is surjective.\ It is an unfolding by the proof of Lemma 3.9(2).

(2) If $G$ is finite, then $\left\vert G\right\vert \geq\left\vert
H\right\vert $. If $\alpha:G\rightarrow K$ is an unfolding, then there is an
unfolding $\beta:K\rightarrow H$ that we define as follows:

$\beta(u):=[x]_{\approx}$ where $u\in V_{K}$ and $\alpha(x)=u;$

$\beta(e):=([x]_{\approx},[y]_{\approx})$ where $e:u\rightarrow v$ is an edge
of $K$, $\alpha(x)=u$, $\alpha(y)=v$ and $x\rightarrow y$ is an edge of $G$.

It is easy to see that $\beta$ is an unfolding. Hence, $\left\vert
K\right\vert \geq\left\vert H\right\vert $. \ If $\left\vert K\right\vert
=\left\vert H\right\vert $, it is an isomorphism.
\end{proof}

 \smallskip
\noindent\textbf{Example 3.12: \ }Figure 3 shows to the left a weighted
digraph\ $G$\ with vertex set $\{s,u,v,w,x,y,z\}$.\ The weights that are not
shown are equal to 1.\ We have $s\approx w\approx y$ and $u\approx v\approx
x\approx z$. The quotient digraph $G/\approx$ \ is shown to the right.
$\square$%

\begin{figure}[h]
\vspace*{1mm}
\begin{center}
\includegraphics[height=1.702in,width=2.7121in]{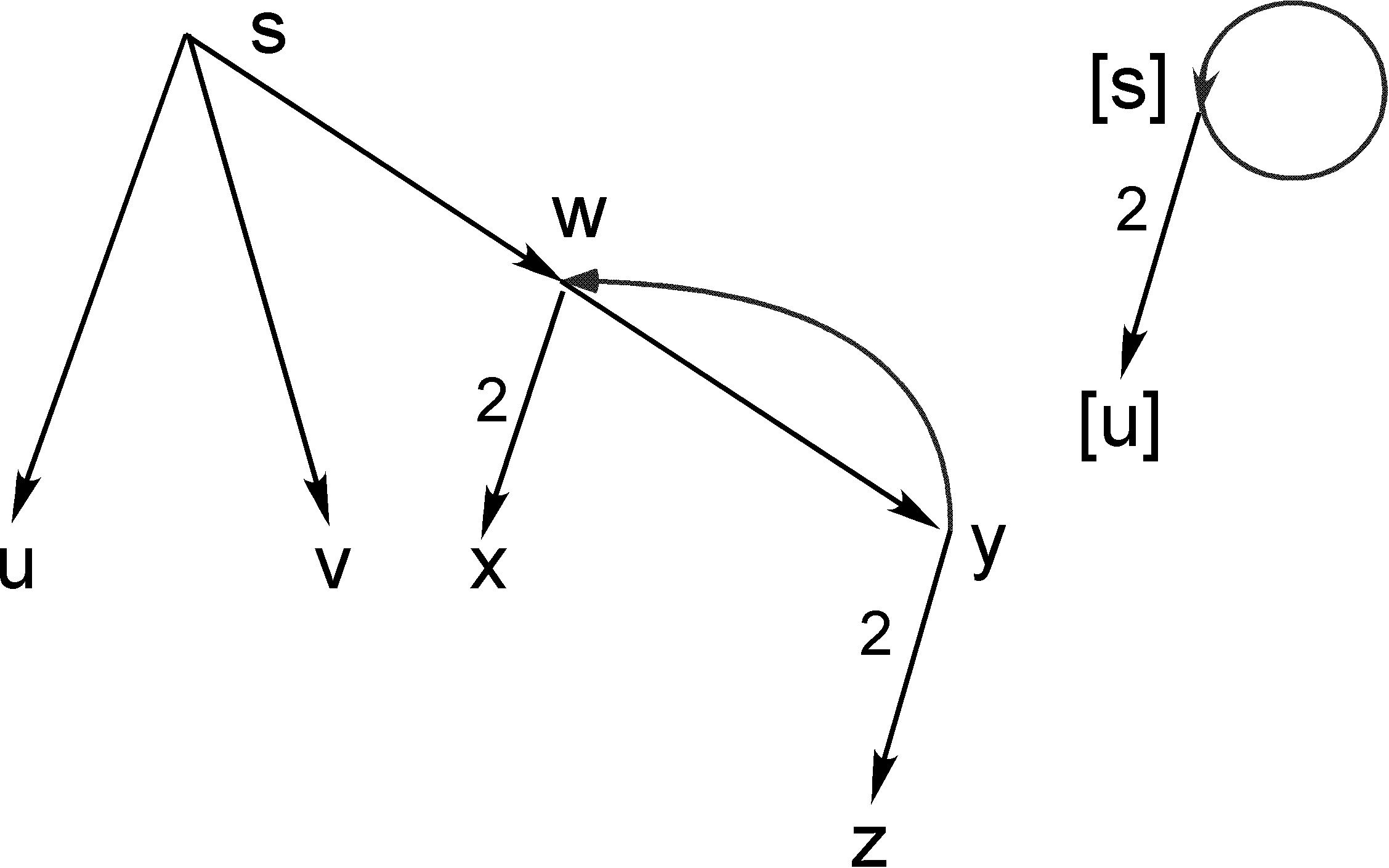}%
\caption{A digraph $G$ and its quotient $G/\approx$, cf.\ Example 3.12.}%
\end{center}\vspace*{-1mm}
\end{figure}

We now consider the case where $R$\ is a regular tree.\ Theorem 3.14\ will
prove that $\approx$ and $G/\approx$ are computable if $G$ is finite.

\bigskip
\noindent\textbf{Theorem 3.13:} (1) A rooted tree $T$\ is regular of index at
most $p$ if it is the complete unfolding of a finite, rooted and weighted
digraph having $p$ vertices.

(2) Conversely, a regular tree $T$\ is the complete unfolding of a unique
rooted and weighted simple digraph having $Ind(T)$ vertices.

(3) If $\eta:T\rightarrow H$ is a complete unfolding of a rooted and weighted
digraph having $p$ vertices ($p\in\mathbb{N}_{+}$), then the labelled rooted
tree $T_{\eta}$ (where each node $u$ is labelled by $\eta(u)$) is regular of
index at most $p$.

\begin{proof}
(1) Let $\eta:T=\mathit{Unf}(H)\rightarrow H$ be the unfolding homomorphism
where $H$ is a rooted and weighted digraph having $p$ vertices. If $u,v$ $\in
N_{T}$ and $\eta(u)=\eta(v)=x$, then $T/u\simeq T/v$ because these two trees
are complete unfoldings of $H/x$ by Proposition 3.4(2). It follows that $T$ is
regular and its index is at most the number of vertices of $H$.

(2) Conversely, let $T$\ be a regular tree of index $p$. Let $\approx$ be the
equivalence relation on $N_{T}$\ such that $u\approx v$ if and only if
$T/u\simeq T/v.$ We have $T/u\simeq\mathit{Unf}(T/u).$\ The quotient
construction of Definition 3.10 shows that $T$ is the complete unfolding of
the finite, rooted and weighted digraph $T/\approx$, that has $p$
vertices.$\ $

(3) Easy extension of (1).
\end{proof}

Finite, rooted and weighted digraphs can be used as finite descriptions of
regular trees. Although an arc of weight $p\in\mathbb{N}_{+}$ can be replaced
(cf.\ Definition 3.7) by $p$ parallel arcs and a loop of weight $q\in
\mathbb{N}_{+}$ by $q$ loops, the use of weights gives more concise
descriptions. Furthermore, the weight $\omega$\ makes it possible to describe
trees of infinite degree in finitary ways, by means of finite arc-labelled
digraphs\footnote{In Section 4, weights on half-edges of graphs will be even
more important, as they will allow us to describe, as universal coverings of
finite weighted graphs, trees of finite degree that are \emph{not} universal
coverings of any finite graph. Furthermore, weights $\omega$\ will yield trees
with nodes of infinite degree.}. The following result shows that this
description is effective.

\bigskip
\noindent\textbf{Theorem}\ \textbf{3.14:} Given a finite weighted digraph $H$
and two vertices $x,y\in V_{H}$, one can decide whether $\mathit{Unf}%
(H/x)\simeq\mathit{Unf}(H/y)$.

\medskip
We need a few technical definitions and lemmas.

\bigskip
\noindent\textbf{Definition 3.15:} \emph{Equivalent weighted sets}. \smallskip

Let $R$ be an equivalence relation on a set $V$ and $X,Y\subseteq V.$

(a) Let $\overline{X}=(X,\lambda)$\ and $\overline{Y}=(Y,\lambda^{\prime}%
)$\ be weighted sets.\

We write $\overline{X}\sim\overline{Y}\ (\operatorname{mod}$ $R)$\ if:
\begin{quote}
(C) For every equivalence class $C$ of $R$, we have $\lambda(C\cap
X)=\lambda^{\prime}(C\cap Y)$.
\end{quote}
Equivalently, for every $x\in X$, there is $y\in Y$ such that $\lambda
([x]_{R}\cap X)=\lambda^{\prime}([y]_{R}\cap Y)$ and $(x,y)\in R,$ and
similarly, for every $y\in Y$, there is $x\in X$ such that $(x,y)$ satisfies
the same property.\ This is an equivalence relation. Condition (C) implies
that $\lambda(X)=\lambda^{\prime}(Y).$

(b) A \emph{witness} of the equivalence $\overline{X}\sim\overline
{Y}\ (\operatorname{mod}$ $R)$\ is a set $S\subseteq X\times Y$ with weight
function $\mu$, that is the (disjoint) union of witnesses of the weight
equalities $\lambda(C\cap X)=\lambda^{\prime}(C\cap Y)$ for all equivalence
classes $C$\ of $R$, (cf. Lemma 2.1(3).$\ \square$

\medskip
We say that an equivalence relation $R$ \emph{refines }an equivalence relation
$R^{\prime}$ on the same set if each class of $R^{\prime}$ is a union of
classes of $R$.\ This is written $R\subseteq R^{\prime}$, by considering
equivalence relations as sets of pairs.

\medskip
In the following two lemmas, $V,R,\overline{X}$ and $\overline{Y}$ are as in
the previous definition.

\bigskip
\noindent\textbf{Lemma 3.16:} If $\overline{X}\sim\overline{Y}%
\ (\operatorname{mod}$ $R)$ and $R\subseteq R^{\prime}$, then $\overline
{X}\sim\overline{Y}\ (\operatorname{mod}$ $R^{\prime})$.

\begin{proof}
Each class $C^{\prime}$ of $R^{\prime}$ is the union of (disjoint) classes
$C_{1},C_{2},\dots$ of $R$.\ Hence, $\lambda(C^{\prime}\cap X)=\lambda(C_{1}\cap
X)+\lambda(C_{2}\cap X)+\dots$ and similarly for $Y$.\ The result follows.
\end{proof}

 \smallskip
\noindent\textbf{Lemma 3.17:} Assume that $\overline{X}\sim\overline
{Y}\ (\operatorname{mod}$ $R)$.\ Let $U$ and $W$ be sets, and $\kappa$ and
$\eta$ be weighted surjections\footnote{A weighted surjection of a set
$X$\ onto a weighted set is well-defined by considering that each element of
$X$ has weight 1.}, respectively $U\rightarrow\overline{X}$ and $W\rightarrow
\overline{Y}$. \ There is a bijection $\ell:U\rightarrow W$ such that
$(\kappa(u),\eta(\ell(u))\in R$ for all $u$ in $U$. Furthermore, for any
$u_{0}\in U$ and $w_{0}\in W$ such that $(\kappa(u_{0}),\eta(w_{0}))\in R,$
one can find $\ell$ as above such that $\ell(u_{0})=w_{0}.$

\begin{proof}
We have a bijection $\gamma:Set(X,\lambda)\rightarrow Set(Y,\lambda^{\prime})$
such that\footnote{We recall that we write $\gamma(x,i)$\ \ for $\gamma
((x,i)).$} $\gamma(x,i)=(y,j)$ implies $(x,y)\in R$, and bijections
$\kappa^{\prime}:U\rightarrow Set(X,\lambda)$ and $\eta^{\prime}:W\rightarrow
Set(Y,\lambda^{\prime})$. We define $\ell:=\eta^{\prime-1}\circ\gamma
\circ\kappa^{\prime}.$

For proving the last assertion, we choose $\gamma$ such that $\gamma
(\kappa(u_{0}),i)=(\eta(w_{0}),j)$ for some $i,j$.
\end{proof}

The bijection $\ell$ is uniquely defined if $\lambda([x]_{R}\cap
X)=\lambda^{\prime}([y]_{R}\cap Y)=1$ for all $x\in X$ and $y\in Y$, but\ not
otherwise. We now prove Theorem 3.14.

\begin{proof}
Let $H$ be a finite weighted digraph\footnote{It is not necessarily rooted.}
and $\approx$ be the equivalence relation on $V_{H}$\ such that $x\approx y$
if and only if $\mathit{Unf}(H/x)\simeq\mathit{Unf}(H/y)$. We recall from
Section 2\ that $N_{H}^{+}(x)$ is the set of heads of the arcs with tail $x$
and that $\overline{N_{H}^{+}}(x):=(N_{H}^{+}(x),\eta)$ where $\eta(y)$ is the
sum of the weights $\lambda_{H}(e)$ of the arcs $e:x\rightarrow y$
(cf.\ Definition 3.2, $H$ may have parallel arcs).\

\medskip
\emph{Claim 1}: The equivalence relation $\approx$\ satisfies the following
property, that we state for an arbitrary equivalence relation $R$ on $V_{H}$: \vspace{2mm}

(E):\ If $xRy$ then $\overline{N_{H}^{+}}(x)\sim\overline{N_{H}^{+}%
}(y)\ (\operatorname{mod}$ $R)$.

 \vspace{2mm}
\emph{Proof}: This follows from Lemma 2.4 and the definitions. $\square$

 \smallskip
\emph{Claim} \emph{2}: If $R$ is an equivalence relation on $V_{H}$\ that
satisfies Property (E), then $R$ $\subseteq$ $\approx$.\

\medskip
\emph{Proof}: We consider $(x,y)\in R$, and we let $\kappa:T\rightarrow H/x$
and $\kappa^{\prime}:T^{\prime}\rightarrow H/y$ be the unfolding homomorphisms
where $T:=\mathit{Unf}(H/x)$ and $T^{\prime}:=\mathit{Unf}(H/y).$

For each $i$, we construct by induction an isomorphism $\eta_{i}%
:T\upharpoonright i\rightarrow T^{\prime}\upharpoonright i$ such that
$(\kappa(u),\kappa^{\prime}(\eta_{i}(u)))\in R$ for every node $u$ of
$T\upharpoonright i$, and $\eta_{i+1}$ extends $\eta_{i}.\ $The common
extension of these isomorphisms will be an isomorphism $T\rightarrow
T^{\prime},$ proving that $x\approx y$.

We let $\eta_{0}$ map $rt_{T}$ to $rt_{T^{\prime}}$.\ We have $(x,y)=(\kappa
(rt_{T}),\kappa^{\prime}(\eta_{0}(rt_{T})))\in R$, as was to be verified.

We now define $\eta_{i+1}$ extending $\eta_{i}$.

Consider $v$ in $T\upharpoonright(i+1)$ at depth $i+1$\ and its father $u$.
Then $w:=\eta_{i}(u)$ is a node of $T^{\prime}\upharpoonright i.$\ Furthermore
$\kappa$ induces a weighted surjection $N_{T}^{+}(u)\rightarrow\overline
{N_{H}^{+}}(\kappa(u)),$ and similarly, $\kappa^{\prime}$ induces a weighted
surjection $N_{T}^{+}(w)\rightarrow\overline{N_{H}^{+}}(\kappa^{\prime
}(w)).\ $By the inductive property of $\eta_{i}$, we have $(\kappa
(u),\kappa^{\prime}(w))\in R$. Hence, by Property (E), we have $\overline
{N_{H}^{+}}(\kappa(u))\sim\overline{N_{H}^{+}}(\kappa^{\prime}%
(u))(\operatorname{mod}$ $R).$\ By Lemma 3.17, there is a bijection $\ell
_{u}:N_{T}^{+}(u)\rightarrow N_{T}^{+}(w)$ such that $(\kappa(s),\kappa
^{\prime}(\ell_{u}(s))\in R$ for each $s$ in $N_{T}^{+}(u)$ ($s$ is in
$T\upharpoonright(i+1)$). We define $\eta_{i+1}(s):=\ell_{u}(s)$ for every son
$s$ of $u$ in $T$ .

We do that for all nodes $v$ at depth $i+1$ in $T$. We obtain the desired
extension with the inductive property $(t,\eta_{i+1}(t))\in R$ for every node
$t$ of $T\upharpoonright(i+1)$. $\square$

\medskip
There are finitely many equivalence relations $R$\ on $V_{H}$.\ For each of
them, one can check if it satisfies Property (E$)$ and contains the pair
$(x,y)$.\ Then $x\approx y$ if and only if one of them has these two properties.
\end{proof}

The following algorithm is similar to the minimization of finite deterministic
automata. It will help to prove Theorem 3.20.

\medskip
\noindent\textbf{Algorithm 3.18:} \emph{Deciding the isomorphism of complete
unfoldings}. \smallskip

\emph{Input}: A finite weighted digraph\footnote{It need not be connected.\ In
order to decide whether $\mathit{Unf}(G/x)\simeq\mathit{Unf}(G^{\prime}/y)$
where $G\neq G^{\prime}$, we can use this algorithm by taking for $H$ the
union of $G$ and\ a disjoint copy of $G^{\prime}$.} $H$.

\emph{Output}: The equivalence relation $\approx$ on $V_{H}$\ such that
$x\approx y$ if and only if $\mathit{Unf}(H/x)\simeq\mathit{Unf}(H/y)$.

 \smallskip
\emph{Method}: We define a decreasing\footnote{It is decreasing for set
inclusion.\ Hence, the equivalence $R_{i+1}$ refines $R_{i}$.} sequence of
equivalence relations $R_{i},i\geq0$ on $V_{H}$ as follows:
\begin{quote}
$R_{0}=V_{H}\times V_{H}$;

$R_{i+1}=R_{i}\cap\{(x,y)\mid\overline{N_{H}^{+}}(x)\sim\overline{N_{H}^{+}%
}(y)\ (\operatorname{mod}$ $R_{i})\}.$
\end{quote}

We have $R_{i+1}=R_{i}$ for some $i:=i_{\max}$, and we output $R_{i}$ as the
desired result.

\bigskip
\noindent\textbf{Proposition 3.19:} Algorithm 3.18 is correct and terminates
with $i_{\max}\leq\left\vert V_{H}\right\vert -1.$

\begin{proof} Let $R$ be the intersection of the relations $R_{i}$. It is
clear that if $R_{i+1}=R_{i}$, then $R_{i+2}=R_{i+1}$ etc...\ so that,
$R_{i}=R$. This guarantees termination.

\medskip
Each step such that $R_{i+1}\neq R_{i}$ splits at least one equivalence class
of $R_{i}$.\ Such a splitting cannot be done more than $\left\vert
V_{H}\right\vert -1$ times.

We now prove the correctness, \emph{i.e.}, that $\approx$ $=R$.

We prove that $\approx$ $\subseteq$ $R_{i}$ for all $i$.\ This is clear for
$i=0$.\ Assume now $\approx$ $\subseteq$ $R_{i}$. If $x\approx y$, then
$\overline{N_{H}^{+}}(x)\sim\overline{N_{H}^{+}}(y)\ (\operatorname{mod}$
\ $\approx),$ hence $\overline{N_{H}^{+}}(x)\sim\overline{N_{H}^{+}%
}(y)\ (\operatorname{mod}$ $R_{i})$\ \ by Lemma 3.16, and so, $(x,y)\in
R_{i+1}$.\ Hence, $\approx$ $\subseteq$ $R$.

\medskip
The relation $R$ satisfies Property (E), hence $R$ $\subseteq$ $\approx$ by
Claim 2 in the proof of Theorem 3.14.
\end{proof}

The following result is similar to a theorem by Norris \cite{Nor} about
universal coverings that we presented in the introduction and that we will
generalize in Section 4 to weighted graphs. See \cite{CouNext}, it implies
that, for every regular tree, there is a first-order sentence using the
generalized quantifier "there exists $\omega$\ elements $x$ that satisfy..."
of which it is the unique model that is a rooted tree.

\bigskip
\noindent\textbf{Theorem 3.20:} Let $H$ be a finite weighted digraph with $p$
vertices. If $x,y\in V_{H}$, then:
\begin{quote}
$\mathit{Unf}(H/x)\upharpoonright(p-1)\simeq\mathit{Unf}(H/y)\upharpoonright
(p-1)$ implies $\mathit{Unf}(H/x)\simeq\mathit{Unf}(H/y)$.
\end{quote}

\begin{proof}
We use the relations $R_{i}$ of Algorithm 3.18.\ We know by Proposition 3.19
that $\approx$ $=R_{p-1}$.\

\medskip
\emph{Claim}: If $\mathit{Unf}(H/x)\upharpoonright(p-1)\simeq\mathit{Unf}%
(H/y)\upharpoonright(p-1),$ then $(x,y)\in R_{p-1}$.

 \smallskip
\emph{Proof}: By using induction, we prove that for every $i$:

$\mathit{Unf}(H/x)\upharpoonright i\simeq\mathit{Unf}(H/y)\upharpoonright i$
implies $(x,y)\in R_{i}$.

If $i=0$, this fact holds because $(x,y)\in R_{0}$ for all $x,y.$

 \smallskip
We prove the case $i+1$ by assuming that we have an isomorphism $\alpha
:\mathit{Unf}(H/x)\upharpoonright(i+1)\rightarrow\mathit{Unf}%
(H/y)\upharpoonright(i+1)$. Hence $\mathit{Unf}(H/x)\upharpoonright
i\simeq\mathit{Unf}(H/y)\upharpoonright i$ and $(x,y)\in R_{i}$ by the
induction hypothesis.

We now check that $\overline{N_{H}^{+}}(x)\sim\overline{N_{H}^{+}%
}(y)\ (\operatorname{mod}$ $R_{i})$\ in order to obtain that $(x,y)\in
R_{i+1}$.

Let $\eta:\mathit{Unf}(H/x)\rightarrow H/x$ and $\eta^{\prime}:\mathit{Unf}%
(H/y)\rightarrow H/y$ be complete unfoldings. We have $\eta(rt_{\mathit{Unf}%
(H/x)})=x$ and $\eta^{\prime}(rt_{\mathit{Unf}(H/y)})=y.$

For each son $u$ of $rt_{\mathit{Unf}(H/x)}$, $\alpha$\ defines an isomorphism:

$\mathit{Unf}(H/x)/u\upharpoonright i\rightarrow\mathit{Unf}(H/y)/\alpha
(u)\upharpoonright i$,

where $\alpha(u)$ is a son of $rt_{\mathit{Unf}(H/y)}.$ But $\mathit{Unf}%
(H/x)/u=\mathit{Unf}(H/\eta(u))$ and $\mathit{Unf}(H/y)/\alpha(u)=\mathit{Unf}%
(H/\eta^{\prime}(\alpha(u))).$ Hence $(\eta(u),\eta^{\prime}(\alpha(u)))\in
R_{i}$ by induction.

Then $N_{H}^{+}(x)$ is the set of such $\eta(u)$ and $N_{H}^{+}(y)$ is that of
such $\eta^{\prime}(\alpha(u)).$ By counting occurrences, we obtain
$\overline{N_{H}^{+}}(x)\sim\overline{N_{H}^{+}}(y)(\operatorname{mod}$
$R_{i}).$ Hence,$\ (x,y)\in R_{i+1}.\ \square$

\medskip
If $\mathit{Unf}(H/x)\upharpoonright(p-1)\simeq\mathit{Unf}%
(H/y)\upharpoonright(p-1)$, we have $(x,y)\in R_{p-1}$ by the claim, hence
$x\approx y$ as was to be proved since $R_{p-1}=$ $\approx$\ by Proposition 3.18.
\end{proof}

 \smallskip
\noindent\textbf{Remark 3.21:} By the proof of Proposition 3.19,
$\mathit{Unf}(H/x)\upharpoonright(p-1)\simeq\mathit{Unf}(H/y)\upharpoonright
(p-1)$ implies $\mathit{Unf}(H/x)\upharpoonright i\simeq\mathit{Unf}%
(H/y)\upharpoonright i$ for all $i$.\ We might think that this implies
$\mathit{Unf}(H/x)\simeq\mathit{Unf}(H/y)$. This argument is correct only if
$\mathit{Unf}(H/x)$ and $\mathit{Unf}(H/y)$ have finite degree, by using
K\"{o}nig's Lemma, as in the proof of Lemma 2.7 of \cite{KreVer}.

\medskip
However, this implication is false for trees with nodes of infinite
degree.\ Let $T$ be the union of the finite paths $0\rightarrow
(1,i)\rightarrow(2,i)\rightarrow\dots\rightarrow(i,i)$ for all $i\in
\mathbb{N}_{+}$, and $T^{\prime}$ be $T$\ together with the infinite path
$0\rightarrow1\rightarrow2\rightarrow\dots\rightarrow i\rightarrow\dots$. They are
not isomorphic, but $T\upharpoonright i\simeq T^{\prime}\upharpoonright i$
\ for each $i$. Theorem 3.14 is used for proving Theorem 3.20.\ To prove its
Claim 2, we cannot use K\"{o}nig's Lemma because the trees $\mathit{Unf}%
(H/x)$\ and $\mathit{Unf}(H/y)$ need not have finite degree.\ Instead, we
construct a sequence of isomorphisms:
\begin{quote}
$\eta_{i}:\mathit{Unf}(H/x)\upharpoonright i\rightarrow\mathit{Unf}%
(H/y)\upharpoonright i$ \ such that $\eta_{i+1}$\ extends $\eta_{i}.$
\end{quote}
Their common extension yields an isomorphism: $\mathit{Unf}(H/x)\rightarrow
\mathit{Unf}(H/y)$.$\ \square$

\medskip
The following theorem is similar to that of Leighton about coverings
(\cite{Lei},\ see below Theorem 4.10), and much easier to prove.

\bigskip
\noindent\textbf{Theorem 3.22: }Given two finite, rooted and weighted digraphs
$G$ and $H$, the following properties are equivalent: \smallskip

1) $G$ and $H$ are unfoldings of a finite rooted and weighted digraph,

2) $G$ and $H$ have isomorphic complete unfoldings,

3) $G$ and $H$ have a common finite unfolding.

They are decidable.

\begin{proof}
Without loss of generality, we assume that $G$ and $H$ are disjoint. \smallskip

1) $\Longrightarrow$2) If $G$ and $H$ are unfoldings of a finite rooted and
weighted digraph $M$, then the complete unfolding of $M$ is a complete one of
both $G$ and $H$ by Theorem 3.5(2). \smallskip

2) $\Longrightarrow$3) Let $\gamma:T\rightarrow G$ and $\eta:T\rightarrow H$
be complete unfoldings of $G$ and $H$. \smallskip

If $u\in N_{T}$, then $T/u\simeq\mathit{Unf}(G/\gamma(u))\simeq\mathit{Unf}%
(H/\eta(u))$ by Proposition 3.4(2).

We define $\approx$ as the equivalence relation on $V_{G}\cup V_{H}$ such that
$x\approx y$ if and only if $\mathit{Unf}((G\cup H)/x)\simeq\mathit{Unf}%
((G\cup H)/y))$, where $\mathit{Unf}((G\cup H)/x)=\mathit{Unf}(G/x)$ if $x\in
V_{G}$\ and similarly for $H$ as $G$ and $H$ are disjoint.

For helping to understand the technical details, we first present the proof
for the special case where there are no two distinct nodes $u,v$ in $T$ with
same father, and such that $T/u\simeq T/v$. This fact implies that all arcs in
$G$ and $H$ have weight 1. In such a case:

 \smallskip
(*) if $u\in N_{T}$, the relation $T/v\simeq\mathit{Unf}(G/y)$ defines by
Lemma 2.4 a bijection between the sons $v$ of $u$ in $T$ and the vertices $y$
in $N_{G}^{+}(\gamma(u))$.\ A similar fact holds for $H$ with the vertices $y$
in $N_{H}^{+}(\eta(u)).$

 \smallskip
We define a digraph $L$\ as follows. Its set of vertices is $V_{L}%
:=\{(x,y)\mid x\in V_{G},y\in V_{H}$ and $\mathit{Unf}(G/x)\simeq
\mathit{Unf}(H/y)\}$. \ For each $(x,y)\in V_{L}$, the relation $\approx
$\ defines, by Fact (*) above, a bijection between $N_{G}^{+}(x)$ and
$N_{H}^{+}(y).$We define in $L$ an arc $(x,y)\rightarrow(x^{\prime},y^{\prime
})$ (of weight 1) if $x^{\prime}\in N_{G}^{+}(x)$ and $y^{\prime}\in N_{H}%
^{+}(y)$ (and of course $x^{\prime}\approx y^{\prime}$).

 \smallskip
We now define $K:=L/(rt_{G},rt_{H})$.\ It is a finite and rooted digraph. The
projection $\pi_{1}$ such that $\pi_{1}(x,y):=x$ is an unfolding $K\rightarrow
G$. The other projection is an unfolding $K\rightarrow H$.

\medskip
We now consider the general case.\ The construction is similar, but the
definition of the arcs $(x,y)\rightarrow(x^{\prime},y^{\prime})$ of $L$\ is
more complicated because the relation $\approx$\ is not necessarily a
bijection between $N_{G}^{+}(x)$ and $N_{H}^{+}(y).$

We define $V_{L}$\ as above.\ For each $(x,y)\in V_{L}$, we have
$\overline{N_{G}^{+}}(x)\sim\overline{N_{H}^{+}}(y)(\operatorname{mod}$
$\approx)$\ by Lemma 2.4. We choose a witness $(S_{x,y},\mu_{x,y})$ of
$\overline{N_{G}^{+}}(x)\sim\overline{N_{H}^{+}}(y)(\operatorname{mod}$
$\approx)$, cf.\ Definition 3.15(b).\ We define in $L$ an arc
$(x,y)\rightarrow(x^{\prime},y^{\prime})$ of weight $\mu_{x,y}(x^{\prime
},y^{\prime})$ for each $(x^{\prime},y^{\prime})$ in $S_{x,y}$. We now define
$K:=L/(rt_{G},rt_{H})$.\ It is rooted and weighted with at most $\left\vert
V_{G}\right\vert $.$\left\vert V_{H}\right\vert $\ \ vertices.

\medskip
\emph{Claim}: $K$ is an unfolding of $G$, and, similarly, of $H$.

 \smallskip
\emph{Proof of claim}: Let $\pi$ map a vertex $(x,y)$ of $K$ to the vertex $x$
of $G$, and an arc $(x,y)\rightarrow(x^{\prime},y^{\prime})$ to the arc
$x\rightarrow x^{\prime}$ of $G$. We make a few observations.

 \smallskip
(1) If $(x,y)\in V_{L}$ and $x-x^{\prime}$ is an arc of $G$, there is an arc
$(x,y)\rightarrow(x^{\prime},y^{\prime})$ in $L$.\ If $(x,y)\in V_{K}$ then
$(x^{\prime},y^{\prime})$ and the arc $(x,y)\rightarrow(x^{\prime},y^{\prime
})$ are in $K$ that is a subgraph of $L$.

 \smallskip
(2) If $x$ is a vertex in $G$, there is a directed path from $rt_{G}$ to $x$
and, by (1), a directed path in $L$\ from the root $(rt_{G},rt_{H})$ to
$(x,y)\in V_{L}$ for some $y\in V_{H}$.\ All vertices and arcs of this path
are in $K$.

 \smallskip
It follows that $\pi$ is a surjective homomorphism: $K\rightarrow G.$ We now
check Definition 3.2.\ We verify the following condition.

 \smallskip
(**) For every $(x,y)\in V_{K}$\ and $x^{\prime}\in N_{G}^{+}(x)$, we have:

 \smallskip
$\lambda_{G}(x,x^{\prime})=\Sigma\{\lambda_{K}((x,y),(x^{\prime},y^{\prime
}))\mid(x^{\prime},y^{\prime})\in V_{K}\}.$

By the definition of $K$, $\lambda_{K}((x,y),(x^{\prime},y^{\prime}%
))=\mu_{x,y}(x^{\prime},y^{\prime})$, and the pairs $(x^{\prime},y^{\prime})$
are in $S_{x,y}$. The weighted set ($S_{x,y},\mu_{x,y})$ is chosen so that
$\lambda_{G}(x,x^{\prime})=\Sigma\{\mu_{x,y}(x^{\prime},y^{\prime}%
)\mid(x^{\prime},y^{\prime})\in$ $S_{x,y}\}.$ This proves (**), the claim and point 3).

 \smallskip
3) $\Longrightarrow$1) Assume that $\gamma:T\rightarrow G$ and $\eta
:T\rightarrow H$ are complete unfoldings.\

Let $\approx$ be the equivalence relation on $N_{T}$\ such that $u\approx v$
if and only if $T/u\simeq T/v$. We define $M\ $ as the weighted graph
$T/\approx$, cf.\ Definition 3.11\ and the proof of Theorem 3.12. There are
unfoldings: $\gamma^{\prime}:G\rightarrow M$ and $\eta^{\prime}:H\rightarrow
M$. We omit details.

\medskip
The decidability follows from Theorem 3.19.
\end{proof}

\noindent\textbf{Remarks 3.23:} In the proof of 2) $\Longrightarrow$3), $T$ is
a complete unfolding of $K$ by Theorem 3.5. Note however that in this proof,
$K$ is not defined in a unique way, in particular because the weighted
relations $(S_{x,y},\mu_{x,y})$ are not uniquely defined. It is however in the
special case we first considered.

\section{Coverings}

In this section and the next two ones, we will consider undirected graphs,
simply called \emph{graphs}, and their coverings. We recall from Section 2.1
that a graph $G$\ is defined as a triple $(V,E,Inc)$ where the elements of
$Inc$ (a subset of $E\times V$) are its \emph{half-edges}.\ This description
allows graphs with parallel edges and loops. An edge $e$ is a loop at a vertex
$x$ if and only if $(e,x)\in Inc$ and there is no pair $(e,y)$ in $Inc$ such
that $y\neq x$. We denote by $Inc(x)$ the set of half-edges $(e,x)$ for some
$e\in E$. Its cardinality is the \emph{degree} of $x$, where a loop at $x$
counts for one.

We will use \emph{trees} (undirected and without root) and \emph{rooted
trees}, in particular the regular trees considered in the previous section.
Trees and graphs may be labelled.

The main contributions of this section are the definition of weighted graphs,
that can be seen as graph interpretations of degree matrices. We extend
coverings to weighted graphs. If two finite graphs have a common (finite)
covering, they cover a common (finite) weighted graph (Theorem
4.10).\ Regarding characteristic polynomials, we obtain an extension of a
known factorization result (Section 4.3). We postpone to Section 5 the study
of universal coverings of weighted graphs.

As in Section 3, equality of trees and graphs is understood in the strict
sense: same nodes or vertices, and same edges or arcs.\ Equality up to
isomorphism is specified explicitely and denoted by~$\simeq$.

\subsection{Coverings of graphs: definitions and known results}

We mainly review known definitions and facts from
\cite{Ang,AngGar,Bod,BodVL,FiaKra,KreVer,Lei,Nor}\textbf{.\ }Our\ main
reference for all assertions is \cite{FiaKra} by Fiala and Kratochv\'{\i}l.

We define the \emph{adjacency matrix} $A_{G}$\ of a finite graph $G$ such that
$V_{G}=[p]:=\{1,\dots,p\}$ for some $p$ as follows: $A_{G}[x,y]=A_{G}[y,x]$ is
the number of edges between $x$ and $y$ and $A_{G}[x,x]$ is the number of
loops at $x$.\

\bigskip
\noindent\textbf{Definition 4.1:} \emph{Covering}. \smallskip

(a) Let $G,H$ be graphs.\ A \emph{covering}\ $\gamma:G\rightarrow H$ is a
surjective homomorphism such that, if $\gamma(x)=y$, then $\gamma$ defines a
bijection $E_{G}(x)\rightarrow E_{H}(y).$ We will also say that $G$ is a\emph{
covering of }$H$.

(b) Let $G,H$ be finite, $V_{G}=[p]$ and $V_{H}=[q]$.\ A surjective mapping
$\gamma:V_{G}\rightarrow V_{H}$ can be represented by a $p\times q$-matrix
$B_{\gamma}$\ such that
$B_{\gamma}[i,j]:=$ \texttt{if} $\gamma(i)=j$ \texttt{then }1 \texttt{else} 0. Each row of this matrix has a unique 1 and each column has at least one
1.\ Then, $\gamma$ defines a covering if and only if $A_{G}B_{\gamma
}=B_{\gamma}A_{H}$. $\square$

\bigskip

An edge covers a loop incident to a single vertex.\ More generally, a
$k$-\emph{regular} graph, \emph{i.e.}\ such that all vertices have degree $k$,
covers $k$ loops incident to a single vertex.

\bigskip
\noindent\textbf{Proposition 4.2:} Let $\gamma:G\rightarrow H$ be a covering. \smallskip

(1) If $\delta:H\rightarrow K$ is a covering, then $\delta\circ\gamma
:G\rightarrow K$ is a covering.

(2) If $G$ and $H$\ are finite and $H$\ is connected, then, either $\gamma$ is
an isomorphism or $\left\vert V_{G}\right\vert >\left\vert V_{H}\right\vert $
and then, $\left\vert V_{G}\right\vert /\left\vert V_{H}\right\vert
=\left\vert Inc_{G}\right\vert /\left\vert Inc_{H}\right\vert $ and this
number is a positive integer.

(3) If $H$ is a\ tree and $G$\ is connected, then $G$\ is a tree and $\gamma$
is an isomorphism.

\begin{proof}
Assertion (2) is due to Reidemeister (see \cite{Bod,BodVL,Red}). Here is a
proof sketch (cf.\ Section 2.1 in \cite{FiaKra}).

Let $T$ be a spanning tree of $H$.\ It does not include the loops that are
irrelevant to connectedness.\ Then, $\gamma^{-1}(E_{T})$ is a set of edges of
$G$.\ By the definition of a covering, it is the union of $k$ pairwise
disjoint trees, all isomorphic to $T$\ by $\gamma$. This union includes all
vertices, hence $\left\vert V_{G}\right\vert /\left\vert V_{H}\right\vert =k.$

We now prove that $\left\vert Inc_{G}\right\vert /\left\vert Inc_{H}%
\right\vert =k.$

Consider the edges $e$ of $G$ such that $\gamma(e)$ is a loop at $x$ in $H$.
Such an edge may link two vertices in different connected component of $G$.
\ We have the pair $(\gamma(e),x)$ in $Inc_{H}$ and a single pair of the form
$(e,u)$ such that $\gamma(u)=x$ in each connected component of $G$. Hence,
there are $k$ such edges $e$.

Consider now the edges $e$ of $G$ such that $\gamma(e):x-y$ is not a loop in
$H$. We have $(\gamma(e),x)$ and $(\gamma(e),y)$ in $Inc_{H}$.\ Each $e$
yields exactly two pairs $(e,u)$ and $(e,v)$ in $Inc_{G}$ such that
$\gamma(u)=x$ and $\gamma(v)=y$. There are exactly $2k$ such pairs $(e,u)$ and
$(e^{\prime},v)$ in $Inc_{G}$.\ Hence, $\left\vert Inc_{G}\right\vert
/\left\vert Inc_{H}\right\vert =k.$

(3) This is known from \cite{FiaKra,Red} if $H$ is finite. Assume now that $H$
is infinite.\ Consider as in (2) the edges of $\gamma^{-1}(E_{H})$ of
$G$.\ They form a union of spanning trees $T$ of $G$.\ There are no other
edges in $G$.\ As $G$ is connected, it is a tree.
\end{proof}

\noindent\textbf{Definition 4.3:} \emph{Degree matrix} \smallskip

(a) For every finite graph $G$, there is a unique partition ($B_{1}%
,\dots,B_{p})$ of $V_{G}$\ having a minimum number of classes, such that for
every $i,j\in\lbrack p]$, every vertex $x$ in $B_{i}$ has the same number of
neighbours, say $r_{i,j}$, in $B_{j}$. It is called the \emph{degree
(refinement) partition}. It can be computed in polynomial time \cite{BBG}.\

(b) Let $\alpha:V_{G}\rightarrow\lbrack p]$ maps a vertex $x$ to the integer
$i$ such that $x\in B_{i}$.\ We call $\alpha$\ a \emph{good indexing} of
$V_{G}$. The numbers $r_{i,j}$ can be organized into a $p\times p$ matrix
$M_{G,\alpha}$ such that \ and $M_{G,\alpha}[i,j]=r_{i,j}.$ It is called the
\emph{degree (refinement) matrix} of $G$. This matrix may not be symmetric.  $\square$

\begin{figure}[!h]
\vspace*{4mm}
\begin{center}
\includegraphics[height=1.5437in,width=2.1257in]{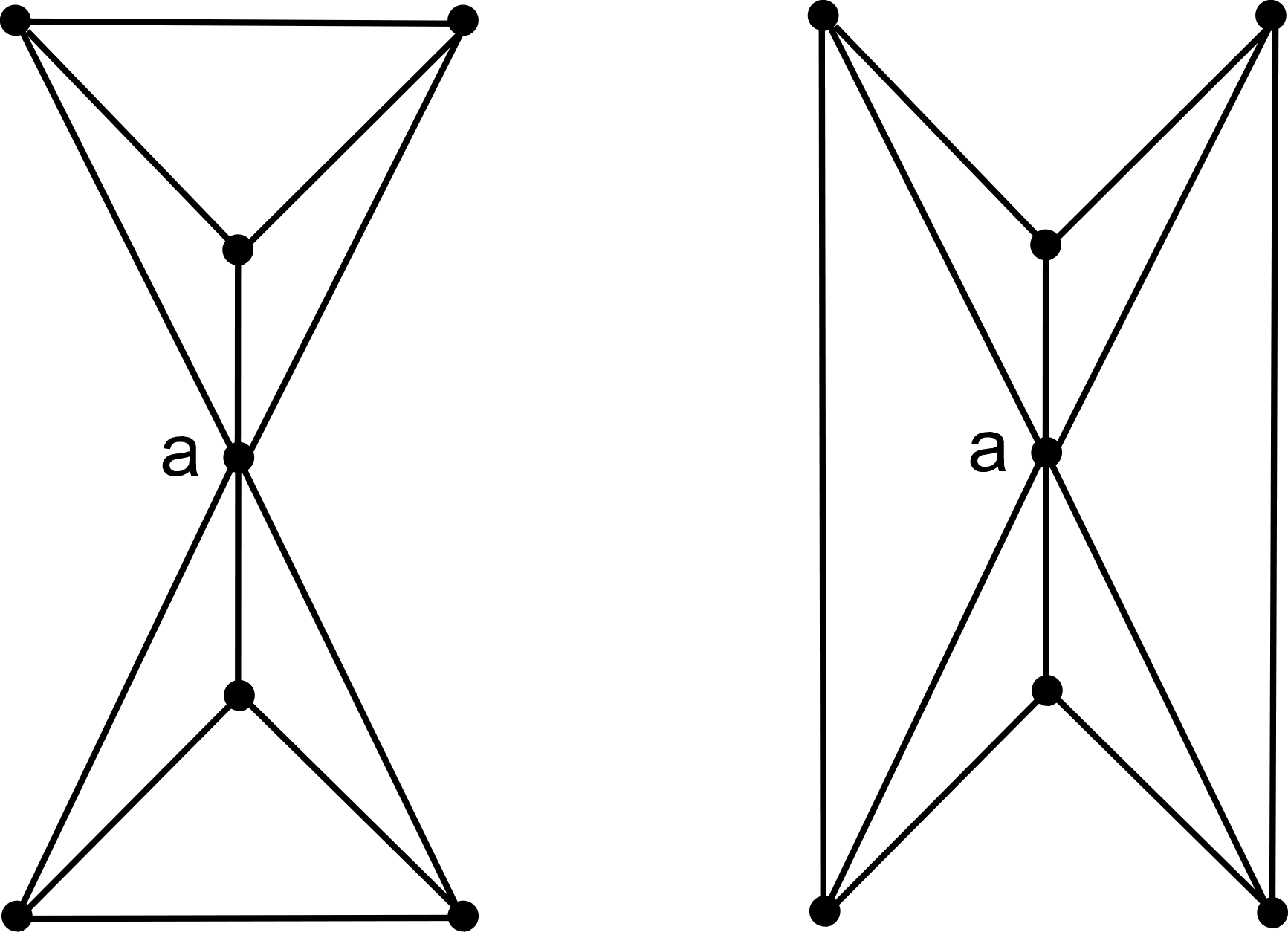}%
\caption{None of these graphs covers any smaller graph. See Example 4.4.}%
\end{center}\vspace*{-3mm}
\end{figure}

\bigskip
\noindent\textbf{Example 4.4:} The two graphs of Figure 4\ have degree
partition ($B_{1},B_{2})$\ where $B_{1}=\{a\}$ and $B_{2}$ consists of the six
other vertices.\ The corresponding matrix is $M:=$ $%
\begin{bmatrix}
0 & 6\\
1 & 2
\end{bmatrix}
.\ $As both have 7 vertices, a prime number, they cannot cover any graph apart
themselves by Proposition 4.2(2) (an observation made by Boldi and Vigna in
\cite{BolVig}). They cover a common weighted graph $H$\ whose weight matrix is
$M$, see Example 4.20(4). $\square$%

\bigskip
\noindent \textbf{Lemma 4.5:} If $G$ and $H$ are finite, if $\alpha$\ is a good indexing
of $V_{H}$ and $\gamma:G\rightarrow H$ is a covering, then $\alpha\circ\gamma$
is a good indexing of $V_{G}$ and $M_{G,\alpha\circ\gamma}=M_{H,\alpha}.$

\begin{proof}
Because of $\gamma$, the graphs $G$ and $H$ have same degree matrices (for
some appropriate numbering of the components of the degree partition, cf.
\cite{FiaKra}, Section 4.1).\ Then $\alpha\circ\gamma$ is a good indexing of
$V_{G}$ and the equality $M_{G,\alpha\circ\gamma}=M_{H,\alpha}$ follows from
the definitions.
\end{proof}

\noindent\textbf{Definition 4.6:} \emph{Universal coverings}\smallskip

(a) A covering of a graph $H$ that is a tree is called a \emph{universal
covering} of $H$ (hence $H$ is connected, cf.\ Remark 4.19).

(b) Every connected graph $H$\ has a universal covering constructed as
follows.\ For a vertex $x$ of $H$, we define $UC(H,x)$ as the rooted tree of
all finite walks in $H$ that start at $x$ and do not use a same edge
(including a loop) twice in a row.\ The tree $Unr(UC(H,x))$ is obtained by
forgetting the root of $UC(H,x)$ and its orientation.\ It is a covering of
$H$, hence a universal one. We have $Unr(UC(H,x))\simeq Unr(UC(H,y))$ for any
two vertices $x$ and $y$ (\cite{FiaKra}, Section 4.2). Examples are given
below.\ $\square$

\bigskip
\noindent\textbf{Examples 4.7:} (1) An edge is the universal covering of a
single loop.\ A path with 4 vertices is that of an edge with a loop at one of
its ends.\

(2) If $H$ consists of two parallel edges, then $Unr(UC(H,x))$ is a
\emph{biinfinite path,} \emph{i.e.}, the union of two infinite paths
originating from a same node.\ (A biinfinite path is somehow isomorphic to
$\mathbb{Z}$).\ Equivalently, it is the unique tree \emph{u.t.i.} (up to
isomorphism) whose nodes have all degree 2.\ It is also the universal covering
of two loops at a same vertex or of any cycle.

(3) The universal covering of a connected $k$-regular graph is the infinite
tree whose nodes have all degree $k$. This is clear from the construction
recalled in Definition 4.6(b). $\square$

\bigskip

We recall that if $u$ is a node of a tree $T$, then $T_{u}$ is the rooted tree
obtained by taking $u$ as the root.

\bigskip
\noindent\textbf{Proposition 4.8:} Let $H,H^{\prime}$ be graphs.\smallskip

(1) If $\gamma:T\rightarrow H$ is a universal covering and $u\in N_{T}$, then
$T_{u}$ $\simeq UC(H,\gamma(u))$.

(2) If there an isomorphism of $H$ to $H^{\prime}$ maps $x$ to $y$, then
$UC(H,x)$ is isomorphic to $UC(H^{\prime},y)$.

\begin{proof}
(1) We will prove below a generalization of this fact for weighted graphs.

(2) This is clear from the descriptions of $UC(H,x)$ and $UC(H^{\prime},y)$ in
terms of walks.
\end{proof}

By Assertion (1) and Definition 4.6(b), all universal coverings of $H$ are
isomorphic.\ One can speak of \emph{the} universal covering of $H,$%
\ denoted\footnote{The use of boldface letters is intended to recall that
$\boldsymbol{UC}(H)$ is only defined up to isomorphism. Most proofs about
universal coverings will be done from the concrete trees $Unr(UC(H,x))$.} by
$\boldsymbol{UC}(H$).

\bigskip
\noindent\textbf{Remark 4.9:} The converse to Assertion (2) does not hold when
$H=H^{\prime}$.\ Take for a counter-example the union of the two graphs of
Figure 4 with an edge between the two vertices marked $a$, that we will call
$x$ and $y$. Then $UC(H,x)\simeq UC(H,y)$ but there exists no automorphism of
$H$\ that maps $x$ to $y$.$\ \square$

\medskip
The relevance to distributed computing can be stated as follows: if $x$ and
$y$ are two nodes of a network represented by a graph $H$ and $UC(H,x)\simeq
UC(H,y),$ then, no computation in $H$ (following certain rules, see
\cite{Ang}) can distinguish $x$ from $y$. It follows that an election
algorithm that would select $x$ would also select $y$, hence would not be correct.

\bigskip
\noindent\textbf{Theorem 4.10:} Let\textbf{ }$G,H$ be finite and connected
graphs. The following properties are equivalent.\smallskip

(i) $G$ and $H$ have a common finite covering,

(ii) $G$ and $H$ have isomorphic universal coverings,

(iii) $M_{G,\alpha}=M_{H,\beta}$ for some good indexings $\alpha$ and $\beta$
of $V_{G}$ and $V_{H}$.

\medskip
The implication (iii)$\Longrightarrow$(i) has a difficult proof by Leighton in
\cite{Lei}.\ We will prove in Theorem 6.1 below is a special case of it from
which follows that of regular graphs, known from Angluin and Gardiner
\cite{AngGar}.

If $G$ and $H$ have the same number of vertices, them (iii) implies that they
are \emph{fractionally isomorphic} by Theorem 6.5.1 of the book \cite{FGT}%
.\ We will not develop this aspect in the present article.\

We will interpret a degree matrix $M_{G,\alpha}$ and a good indexing $\alpha$
of a graph $G$\ as a covering $\alpha:G\rightarrow M$ where $M$\ is a finite
\emph{weighted graph}. Furthermore, we will allow infinite weights and obtain
universal coverings that are trees of infinite degree, as in Section 3 for unfoldings.

\bigskip
\noindent\textbf{Definition 4.11:} \emph{Equivalences on graphs that yield
coverings.}\smallskip

We recall from Section 1 that an equivalence relation $\sim$ on a graph
$G=(V,E,Inc)$ is an equivalence relation on $V\cup E$ such that each
equivalence class is a set of vertices or of edges, and, if $e$ and
$e^{\prime}$ are equivalent edges, then each end of $e$ is equivalent to an
end of $e^{\prime}$. The \emph{quotient graph} is then defined as
$G/\sim:=(V/\sim,E/\sim,Inc_{G/\sim})$ such that $([e]_{\sim},[v]_{\sim})\in$
$Inc_{G/\sim}$ if and only if $(e^{\prime},v^{\prime})\in Inc_{G}$\ for some
$e^{\prime}\sim e$ and $v^{\prime}\sim v$.

\medskip
We say that such an equivalence $\sim$\ is \emph{strong} if, whenever $x$ and
$x^{\prime}$ are equivalent vertices, it defines a bijection between $E(x)$
and $E(x^{\prime})$. $\square$

\bigskip
\noindent\textbf{Proposition 4.12:} (1) If $\sim$ is a strong equivalence on a
graph $G$, then the surjection $\alpha:V\cup E\rightarrow(V/\sim)\cup(E/\sim)$
that maps $x$ to its equivalence class $[x]_{\sim}$ is a covering
$G\rightarrow G/\sim$.

(2) Every connected graph $H$\ is isomorphic to $T/\sim$ where $T$\ is its
universal covering and $\sim$ is\ a strong equivalence relation on $T$.

\begin{proof}
(1) The proof is straightforward.

(2) We let $\gamma:T\rightarrow H$ be a universal covering where
$T=(N,E,Inc)$.\ We define $x\sim y$ for $x,y\in N\cup E$ if and only if
$\gamma(x)=\gamma(y)$.\ Then $T/\sim$ is isomorphic to $H$.
\end{proof}

Quotients of trees will be studied in Sections 5.2 and 6.

\subsection{Coverings of weighted graphs}

We extend to weighted graphs the notion of covering. The two graphs of Example
4.4\ cover a same weighted graph but no same graph.\ The case of finite
weighted graphs will be of particular interest, because they provide us with
finite descriptions of certain regular trees.

\bigskip
\noindent\textbf{Definitions 4.13:} \emph{Weighted graphs and weight
matrices.}\smallskip

(a) A \emph{weighted graph} is a quadruple $G=(V,E,Inc,\lambda)$ such that
$(V,E,Inc)$ is a simple graph (it has no two parallel edges and no two loops
at a same vertex) and $\lambda$ is a \emph{weight function}: $Inc\rightarrow
\mathbb{N}_{+}\cup\{\omega\}$. The two halves of an edge may have different weights.

A graph $G$ is made into a weighted graph $W(G)$\ as follows: $p$ parallel
edges between $x$ and $y$ are fused into a single edge whose two half-edges
have weight $p$; similarly, $p$ loops at $x$ are fused into a single one at
$x$ of weight $p.$ A simple graph is a weighted graph whose weights are all 1.

(b) A finite weighted graph $G$ with vertex set equal to (or indexed by) $[p]$
can be represented by the \emph{weight matrix} $M_{G}:[p]\times\lbrack
p]\rightarrow\mathbb{N}\cup\{\omega\}$ such that $M_{G}[x,y]:=\lambda
_{G}(e,x)$ if $e:x-y$. Then the sum of weights of the half-edges is the sum of
coefficients of $M_{G}$.

\eject
\noindent\textbf{Definition 4.14:} \emph{Coverings of weighted graphs}\smallskip

Let $H$\ and $G$\ be a weighted graphs. A \emph{covering}\ $\gamma
:G\rightarrow H$ is a surjective homomorphism of unweighted graphs such that,
if $x\in V_{G}$, $\gamma(x)=y$ and $e\in E_{H}(y)$, then:
\begin{quote}
$\lambda_{H}(e,y)=\Sigma\{\lambda_{G}(e^{\prime},x)\mid e^{\prime}\in
E_{G}(x),\gamma(e^{\prime})=e\},$

or equivalently, $\gamma$ induces a weighted surjection $(Inc_{G}%
(x),\lambda_{G})\rightarrow(Inc_{H}(y),\lambda_{H}).$
\end{quote}
We will say that $G$ is a\emph{ covering of }$H$.

\bigskip
\noindent\textbf{Remarks 4.15:} (1) If in Definition 4.14, $G$ be a graph,
then $\lambda_{H}(e,y)=\left\vert \{e^{\prime}\mid e^{\prime}\in
E_{G}(x),\gamma(e^{\prime})=e\}\right\vert ,$ and, equivalently, $\gamma$
induces a weighted surjection $Inc_{G}(x)\rightarrow(Inc_{H}(y),\lambda_{H}).$
The degree of $x$ in $G$ is the sum of weights of the half-edges in
$Inc_{H}(\gamma(x)).$

(2) If $H$ is a simple graph, then $G$ is a graph and the condition implies
that $\gamma$ is injective on each set $Inc_{G}(x)$, whence bijective: we get
the notion of covering of Section 4.1.

(3) Each graph $G$\ covers the weighted graph $W(G)$. $\square$

\medskip
Coverings of finite weighted graphs, even having infinite weights, can also be
expressed in terms of weight matrices (as for graphs in terms of adjacency
matrices, cf.\ Definition 4.1).

\medskip
Let $G$ and $H$\ be finite weighted graphs and $\alpha:V_{G}\rightarrow V_{H}$
be surjective, where $V_{G}=[p]$ and $V_{H}=[q].$ This mapping is represented
by the matrix $B_{\alpha}$\ (as in Definition 4.1) such that:
\begin{quote}
$B_{\alpha}[i,j]:=$ \texttt{if} $\alpha(i)=j$ \texttt{then }1 \texttt{else} 0.
\end{quote}
The following proposition is straightforward from the definition.\ For
defining the product of two matrices, we use the rules $\omega+x=\omega$ for
every $x$, $\omega.x=\omega$ if $x>0$ and $\omega.0=0.$ We need no substractions.

\bigskip
\noindent\textbf{Proposition 4.16:} A homomorphism $\alpha:G\rightarrow H$ is
a covering if and only if $M_{G}B_{\alpha}=B_{\alpha}M_{H}$.

\bigskip
\noindent\textbf{Remark 4.17: }Here is a method to build a graph $G$\ that
covers a finite or infinite weighted graph $H$. It is similar to the
construction of the proof of Proposition 4.2(2). Given $H=(V,E,Inc,\lambda)$,
we construct $G=(V^{\prime},E^{\prime},Inc^{\prime})$, as follows (it is unweighted).

We choose a set $V^{\prime}$ and a surjective mapping $\alpha:V^{\prime
}\rightarrow V$. For each $x\in V^{\prime}$ and $(e,\alpha(x))\in Inc$, we
create $\lambda(e,\alpha(x))$ (yet abstract) half-edges incident with $x$,
defined as pairs $((e,i),x)$ for $i=1,2,\dots,\lambda(e,\alpha(x))$. In this
way, we have defined $Inc^{\prime}$. We let $\alpha$ map $((e,i),x)$ to
$(e,\alpha(x)).$

\medskip
We choose a partial matching $M$\ on $Inc^{\prime}$ satisfying the following property:
\begin{quote}
A pair in $M$\ is of the form ($((e,i),y)$,$((e,j),z))$ such that $y\neq z$
and $e:\alpha(y)-\alpha(z)$ is an edge of $H$, and this pair defines an edge
$f$ in $G$; we define $\alpha(f):=e.$

If $((e,i),y)$ is not matched in $M$, then $e$ is a loop in $H$ incident with
$\alpha(y)$; $((e,i),y)$ is a loop $f$ of $G$ and define $\alpha(f):=e.$
\end{quote}
There are numerical constraints on $V^{\prime}$ and $\alpha$, as we will see
in Theorem 4.24. $\ $
\eject

\begin{figure}[h]
\vspace{2mm}
\begin{center}
\includegraphics[height=2.4647in,width=3.1713in]{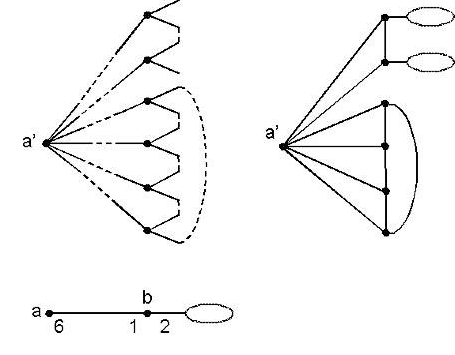}\vspace*{-4mm}
\caption{See Remark 4.17}%
\end{center}\vspace*{-2mm}
\end{figure}

For an example illustrating this construction, Figure 5 shows the weighted
graph $H$\ of Example 4.4 with vertices $a$ and $b$.\ It shows above an
intermediate step $H$\ in the construction of $G$, where $\alpha
^{-1}(a)=\{a^{\prime}\}$ and $\alpha^{-1}(b)$ consists of the six other
vertices. The half-edges are solid lines.\ The matching is shown by dotted
lines.\ The two half-edges that are not matched yield loops in the final graph
shown to the right. Their drawing recalls that they count for one in the
degree of their vertices. \ The graph $G$ covers the same weighted graph $H$
as the two graphs of Figure 4. $\square$

\medskip
As for graphs (Proposition 4.2(1)), we have:%

\bigskip
\noindent\textbf{Proposition 4.18:} Let $G,H,K$ be weighted graphs. If \ $\gamma
:G\rightarrow H$ and $\delta:H\rightarrow K$ are coverings, then so is
$\delta\circ\gamma:G\rightarrow K$. The same holds if $G$ is a graph, or if
$G$ and $H$\ are graphs.

\bigskip
\noindent\textbf{Remark 4.19:} If two disjoint weighted graphs are coverings
of $H$, then, their union is a covering of $H$.\ If $\gamma:G\rightarrow H$ is
a covering and $G$ is connected, then $H$ is connected because $\gamma$\ maps
every path in $G$ to a walk in $H$.\ If $H$ is not connected, then $G$ is the
union of (disjoint) coverings of its connected components. It follows that we
need only consider connected coverings of connected weighted \linebreak graphs.

\bigskip
\noindent\textbf{Examples 4.20:} 1) The complete bipartite graph $K_{3,4}$
(with 3+4 vertices) covers $H$ consisting of one edge whose half-edges have
weights 4 and 3. Although $H$ is a tree, Proposition 4.2(3) does not hold.
Proposition 4.2(2) does not either: $\left\vert V_{G}\right\vert $ need not be
a multiple of $\left\vert V_{H}\right\vert $ when $G$ is a covering of a
weighted graph $H$.

2) A graph $G$\ consisting of 3\ parallel edges covers the graph $W(G)$
consisting of an edge whose two half-edges have weight 3, that itself covers a
loop of weight 3.

3) If $H$\ has a loop of weight $p$ at a vertex $x$, then $H$ is covered by
the weighted graph built as follows: we remove the loop at $x$, obtaining thus
$H^{\prime}$; we take the union of $H^{\prime}$ and a disjoint copy of it
where $x^{\prime}$ is the copy of $x$ and we add one edge between $x$ and
$x^{\prime}$ whose two half-edges have weight $p$.

4) The two graphs of Figure 4, Example 4.4 cover both the weighted graph $H$
shown in Figure 5.\

5) The graph $G$\ consisting of two vertices, $x$ and $y$, an edge $e:x-y$ and
loops $f$ and $g$ at $x$ and $y$ with weights $\lambda(e,x)=3,$ $\lambda
(e,y)=2,$ $\lambda(f,x)=4$ \ and $\lambda(g,y)=5,$ covers $H$ consisting of a
single vertex with a loop of weight 7.

\medskip
The Kronecker product of a weighted graph $H$\ by an edge is a weighted
bipartite graph, whose universal covering is that of $H$.\ We will use this
notion in Section 6.

\bigskip
\noindent\textbf{Definition 4.21:} \emph{Kronecker product} \emph{by an edge.}\smallskip

Let $H$ be a weighted graph.\ Its \emph{Kronecker (or categorical) product} by
$K_{2}$ (a single edge) is the weighted bipartite graph $H\times K_{2}%
$\ defined as follows.\ Its vertex set is $V_{H\times K_{2}}:=V_{H}%
\times\{1,2\}$, partitioned into $(V_{H}\times\{1\},V_{H}\times\{2\})$. For
each edge $e$ of $H$\ between $x$ and $y\neq x$, $H\times K_{2}$ we have the
edge $e_{x,y}:(x,1)-(y,2)$ (and also $e_{y,x}:(y,1)-(x,2)).\ $A loop $e$ at
$x$ yields a unique non-loop edge $e_{x,x}:(x,1)-(x,2).$ The weight
$\lambda_{G}(e_{x,y},(x,i))$ is $\lambda_{H}(e,x)$ for $i=1,2$.\ $\square$

\bigskip
\noindent\textbf{Lemma 4.22:} Let $G$ and $H$ be weighted graphs.\smallskip

(1) There is a covering $H\times K_{2}\rightarrow H.$

(2) From a covering $\alpha:H\rightarrow G$ one can define a covering
$\alpha^{\prime}:H\times K_{2}\rightarrow G\times K_{2}$.

\begin{proof}
(1) The mapping $\pi$: $(x,i)\longmapsto x$, $e_{x,y}\longmapsto e$ is a
covering: $H\times K_{2}\rightarrow H$. If $H$ is connected and bipartite,
then $H\times K_{2}$ has two connected components, that are isomorphic.\ Each
of them is a covering of $H$.\

(2) From $\alpha:H\rightarrow G$, we define $\alpha^{\prime}:H\times
K_{2}\rightarrow G\times K_{2}$ by $\alpha^{\prime}(x,i):=(\alpha(x),i)$ and
$\alpha^{\prime}(e_{x,y}):=\alpha(e)_{\alpha(x),\alpha(y)}$.
\end{proof}

In particular, every (finite) weighted graph\ is covered by a (finite)
weighted bipartite graph.

\bigskip
\noindent\textbf{Example 4.23: }\emph{Weighted graphs, weight matrices and
coverings.}\smallskip

Every matrix $W:[p]\times\lbrack p]\rightarrow\mathbb{N}\cup\{\omega\}$ such
that $W[x,y]=0$\ implies $W[y,x]=0$\ is the weight matrix of a finite weighted
graph with $p$ vertices. The matrix $M:=$ $%
\begin{bmatrix}
1 & 3\\
2 & 0
\end{bmatrix}
$ is the weight matrix of $H$\ having one edge $e:x-y$, weights $\lambda
(e,x)=3$, and $\lambda(e,y)=2$ and a loop at $x$ of weight 1.\ It is covered
by the graph $G$ equal to $K_{2,3}$ with an additional edge between the two
vertices of degree 3. Then we have:
\begin{quote}
$B_{\alpha}=%
\begin{bmatrix}
1 & 0\\
1 & 0\\
0 & 1\\
0 & 1\\
0 & 1
\end{bmatrix}
$, $\ \ \ \ \ \ M_{G}=%
\begin{bmatrix}
0 & 1 & 1 & 1 & 1\\
1 & 0 & 1 & 1 & 1\\
1 & 1 & 0 & 0 & 0\\
1 & 1 & 0 & 0 & 0\\
1 & 1 & 0 & 0 & 0
\end{bmatrix}
$ \ \ \ \ \ \ and \hspace{2mm}
$M_{G}B_{\alpha}=%
\begin{bmatrix}
1 & 3\\
1 & 3\\
2 & 0\\
2 & 0\\
2 & 0
\end{bmatrix}
=B_{\alpha}M_{H}\ \ \ \ \ \ \ \ $
\end{quote}
where $\alpha(1)=\alpha(2)=1$ and $\alpha(3)=\alpha(4)=\alpha(5)=2$%
.$\ \square$

\medskip
The following theorem is stated without proof in \cite{Lei} but is\ essential
in this article (which proves a part of Theorem 4.10 ; see also \cite{FiaKra},
Section 4.1).

\bigskip
\noindent\textbf{Theorem 4.24:} Given a finite weighted graph $H$ with finite
weights, one can decide if it is covered by a finite unweighted graph $G$. If
this is the case, one can construct $G$ loop-free.

\begin{proof}
Let $H=([p],E,Inc,\lambda)$ be a finite weighted graph with finite
weights.\ We first assume that $H$ has no loops.

\smallskip
Assume that $\gamma:G\rightarrow H$ is a covering where $G$ is a finite graph.

For each $i$, let $w_{i}:=\left\vert \gamma^{-1}(i)\right\vert $. Let
$e_{i,j}:i-j$ be an edge of $H$, with $i<j$.\ Let $m_{i,j}=\lambda(e_{i,j},i)$
and $m_{j,i}=\lambda(e_{i,j},j$). We have $\left\vert \gamma^{-1}%
(e_{i,j})\right\vert =m_{i,j}.w_{i}=m_{j,i}.w_{j}.$

Consider the system $\Sigma_{H}$ of equations of the form $m_{i,j}%
.x_{i}=m_{j,i}.x_{j}$, with one equation for each edge $e_{i,j}$. It is
satisfied by the numbers ($w_{1},\dots,w_{p})$.\ This system may have no
solution. We give an example after the proof.

\medskip
\emph{Claim}\ \emph{1}: If $\Sigma_{H}$ has a solution ($w_{1},\dots,w_{p})$ in
positive integers, then this $p$-tuple is equal to ($\left\vert \gamma
^{-1}(1)\right\vert ,\dots,\left\vert \gamma^{-1}(p)\right\vert $)\ for some
finite covering $\gamma$ of $H$ by a graph $G$.

\smallskip
\emph{Proof }: We define $G$ from ($w_{1},\dots,w_{p})$.\ \ Its vertices are the
pairs $(i,s)$ where $i\in\lbrack p]=V$ and $s\in\lbrack w_{i}].$ For an edge
$e_{i,j}$ of $H$, we let $m:=m_{i,j}.w_{i}=m_{j,i}.w_{j}.$ We define as
follows $m$ edges $f_{1},\dots,f_{m}$ between the vertices $(i,s)$\ and
$(j,s^{\prime})$ where $s\in\lbrack w_{i}]$\ and $s^{\prime}\in\lbrack w_{j}]$.\

\medskip
We partition $[m]$ into pairwise disjoint intervals\footnote{We use intervals
to be easy and concrete, but any two partitions will work.\ They yield
different nonisomorphic graphs.} ;

$[m]=I_{1}\cup I_{2}\cup\dots\cup I_{w_{i}}$, where all intervals $I_{q}$ have
size $m_{i,j},$ and also

$[m]=J_{1}\cup J_{2}\cup\dots\cup J_{w_{j}}$, where all intervals $J_{q}$ have
size $m_{j,i}.$

For $k\in\lbrack m]$, we define an edge $f_{i,j,k}$ between $(i,s)$ and
\ $(j,s^{\prime})$ if and only if $k\in I_{s}\cap J_{s^{\prime}}.$ We define
$\gamma(f_{i,j,k}):=e_{i,j}.$ Hence, $\gamma$ is a surjective homomorphism.

For each vertex $(i,s$), if $e_{i,j}$ is an edge in $H$, then the edges
$f_{i,j,k}$ such that $\gamma(f_{i,j,k}):=e_{i,j}$ are those such that $k\in
I_{s}\cap J_{s^{\prime}}$ for some $s^{\prime}.$ There are $m_{i,j}$ such
edges. Similarly for each vertex $(j,s^{\prime}$) such that $e_{i,j}$ is an
edge in $H$ (hence where $i<j$) , there are $m_{j,i}$ edges $f_{i,j,k}$ such
that $\gamma(f_{i,j,k}):=e_{i,j}$: they are those such that $k\in I_{s}\cap
J_{s^{\prime}}$ for some $s.$ Hence, $G$\ is a finite covering of $H$.
$\square$

\medskip
\emph{Claim 2}: A system\ $\Sigma_{H}$ has a solution in positive integers if
and only if it has one in rational numbers. This is decidable and a solution
in positive integers can be computed if there is one. If $H$ is a tree, then
$\Sigma_{H}$ has a solution.

\smallskip
\emph{Proof :} We first decide if \ $\Sigma_{H}$ has a solution in real
numbers.\ We eliminate unknowns one by one.\

To eliminate an unknown $x$, we list the equations where it occurs: say
$ax=by,cx=dz,\dots,ex=fu$. Then, any solution must satisfy $ba^{-1}%
y=dc^{-1}z=\dots=fe^{-1}u$. We replace the equations containing $x$ by the new
equations $ba^{-1}y=dc^{-1}z=\dots=fe^{-1}u.$ The new system has one less
unknown and has a solution if and only if \ $\Sigma_{H}$\ has one.\ From it,
we get the value of $x$.\ We may obtain two equations concerning the same
variables, say $gy=hz$, and $g^{\prime}y=h^{\prime}z$, where $g,h,g^{\prime
},h^{\prime}$ are positive rational numbers.\ We have no solution if
$gh^{\prime}\neq g^{\prime}h$: we can stop the construction and report a
negative answer.\ Otherwise, we discard one of these two equations.

If there is a solution, there is one in positive rational numbers.\ To obtain
one in positive integers, it suffices to multiply all its components by the
least common multiple of the denominators.

If $H$ is a tree, then, at each step, we can eliminate an unknown that belongs
to a single equation, equivalently, that corresponds to a leaf.\ Hence, this
step does not create any new equation.\ The resulting system still corresponds
to a tree.\ We continue in the same way and we get a solution. $\square$\

\medskip
We now complete the main proof for weighted graphs with loops. Loops do not
create constraints: if we add to a weighted graph $L$ a loop of weight $q$
incident with a vertex $x$, and if a covering $\gamma$ of $L$\ by a graph $G$
has been found, then we need only add $q$ loops to $G$, incident to each
vertex in $\gamma^{-1}(x)$.\ We do that for all loops of the given graph $H$
and we get a covering as wanted.

\smallskip
If we replace the obtained graph $G$ by $G\times K_{2}$ of Lemma 4.22, we
obtain a loop-free graph that covers $G$ hence also $H$.
\end{proof}

\smallskip
\noindent\textbf{Example 4.25:} Let $H$ be the cycle $C_{3}$\ with vertices
1,2,3 and weights on its half-edges such that we get the equations
$2x_{1}=3x_{2},4x_{2}=3x_{3},x_{3}=5x_{1}$.\ This system has no solution in
positive integers.\ This means that $H$ is not covered by any finite
graph.\ It is covered by the infinite tree described as follows.\ Its set of
nodes is $N_{1}\cup N_{2}\cup N_{3}$ where $N_{1},N_{2},N_{3}$\ are infinite
and pairwise disjoint; each node in $N_{1}$ has 2 neighbours in $N_{2}$ and 5
in $N_{3}$, each node in $N_{2}$ has 3 neighbours in $N_{1}$ and 4 in $N_{3}$,
and each node in $N_{3}$ has 1 neighbour in $N_{1}$ and 3 in $N_{3}.$ This
tree does not cover any finite graph. $\square$

\medskip
The following corollary is a key fact in the proof of Theorem 4.10 by Leighton
\cite{Lei}. It is an immediate consequence of the proof of Theorem 4.24.\ If
$H$ is a graph, then the corresponding $p$-tuple is (1,\dots,1) by Proposition 4.2(2).\

\bigskip
\noindent\textbf{Corollary 4.26:} Let $H$ be a weighted graph with finite
weights and vertex set $[p]$.\ If it has finite coverings by graphs, then
there is a unique $p$-tuple $(n_{1},\dots,n_{p})\in(\mathbb{N}_{+})^{p}$\ such
that $\{(kn_{1},\dots,kn_{p})\mid k\in\mathbb{N}_{+}\}$ is the set of $p$-tuples
$(\left\vert \gamma^{-1}(1)\right\vert ,\dots,\left\vert \gamma^{-1}%
(p)\right\vert )$ such that $\gamma:G\longrightarrow H$ is a covering where
$G$ is a finite graph.

\subsection{Characteristic polynomials}

It is known that if $G$ is a covering of $H$ where $G$ and $H$ are finite
graphs, then the characteristic polynomial of $H$\ is a factor of that of $G$
(\cite{FiaKra}, Theorem 4). We extend this result to finite weighted graphs.

\bigskip
\noindent\textbf{Definitions 4.27: }\emph{Characteristic polynomials.}\smallskip

(a) The \emph{characteristic polynomial }$P_{M}$\emph{\ }of a $p\times p$
matrix $M$\ with coefficients in a ring with multiplicative unit, typically
$\mathbb{Z}$,$\mathbb{R}$ or $\mathbb{C}$, is defined as the determinant of
the matrix $M-xI_{p}$ where $I_{p}$ is the $p\times p$ (diagonal) unity
matrix, denoted by $\det(M-xI_{p})$.\ It is a polynomial in $x$ of degree $p$.
The \emph{characteristic polynomial }$P_{G}$ of a finite graph $G$ is defined
as that of its adjacency matrix $A_{G}$ that is symmetric with coefficients in
$\mathbb{N}$.\ The coefficients of $P_{G}$ are in $\mathbb{Z}$.

(b) We define the \emph{characteristic polynomial} of a finite weighted graph
$H$ with finite weights as $P_{H}:=\det(M_{H}-xI_{p})$ where $M_{H}$ is its
weight matrix, having coefficients in $\mathbb{N}$.\ \ For an example, if $H$
is as in Example 4.4, Remark 4.15(3) and Example 4.20(4), then $P_{H}%
=-x(2-x)-6=x^{2}-2x-6.$

\bigskip
\noindent\textbf{Theorem 4.28:} If $G$ and $H$ are finite weighted graphs with
finite weights and $G$ covers $H$, then $P_{H}$ is a factor of $P_{G}$.

\begin{proof}
Immediate consequence of Proposition 4.16 and the following one.
\end{proof}

The representation of a surjective map $\alpha:[q]\longrightarrow\lbrack p]$
by a $q\times p$ matrix $B_{\alpha}$ is in Definition 4.1(b).

\bigskip
\noindent\textbf{Proposition 4.29:} Let $M$ and $N$ be, respectively, $q\times
q$ and $p\times p$ matrices over a ring with multiplicative unit.\ Let
$\alpha:[q]\longrightarrow\lbrack p]$ be a surjective mapping.\ If
$MB_{\alpha}=B_{\alpha}N$, then $P_{N}$ is a factor of $P_{M}$.

\begin{proof}
We transform the matrix $M-xI_{q}$ by row and column operations into a matrix
$M^{\prime\prime}$ such that $\det(M-xI_{q})=\det(M^{\prime\prime})$.\

We do that in such a way that $M^{\prime\prime}$\ has the block structure $%
\begin{bmatrix}
N-xI_{p} & R\\
0 & S
\end{bmatrix}
.$ It follows that $\det(M-xI_{q})=\det(N-xI_{p}).\det(S)$, hence $P_{M}%
=P_{N}.\det(S)$.

\medskip
We can organize $M$\ in such a way that $i\in\alpha(i)$ for each $i\in\lbrack
p]$. This means that $i$ is the smallest element of each set $\alpha(i)$.\ For
each such $i$, we add to the $i$-th column of $M,$\ all its $j$-th columns,
for $j\in\alpha(i),j>i$.\

We obtain a matrix $M^{\prime}$ with same determinant as $M-xI_{q}.$ Since
$MB_{\alpha}=B_{\alpha}N,$ the first $p$ elements of the $j$-th line of
$M^{\prime}$ are the same as those of the $\alpha(j)$-th one . By substracting
the $i$-th line from each $j$-th line, for all $i\in\lbrack p],$ $j\in
\alpha(i),j>i$, we get a matrix $M^{\prime\prime}$\ of the desired form, with
same determinant as $M-xI_{q}$ and $M^{\prime\prime}$.\ This concludes the proof.
\end{proof}

\smallskip
\noindent\textbf{Example 4.30:} (1) For the matrices $N\ =M_{H}$ and $M=M_{G}$
of\ Example 4.23, we have $q=5,p=2,$\ and:

\begin{quote}
$M-xI_{5}=$ $%
\begin{bmatrix}
-x & 1 & 1 & 1 & 1\\
1 & -x & 1 & 0 & 0\\
1 & 1 & -x & 1 & 1\\
1 & 0 & 1 & -x & 0\\
1 & 0 & 1 & 0 & -x
\end{bmatrix}
$, $M^{\prime}=$ $%
\begin{bmatrix}
1-x & 3 & 1 & 1 & 1\\
2 & -x & 1 & 0 & 0\\
1-x & 3 & -x & 1 & 1\\
2 & -x & 1 & -x & 0\\
2 & -x & 1 & 0 & -x
\end{bmatrix}
$,

\bigskip
$M^{\prime\prime}=$ $%
\begin{bmatrix}
1-x & 3 & 1 & 1 & 1\\
2 & -x & 1 & 0 & 0\\
0 & 0 & -1-x & 0 & 0\\
0 & 0 & 0 & -x & 0\\
0 & 0 & 0 & 0 & -x
\end{bmatrix}
=%
\begin{bmatrix}
N-xI_{2} & R\\
0 & S
\end{bmatrix}
$
\end{quote}

so that $\det(M-xI_{5})=\det(N-xI_{2}).\det(S).$ One can check\footnote{By
using for instance https://www.dcode.fr/matrix-characteristic-polynomial
\par
{}} that:
\begin{quote}
$\det(N-xI_{2})=(x+2)(x-3),\det(S)=-x^{2}(x+1)$ and

$\det(M-xI_{5})=-x^{2}(x+1)(x+2)(x-3).$
\end{quote}

\smallskip
(2) If $G$ is a weighted graph with $p$ vertices, then $P_{G\times K_{2}%
}(x)=(-1)^{p}P_{G}(x).P_{G}(-x)$ where $G\times K_{2}$ is the Kronecker
product (Definition 4.21).\ This fact can be proved by using the algorithm of
the previous proposition.

\section{Universal coverings of weighted graphs}

We will construct the universal covering of a weighted graph from an unfolding
of an associated weighted and rooted digraph.\ This construction will
enlighten the relationships between universal coverings and complete
unfoldings.\ It extends the description given for graphs in Definition 4.6(2),
based on walks that do not traverse an edge twice in a row.\ Because of
weights, this construction is no longer convenient.\

Furthermore, we will use in a straighforward manner the results of Section 3.2
about complete unfoldings, in particular our adaptation of Norris's Theorem
(Theorem 3.20), to obtain a corresponding result about universal coverings of
finite weighted graphs. We will also define \emph{strongly regular graphs}, a
new notion linked with coverings of finite weighted graphs.

\subsection{Universal coverings of weighted graphs}

\noindent\textbf{Definition 5.1:} \emph{Universal coverings of weighted
graphs.}\smallskip

A covering of weighted graphs $\gamma:G\rightarrow H$ is \emph{universal} if
$G$ is a tree (without weights), which implies that $H$ is connected. We also
say that $G$ is a \emph{universal covering of} $H$.\

\bigskip

We will prove that any two universal coverings of a connected and weighted
graph are isomorphic. We first give some examples.

\bigskip
\noindent\textbf{Examples 5.2:} 1) An infinite tree whose nodes all have
degree $p$ where $1<p\leq\omega$ is a universal covering\ of a loop of weight
$p>1.$ All nodes of the tree are mapped to the vertex at the loop.\ It is also
a universal covering of an edge whose half-edges both have weight $p$.

2) A tree such that every node of degree 3 is adjacent to a node of degree 4
and vice-versa is a universal covering of $K_{3,4}$ and also, of an edge whose
half-edges have weights 4 and 3.

3) A tree consisting of one node adjacent to $\omega$ leaves is a universal
covering of an edge whose half-edges have weights 1 and $\omega$.

4) A universal covering $\gamma$\ of the graph $H$ consisting of a path
$x-y-z$ with a loop at $x$, all weights being 1, is the path $z_{1}%
-y_{1}-x_{1}-x_{2}-y_{2}-z_{2}$ with $\gamma(x_{1})=\gamma(x_{2})=x$,
$\gamma(y_{1})=\gamma(y_{2})=y$ and $\gamma(z_{1})=\gamma(z_{2})=z$.

5) A biinfinite path (cf.\ Example 4.9(2)), is a universal covering of the
following weighted graphs:

\vspace{1.5mm}(a) a cycle (in particular two parallel edges) whose half-edges have weight 1,
or an edge with both half-edges of weight 2,

(b) the weighted graph $H$ as in 4) except that the weight of the half-edge at
$z$ is 2,

(c) one loop of weight 2 or two loops of weight 1 incident to a same vertex,

(d) a path $P$\ with ends $x$ and $y\neq x$ such that, either $x$ and $y$ have
both a loop of weight 1, or $x$ has a loop of weight 1 and the half-edge
$(f,y)$ on\footnote{We mean that $f$ belongs to the path $P$.} $P$\ has weight
2, or the half-edges $(e,x)$ and $(f,y)$ on $P$\ has both weight 2. $\square$

\medskip
We will describe a construction of a universal covering for weighted graphs
and prove a characterization similar to that of complete unfoldings of Theorem
3.5, that entails unicity, $\emph{u.t.i.}$, of universal coverings.

\bigskip
\noindent\textbf{Definition 5.3:} \emph{The symmetric weighted digraph of a
weighted graph and its expansion.}

\smallskip
(a) Let $H=(V,E,Inc,\lambda)$ be a connected and weighted graph, for which we
fix a linear order $\leq$ on $V$.\ The associated \emph{symmetric weighted
digraph }is $Sym(H):=(V,E^{\prime},\lambda^{\prime})$ defined as follows. For
each edge $e:x-y$ of $E,$\ we define the following arcs of $E^{\prime}$ and
their weights:
\begin{quote}
if $x<y$ ($e$ is not a loop), we define\footnote{The purpose of the order on
vertices is to differenciate without ambiguity $e^{+}$ from $e^{-}$.}
$e^{+}:x\rightarrow y$ and $e^{-}:y\rightarrow x$, of respective weights
$\lambda(e,x)$ and $\lambda(e,y)$,

if $x=y$ ($e$ is a loop), we define $e^{\ell}:x\rightarrow x$ of weight
$\lambda(e,x).$
\end{quote}

(b) We define $ES(H)$ as the expansion of $Sym(H)$ (cf.\ Definition 3.7).\ It
is the (unweighted) digraph $(V,E^{\prime\prime})$ defined as follows,
directly from $H$. For each edge $e:x-y$ of $E,$ we define the following arcs
of $E^{\prime\prime}$:
\begin{quote}
($e^{+},i):x\rightarrow y$ if $x<y$, for $i\in\mathbb{N}_{+},$ $1\leq
i\leq\lambda(e,x),$

($e^{-},i):y\rightarrow x$ if $x<y$, for $i\in\mathbb{N}_{+}$, $1\leq
i\leq\lambda(e,y),$

($e^{\ell},i):x\rightarrow x$ if $x=y$ ($e$ is a loop) for $i\in\mathbb{N}%
_{+}$, $1\leq i\leq\lambda(e,x).$
\end{quote}

The digraphs $Sym(H)$\ and $ES(H)$\ are strongly connected as $H$ is
connected.$\ \ \ \ $ \smallskip

(c) Let $\iota:ES(H)\rightarrow H$\ be the homomorphism\footnote{A
homomorphism can map a digraph to a graph, cf.\ Section 2.} that is the
identity on $V_{H}=V_{ES(H)}$ and is defined as follows on the arcs of $ES(H)$:
\begin{quote}
$\iota(e^{+},i):=e,$ $\iota(e^{-},i):=e\ $and \ $\iota(e^{\ell},i):=e.$
\end{quote}
For each $x\in V_{H}$, it induces a weighted surjection of the set
$E_{ES(H)}^{+}(x)$ onto $(Inc_{H}(x),\lambda_{H}).\ \square$%

\begin{figure}[h]
\vspace*{-2mm}
\begin{center}
\includegraphics[height=1.55in,width=3.8536in]{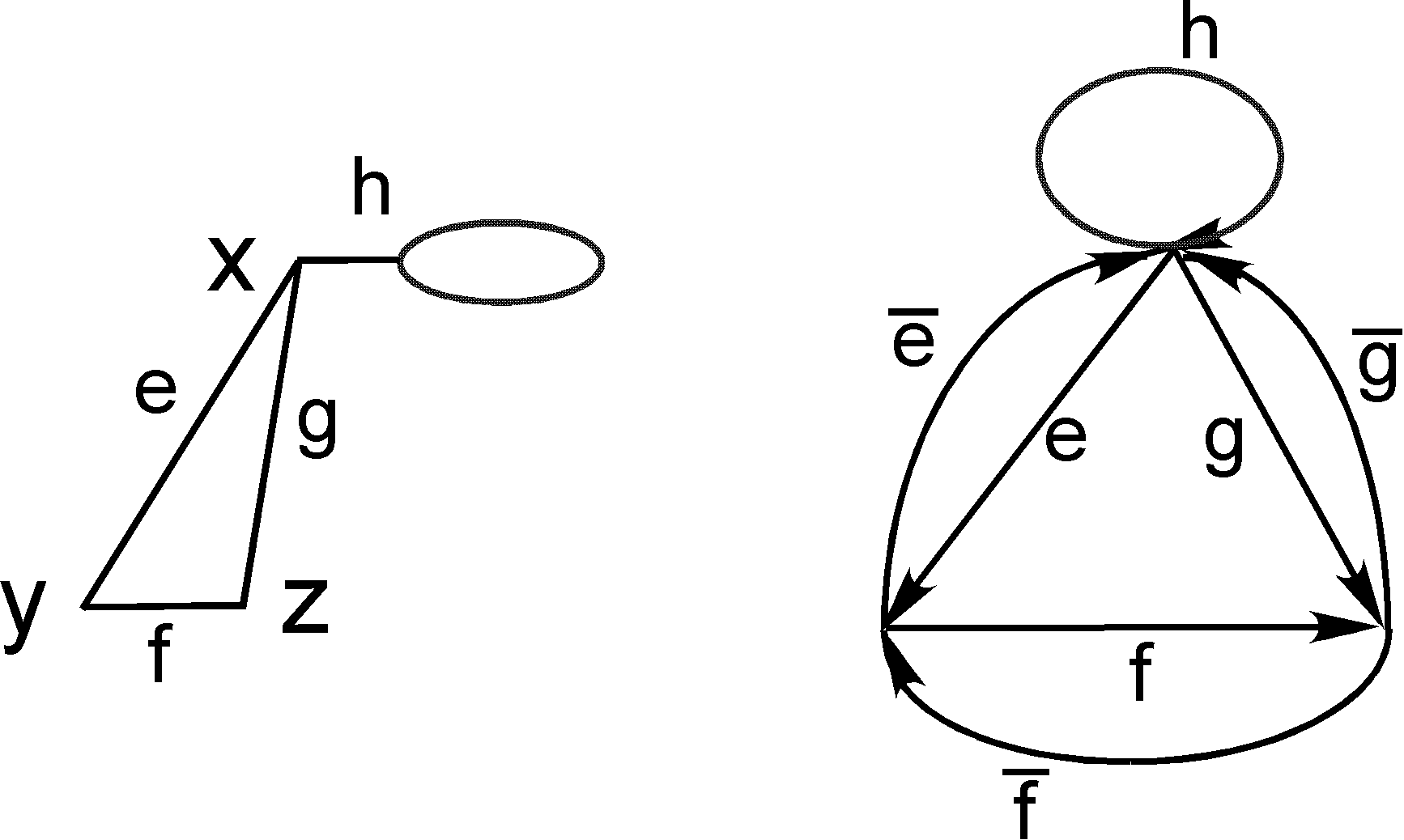}%
\caption{The graph $H$\ and the digraph $Sym(H)$ of Example 5.4.}%
\end{center}\vspace*{-3mm}
\end{figure}

Any vertex $x$ of the weighted digraph $Sym(H)$ can be taken as a \emph{root}.
We obtain a \emph{rooted} digraph denoted by $Sym(H)_{x}$, similarly as for
$T_{x}$, Section 2.3. The accessibility condition of Section 2\ is satisfied
because $Sym(H)$\ is strongly connected.\ We define $ES(H)_{x}$ in the same way.

\bigskip
\noindent\textbf{Example 5.4:} Figure 6 shows a graph $H$\ and the digraph
$Sym(H)$ defined from the ordering $x<y<z$. The drawing of the loop $h$ of
$H$\ recalls that it counts for 1 in the degree of $x$. For readability, we
denote in $Sym(H)$ the arc $e^{+}$ by $e$, the arc $e^{-}$ by $\overline{e}$
(and similarly for $f$ and $g$), and the loop $h^{\ell}$\ by $h$.\ As $H$ has
no weights, $i.e.$ all weights are 1, $ES(H)=Sym(H)$, with the arc $(e^{+},1)$
identified with $e^{+}$, and similarly for the other arcs.\

\medskip
Figure 7 shows the rooted tree $\mathit{Unf}(Sym(H)_{x})\upharpoonright3$ that
consists in the first three levels of $\mathit{Unf}(Sym(H)_{x})$. Its root is
denoted by $\overline{x}$. $\square$%

\begin{figure}[!h]
\begin{center}
\includegraphics[height=2.3099in,width=3.3382in]{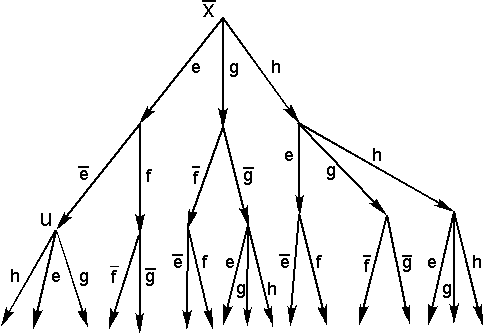}%
\caption{The rooted tree $\mathit{Unf}(Sym(H)_{x})\upharpoonright3$, see
Example 5.4.}\vspace*{-4mm}
\end{center}
\end{figure}

\medskip
For defining the universal covering of a weighted graph, we generalize, by the
following definition, the condition of Definition 4.6(2) requiring that the
walks defining nodes do not traverse twice in a row a same edge or loop. That
is, we eliminate from $\mathit{Unf}(ES(H)_{x})$\ the walks that violate this condition.

\bigskip
\noindent\textbf{Definition 5.5:} \emph{The pruning operation.} \smallskip

(a) Let $H$ be a weighted graph and $x\in V_{H}$.\ Then, $\mathit{Unf}%
(ES(H)_{x})$ is a rooted tree, whose root is denoted by $\overline{x}$ rather
than by $rt_{\mathit{Unf}(ES(H)_{x})}$.\ The \emph{pruned} rooted tree
$Pr(\mathit{Unf}(ES(H)_{x}))$ is obtained by deleting nodes and arcs as follows:

if a node $u$ of $\mathit{Unf}(ES(H)_{x})$\ is a walk $(e_{1},e_{2}%
,\dots,e_{n})$ in $ES(H)$\ (that starts from $x)$, $n>1$,\ and, for some $f\in
E$,
\begin{quote}
either $e_{n-1}=(f^{+},i),e_{n}=(f^{-},1),$

or $e_{n-1}=(f^{-},i),e_{n}=(f^{+},1),$

or $e_{n-1}=(f^{\ell},i)$, $e_{n}=(f^{\ell},1),$
\end{quote}
then, we remove from $\mathit{Unf}(ES(H)_{x})$ the arc from $w:=(e_{1}%
,e_{2},\dots,e_{n-1})$ to $u$ and the subtree issued from $u$.\

(b) We denote by $UC(H,x)$ the rooted tree $Pr(\mathit{Unf}(ES(H)_{x}%
))$.\ $\square$

\medskip
If $H$ is a graph, \emph{i.e.}, all weights are 1, then $UC(H,x)$ is as in
Definition 4.6. \

\bigskip
\noindent\textbf{Example 5.6: }We continue Example 5.4. Figure 8 shows the
rooted tree \ $Pr(\mathit{Unf}(Sym(H)_{x}))\upharpoonright
3=UC(H,x)\upharpoonright3.$\ The first case of pruning removes the subtree
$\mathit{Unf}(Sym(H)_{x})/u$ where $u$ is the head of the arc labelled by
$\overline{e}$ at level 2 in the tree of Figure 7. $\square$%

\begin{figure}[h]
\vspace*{2mm}
\begin{center}
\includegraphics[height=2.1in,width=1.7876in]{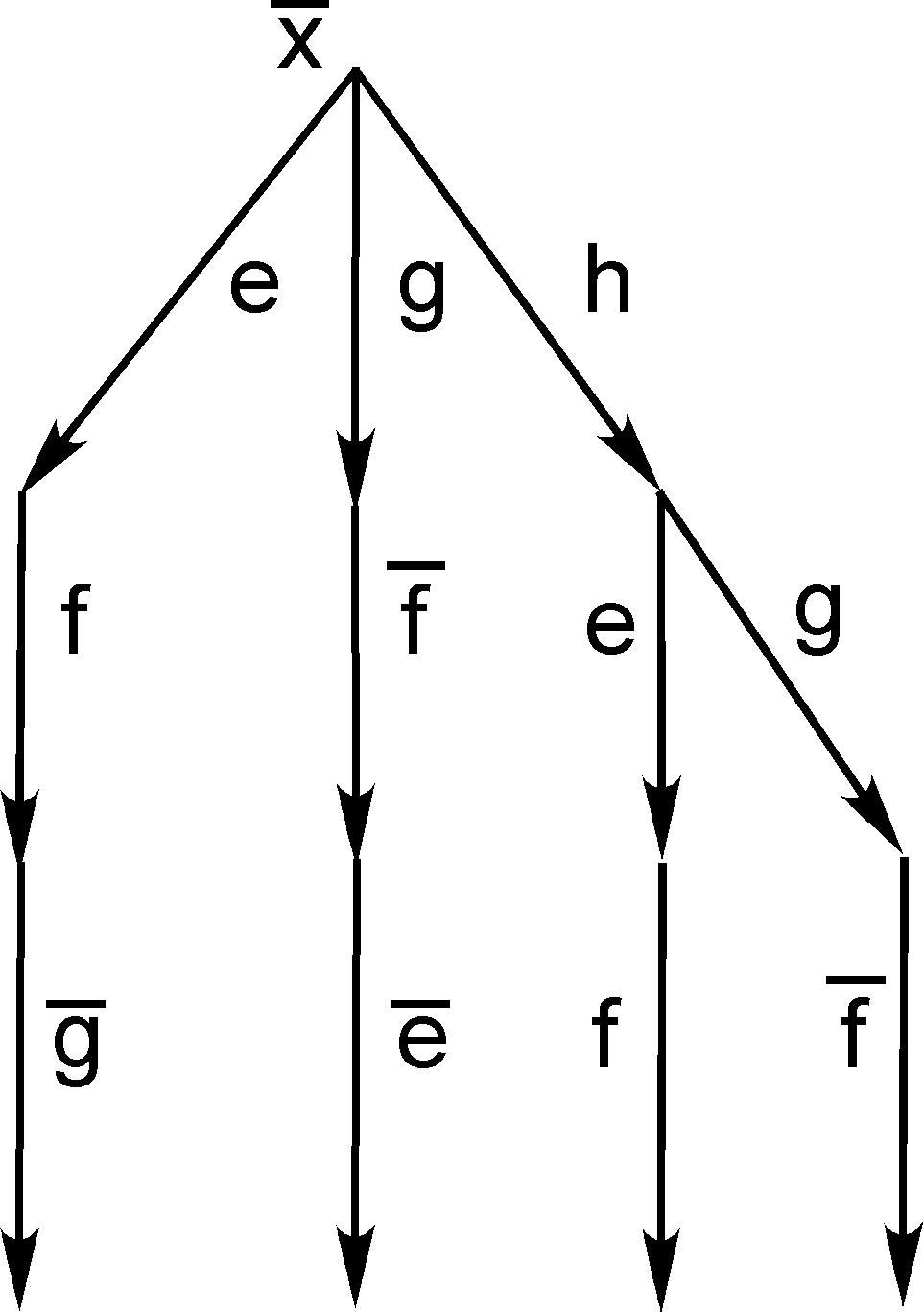}%
\caption{The rooted tree $UC(H,x)\upharpoonright3$ $=Pr(Unf(Sym(H)_{x}%
))\upharpoonright3.$ See Example 5.6.}%
\end{center}\vspace*{-1mm}
\end{figure}

The following theorem is similar to Theorem 3.5. We recall that $Unr$ forgets
the root and removes the orientations of a rooted tree.

\bigskip
\noindent\textbf{Theorem 5.7:} Let $H$ be a connected and weighted graph.

1) For each $x\in V_{H}$,\ the tree $Unr(UC(H,x))$ is a universal covering of
$H$.

2) If $\mu:T\rightarrow H$\ is a universal covering, then:
\begin{quote}
(C) For every covering $\kappa:G\rightarrow H$, where $G$ is connected and
weighted, there is a universal covering $\eta:T\rightarrow G$ such that
$\mu=\kappa\circ\eta$.
\end{quote}

3) Any two universal coverings of $H$ are isomorphic.\

4) If $\mu:T\rightarrow H$ is a covering such that Condition (C) holds, then
$T$ is a tree, hence a universal covering of $H$.

  \smallskip
  \begin{proof}
1) By Definition 5.3(c), we have a homomorphism\footnote{See Section 2.2 for
homomorphisms from digraphs to graphs.} $\iota:ES(H)\rightarrow H.$ Let $x\in
V_{H}$.\ We have an unfolding homomorphism $\alpha:\mathit{Unf}(ES(H)_{x}%
)\rightarrow ES(H)_{x}.$ It maps the root $\overline{x}$ of $\mathit{Unf}%
(ES(H)_{x})$ to $x$. We will prove that the homomorphism $\iota\circ\alpha:$
$\mathit{Unf}(ES(H)_{x})\rightarrow H/x$ induces a covering $Unr(UC(H,x))$
\ $=Unr(Pr(\mathit{Unf}(ES(H)_{x})))$ of $H.$

We let $W$ be the (unrooted tree)\ $Unr(Pr(\mathit{Unf}(ES(H)_{x})))$.\ We
claim that it is a universal covering of $H$, with covering homomorphism
induced by $\iota\circ\alpha.$

\medskip
First, we prove that $\alpha$\ is surjective on $Pr(\mathit{Unf}(ES(H)_{x})).$
Let $y\in V_{H}$.\ \ There is a path $P$\ in $H$ from $x$ to $y$. There is a
path $P^{\prime}$ in $\mathit{Unf}(ES(H)_{x})$ from $\overline{x}$ to some
$y^{\prime}$ whose image by $\iota\circ\alpha$ is $y$.\ This path neither uses
an arc of type $(f^{+},i)$ after one of type $(f^{-},j)$ or vice-versa, nor an
arc of type $(f^{\ell},i)$, otherwise $P$ would have an edge occurring twice
or a loop.

Hence, the path $P^{\prime}$ is not deleted by the pruning operation, so that
$y^{\prime}$ is in $Pr(\mathit{Unf}(ES(H)_{x}))$ and yields $y$ by $\iota
\circ\alpha$. Similarly, any edge $e$ of $H$ is on a path $P$\ from $x$ with
corresponding path $P^{\prime}$ in $\mathit{Unf}(ES(H)_{x})$ and $e$ is the
image under $\iota\circ\alpha$ of an arc in $P^{\prime}$.\ Hence, $\iota
\circ\alpha$\ is surjective on $Pr(\mathit{Unf}(ES(H)_{x})).$

\medskip
Next, we check the condition of Definition 4.14.\ Let $u$ be a node of
$Pr(\mathit{Unf}(ES(H)_{x})).$ Let $e$ be an edge of $H$\ incident to
$\alpha(u).$ The arcs of $\mathit{Unf}(ES(H)_{x})$ incident to $u$ whose image
by $\alpha$ is $e$ are as follows, according to different cases.

\medskip
\emph{Case 1}: $u$ is the root.\ There are $\lambda_{H}(e)$ such arcs.\ They
are all of type $(e^{+},i)$ (cf.\ the proof of Theorem 3.5(1) for the notion
of type), or all of type $(e^{-},i),$ or all of type $(e^{\ell},i)$ and they
are in $Pr(\mathit{Unf}(ES(H)_{x})).$ Hence $\lambda_{H}(e,\alpha
(u))=\left\vert \{e^{\prime}\mid e^{\prime}\in E_{T}(u),\iota\circ
\alpha(e^{\prime})=e\}\right\vert .$

 \smallskip
\emph{Case 2}: $u$ is not the root and is the head of an arc of type
$(f^{+},i)$ or $(f^{-},i)$ or $(f^{\ell},i)$\ where $f\neq e$. We are exactly
as in Case 1.

 \smallskip
\emph{Case 3}: $u$ is not the root and is the head of an arc of type
$(e^{+},i)$ for some $i$. The arcs of $\mathit{Unf}(ES(H)_{x})$ we are
considering are the $\lambda_{H}(e,\alpha(u))$ arcs of types $(e^{-},j)$
together with the arc with head $u$. Hence, we seem to have one arc too much.
But the pruning operation eliminates the arc $(e^{-},1).$ Hence, we still have
$\lambda_{H}(e,\alpha(u))=\left\vert \{e^{\prime}\mid e^{\prime}\in
E_{T}(u),\iota\circ\alpha(e^{\prime})=e\}\right\vert .$

 \smallskip
\emph{Case 4}: As in Case 3 with an incoming arc of type $(e^{-},i)$ or
$(e^{\ell},i)$ for some $i$. The argument is as in Case 3.

\medskip
Hence, $\iota\circ\alpha$\ induces (via the restriction to $Pr(\mathit{Unf}%
(ES(H)_{x}))$) a covering from the tree $W:=Unr(UC(H,x))$ to $H$.

 \smallskip
2) We prove the assertion for $W:=Unr(UC(H,x))\ $and\ $\iota\circ
\alpha:W\rightarrow H$ as in 1).\ We let $\kappa:G\rightarrow H$ be a covering
where $G$ is connected and weighted.\ Let $x^{\prime}\in V_{G}$ $\ $be such
that $\kappa(x^{\prime})=x.\ $For each $i$, we construct an $i$%
-\emph{covering} $\eta_{i}:UC(H,x)\upharpoonright i\rightarrow G_{x}$,
\emph{i.e.}, a homomorphism such that Condition (S')\ holds for all nodes of
\ $UC(H,x)\upharpoonright i$\ at depth less than $i$.\ This is similar to the
notion of $i$-unfolding in Definition 3.6. We want that $\eta_{i+1}$\ extends
$\eta_{i}$ and that $\kappa\circ\eta_{i}$ is the restriction of $\iota
\circ\alpha$ to $UC(H,x)\upharpoonright i.$

 \smallskip
For $i=0$, we define $\eta_{0}(\overline{x}):=x^{\prime}.$

\medskip
We now define $\eta_{i+1}$\ extending $\eta_{i}$ . Let $u$ be at depth $i.$
\ We have a weighted surjection from the set $E_{W}(u)$ to $(Inc_{H}%
(x),\lambda_{H})$ and a weighted surjection $(Inc_{G}(\eta_{i}(u)),\lambda
_{G})$ to $(Inc_{H}(x),\lambda_{H})$. Lemma 2.1(2) shows that we have a
weighted surjection $\beta$ from $E_{W}(u)$ to $(Inc_{G}(\eta_{i}%
(u)),\lambda_{G})$\ such that $\kappa\circ\beta=\ \iota\circ\alpha$\ \ on
$E_{W}(u).$ Furthermore, we can choose $\beta$ such that $\beta(e)=\eta
_{i}(e)$ where $e$ is the arc in $UC(H,x))\ $with head $\eta_{i}(u).$ (The
node $u$ is not the root of $UC(H,x)).$

If $v$ is the head of an arc with tail $u$ of type $(f^{+},i)$ or $(f^{-},i)$,
then $\beta(v)$ is the end of $f$ different from $\eta_{i}(u)$; if the type is
$(f^{\ell},i),$ then $\beta(v):=\eta_{i}(u).$

 \smallskip
We define $\eta_{i+1}$\ as $\eta_{i}$ \ extended by all such mappings $\beta$.
The union of the mappings $\eta_{i}$ yields a universal covering

\medskip
$\eta:W\rightarrow G$, where $W:=Unr(UC(H,x))$.

\medskip
If $G$ is a tree, then $\eta$ is an isomorphism $UC(H,x)\rightarrow G$ by
Proposition 4.2(3). This completes the proof of 2) and proves 3).

4) As in Theorem 3.5.
\end{proof}

\noindent\textbf{Corollary 5.8:} (1) If $\gamma:T\rightarrow H$\ is a
universal covering and $x\in N_{T}$, then $T_{x}\simeq UC(H,\gamma(x)).$

(2) If $x,y\in N_{T}$ and $\gamma(y)=\gamma(x),$\ then $T_{x}\simeq T_{y}$.

\begin{proof}
(1) Follows from the proof of Theorem 5.7(2).\

(2) If $x,y\in N_{T}$ and $\gamma(y)=\gamma(x),$\ then $T_{x}\simeq
UC(H,\gamma(x))=UC(H,\gamma(x))\simeq T_{y}$.
\end{proof}

As in Definition 4.6, we denote by $\boldsymbol{UC}(H)$ \emph{the} universal
covering of $H$, that is the isomorphism class of the trees $Unr(UC(H,x)).$

\bigskip
\noindent\textbf{Example 5.9: }Figure 9\ shows a weighted graph $H$\ and, to the right,
the digraph $ES(H)$. Figure~10\ shows the first two levels of $\mathit{Unf}%
(ES(H)_{x})$.\ The dotted arcs are eliminated by pruning. $\square$%

\begin{figure}[h]
\vspace*{3mm}
\begin{center}
\includegraphics[height=1.4in,width=4.5567in]{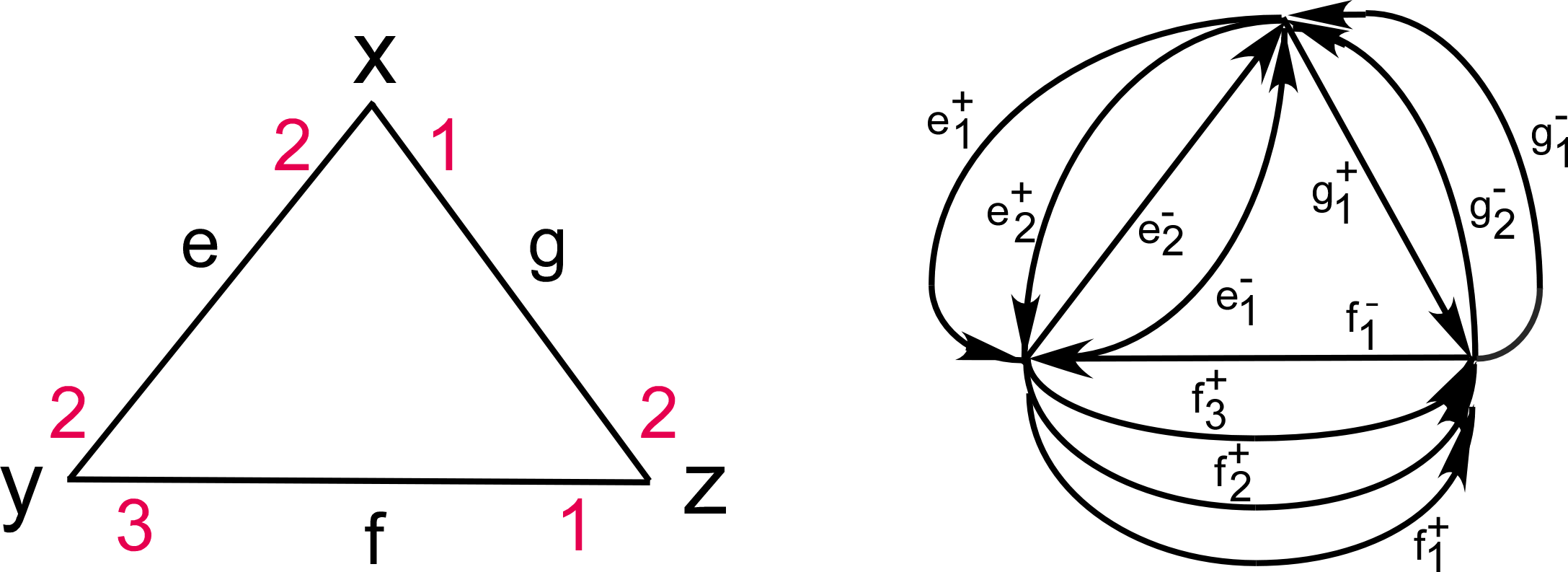}%
\caption{A weighted graph $H$ and the digraph $ES(H)$, see Example 5.9.}%
\end{center}\vspace*{-3mm}
\end{figure}
%

\medskip
If $R$ is a rooted tree, we define $Sym(R)$ by adding to $R$ an "up-going" arc
$v\rightarrow u$ for each arc $u\rightarrow v$. It is nothing but
$Sym(Unr(R))$ constructed by Definition 5.3 with all weights equal to 1 and a
linear order such that $x<y$ if $x\rightarrow y$ in $R$.\ We obtain a strongly
connected rooted digraph with root $rt_{R}$. See Figure 11 for an example.

\begin{figure}[ht]
\vspace*{1mm}
\begin{center}
\includegraphics[height=1.2642in,width=4.222in]{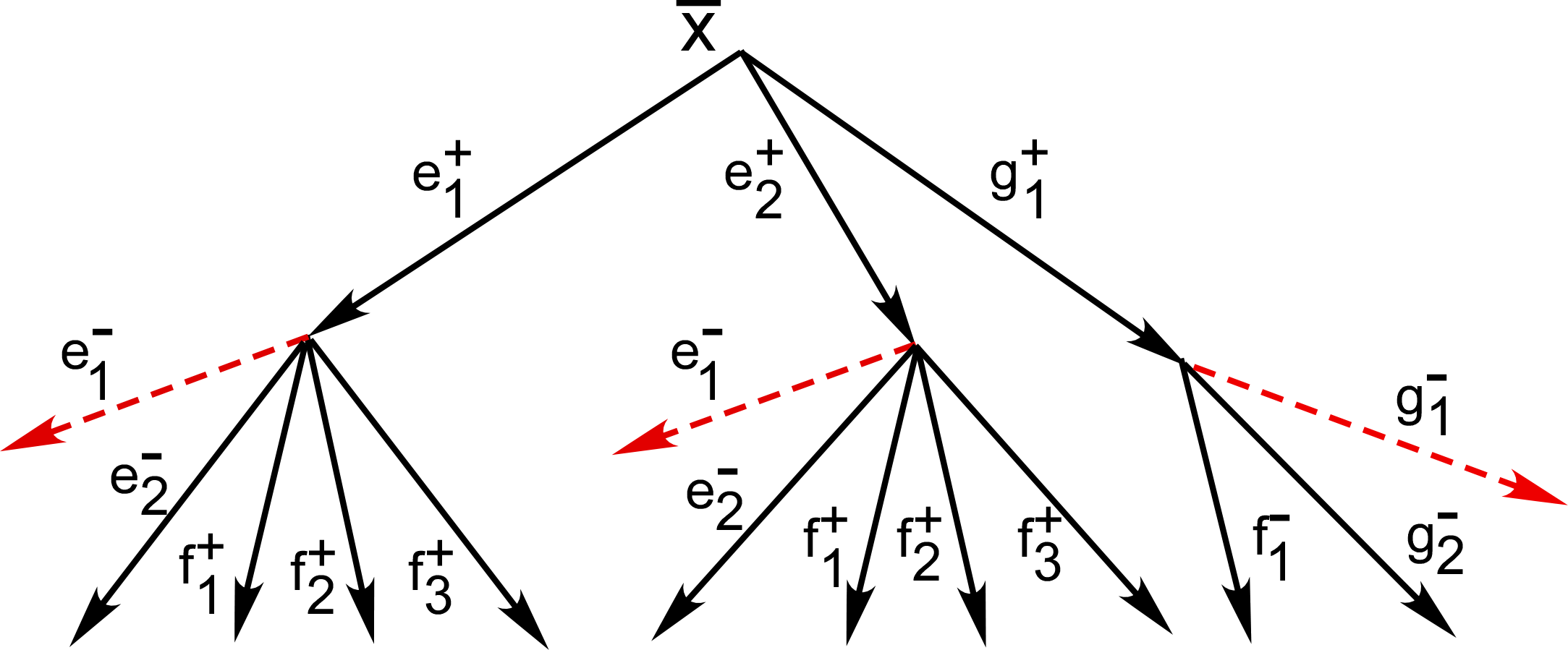}%
\caption{The tree $Unf(ES(H),x)\upharpoonright2,$ cf.\ Example 4.11. }%
\end{center}\vspace*{-5mm}
\end{figure}

\begin{figure}[!h]
\vspace*{3mm}
\begin{center}
\includegraphics[height=1.9908in,width=1.9501in]{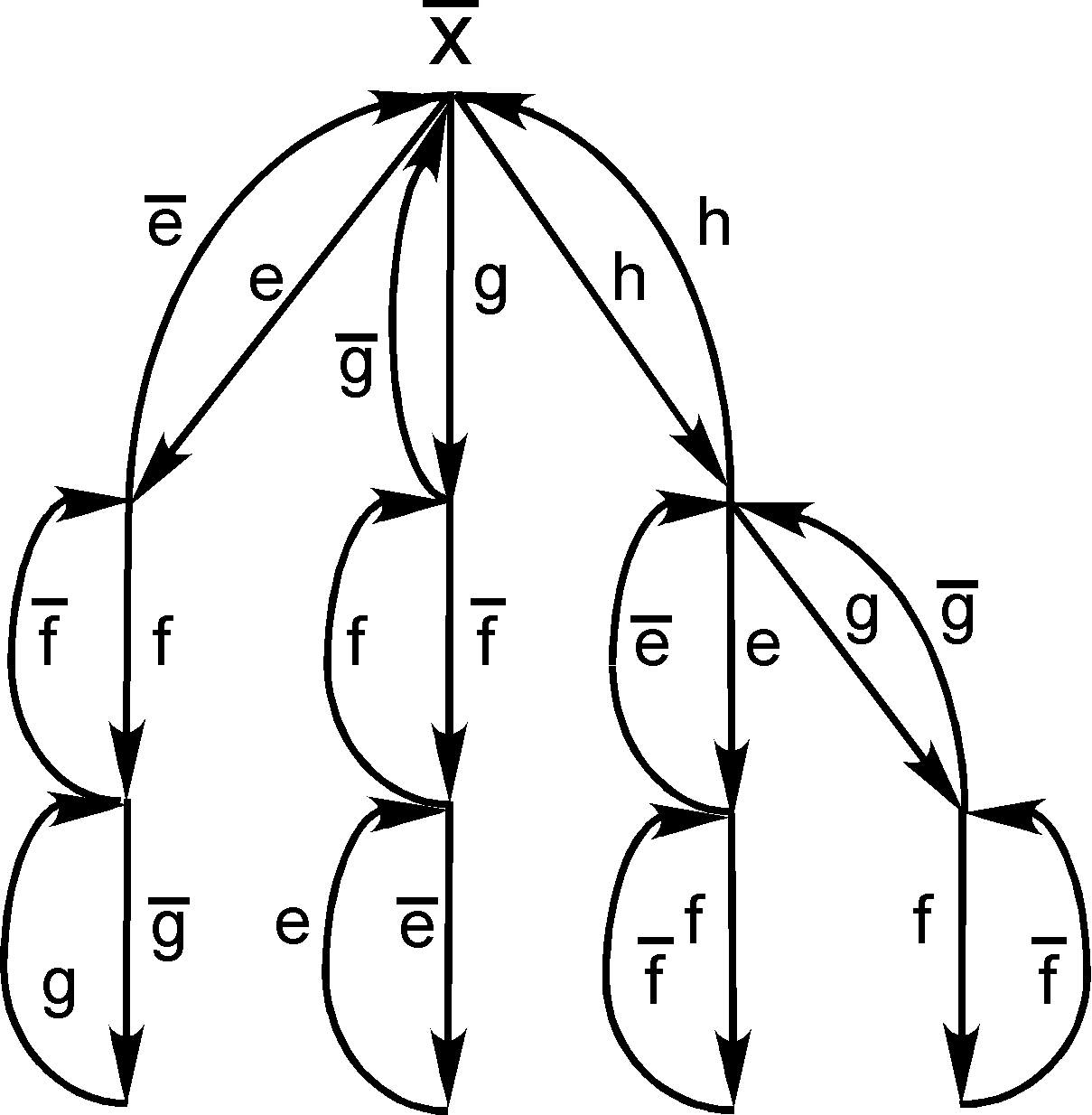}%
\caption{The top part of the digraph $Sym(UC(H,x))$, cf.\ Examples 5.4 and 5.9.}
\end{center}\vspace*{-2mm}
\end{figure}

\bigskip
\noindent\textbf{Proposition 5.10:} Let $H$ be a weighted connected graph and
$x\in V_{H}.$ We have \\ $\mathit{Unf}(Sym(UC(H,x)))\simeq\mathit{Unf}(ES(H),x).$

\begin{proof}
Let $\gamma:UC(H,x)\rightarrow H$ be the covering homomorphism\footnote{It
maps the root $\overline{x}$ of $UC(H,x)$ \ to $x$.}. It is actually a
homomorphism: $UC(H$, $x)\rightarrow ES(H)$ that is not surjective on arcs
because of pruning.\ We extend $\gamma$ into a surjective homomorphism
$\gamma^{\prime}:Sym(UC(H,x))\rightarrow ES(H)$ by defining $\gamma^{\prime
}(e)$ for $e\in E_{Sym(UC(H,x))}-E_{UC(H,x)}$ as follows:

\vspace{2mm}
if $e:u\rightarrow v$ is of the form\ $(f^{+},i)\in E_{ES(H)}$ (this means
that we have $f:\gamma(u)-\gamma(v)$ in $H$ and $\gamma(u)<\gamma(v))$,\ then,
$\gamma^{\prime}(e):=(f^{-},1)$,

if it is of the form $(f^{-},i)$, then, $\gamma^{\prime}(e):=(f^{+},1),$

if it is of the form $(f^{\ell},i)$, then, $\gamma^{\prime}(e):=(f^{\ell},1).$

\vspace{2mm}
These "up-going" arcs restablish some arcs deleted by pruning, but not the
deleted subtrees.

\smallskip
Then $\gamma^{\prime}:Sym(UC(H,x))\rightarrow ES(H)_{x}$ is an unfolding.\ It
follows from Theorem 3.5 that $\mathit{Unf}(Sym(UC(H,x)))$ is a complete
unfolding of $ES(H)_{x}$. Hence, $\mathit{Unf}(Sym(UC(H,x)))\simeq$
$\mathit{Unf}(Sym(ES(H)_{x})).$
\end{proof}

\noindent\textbf{Example 5.11: }We continue Example 5.4 illustrated in Figure
6.\ Figure 7 shows $\mathit{Unf}(ES(H)_{x})$.   Figure 8 shows $Pr(\mathit{Unf}(ES(H)_{x}))=UC(H,x)$.\ Figure 11\ shows $Sym(UC(H,x))$  and its up-going
arcs.\ The deleted arc labelled by $\overline{e}$  that reaches $u$(see Figure 7) is restablished towards $\overline{x}$.\ The three arcs outgoing from
$u$ and labelled by $h,e$ and $g$ are not. $\square$%

\medskip
The following theorem relates universal coverings to complete unfoldings.\ We
will use it for proving Theorem 5.15\ from Theorem 3.20.

\bigskip
\noindent\textbf{Theorem 5.12:} Let $H$be a weighted graph.\ For every two
vertices $x$and $y$, we have:
\begin{quote}
$UC(H,x)\simeq UC(H,y)$ if and only if $\mathit{Unf}(ES(H)_{x})\simeq
\mathit{Unf}(ES(H)_{y}).$
\end{quote}

If $H$ is finite, this property is decidable.

\begin{proof}
If $UC(H,x)\simeq UC(H,y)$, we have $Sym(UC(H,x))\simeq$\ \ \ $Sym(UC(H,y))$.
Hence, $\mathit{Unf}(ES(H)_{x})\simeq\mathit{Unf}(ES(H)_{y})$ by Proposition 5.10.\smallskip

For the converse, observe that $UC(H,x):=Pr(\mathit{Unf}(ES(H)_{x}))$, hence
is defined by using the definition of nodes as walks.\

The definition of $Pr(R)$ where $R:=\mathit{Unf}(ES(H)_{x})$\ uses a mapping
$s$ such that any node $u$ of $R$\ that is not the root is mapped by $s$ to
one of its sons such that $R/s(u)\simeq R/w$ where $w$ is the father of~$u$.

Let $S:=\{s(u)\mid u\in N_{R},u\neq rt_{R}\}$. Then $Pr(R)$ is obtained from
$R$\ by deleting the subtrees $R/v$ for all $v\in S$.

\medskip
Assume now that $R$ is any rooted tree isomorphic to\ $\mathit{Unf}%
(ES(H)_{x})$ and that $S^{\prime}$ is any subset of $N_{R}$\ such that:

\smallskip
each node $v$ in $S^{\prime}$ is at depth at least 2,

\smallskip
each node $u\neq rt_{R}$ has a unique son $v$ in $S^{\prime}$,

\smallskip
and $R/v\simeq R/w$ where $w$ is the father of $u$ that is itself the father
of $v$.

\medskip
Then, the labelled rooted trees $(\mathit{Unf}(ES(H)_{x}),S)$ and
$(R,S^{\prime})$ are isomorphic\footnote{A node\ is labelled by 1 if it is in
$S$\ or in $S^{\prime}$ and by 0 otherwise.}. It follows that $Pr(\mathit{Unf}%
(ES(H)_{x}))$ is isomorphic to the tree obtained from $R$ by deleting the
subtrees $R/v$ for all $v\in S^{\prime}$. Hence $UC(H,x)$ can be constructed,
\emph{u.t.i}, from any rooted tree isomorphic to $\mathit{Unf}(ES(H)_{x})$ and
any appropriate set $S^{\prime}$, without using the concrete description of
the nodes of $\mathit{Unf}(ES(H)_{x})$ by walks.\ It follows that
$UC(H,x)\simeq UC(H,y)$ if $\mathit{Unf}(ES(H)_{x})\simeq\mathit{Unf}%
(ES(H)_{y})$.

The last assertion follows from Theorem 3.14 applied to $Sym(H)$ by using Algorithm 3.18.
\end{proof}

The next proposition defines from a tree a canonical weighted graph of which
it is a universal covering.

\bigskip
\noindent\textbf{Proposition 5.13:} Let $T$ be a tree and $\sim$ be an
equivalence relation on $N_{T}$ satisfying the following condition:
\begin{quote}
(N): if $v\sim v^{\prime}$, $w$ is a neighbour of $v$, and $v$ has exactly $p$
($p$ may be $\omega$) neighbours equivalent to $w$, then $v^{\prime}$ has
exactly $p$ neighbours equivalent to $w$,
\end{quote}
then $T$ is a universal covering of the weighted graph $H:=T/\sim$\ defined as follows:
\begin{quote}
-- $\ V_{H}:=N_{T}/\sim$,

-- $E_{H}$ contains the edge $e:[v]_{\sim}-[w]_{\sim}$ if and only if $v$ is
adjacent to some vertex in $[w]_{\sim}$ if and only if, by Condition (N), each
vertex of $[v]_{\sim}$ is adjacent to some vertex in $[w]_{\sim}$,

-- the weight $\lambda(e,[v]_{\sim})$ is the number of edges of $T$ linking $v$
and a vertex in some $[w]_{\sim}$ such that $w$ is adjacent to $v$.
\end{quote}

\begin{proof}
Condition (N) implies that an edge $[v]_{\sim}-[w]_{\sim}$ is defined from an
edge $v-w$ of $T$, and that $\lambda(e,[v]_{\sim})$ is well-defined.\ The
mapping $\gamma$\ such that $\gamma(v)=[v]_{\sim}$ and $\gamma(e)$\ is the
edge $[v]_{\sim}-[w]_{\sim}$ if $e:v-w$ is a universal covering of $H$. If
$v\sim w$, then the edge $[v]_{\sim}-[w]_{\sim}$ is a loop.
\end{proof}

\noindent\textbf{Corollary 5.14:} Every tree is a universal covering of a
weighted graph.

\begin{proof}
If $T$ is a tree and $\approx$ is the equivalence relation on $N_{T}$\ defined
by $x\approx y$ if and only if $T_{x}\simeq T_{y}$ , then $H:=T/\approx$ is a
weighted graph and $T$ is a universal covering of it.
\end{proof}

\subsection{Universal coverings of finite weighted graphs}

We will call \emph{strongly regular} the universal coverings of finite
weighted graphs. These regular trees have not been previously identified to
our knowledge. We first extend a result proved by Norris \cite{KreVer,Nor} for
graphs without weights.\ Our proof will use Theorem 3.20, the similar result
for complete unfoldings, by means of Theorem 5.12.

\bigskip
\noindent\textbf{Theorem 5.15:} Let $T$\ be a universal covering of a finite
weighted graph $H$ with $p$ vertices. For every two nodes\ $x,y$ of $T$, we
have $T_{x}\simeq T_{y}$ if and only if $T_{x}\upharpoonright(p-1)\simeq
T_{y}\upharpoonright(p-1)$.

\begin{proof}
We prove the property for $T_{x}:=UC(H,x)$\ and $T_{y}:=UC(H,y)$. \ The "only
if " direction is clear.\smallskip

For proving the converse, assume $UC(H,x)\upharpoonright(p-1)\simeq
UC(H,y)\upharpoonright(p-1)$.\ We have also $\ Sym(UC(H,x)\upharpoonright
(p-1))\simeq Sym(UC(H,y)\upharpoonright(p-1))$.\ The directed walks of length
$p-1$ in $ES(H)$ that start from $x$ are in bijection with the directed paths
of length $p-1$ in $Sym(UC(H,x)\upharpoonright(p-1))$ that start from
$\overline{x}$\ , the root of $UC(H,x).$ It follows that $\mathit{Unf}%
(ES(H)_{x})\upharpoonright(p-1)\simeq\mathit{Unf}(ES(H)_{y})\upharpoonright
(p-1).$ Hence, by Theorem 3.20,\ we get $\mathit{Unf}(ES(H)_{x})\simeq
\mathit{Unf}(ES(H)_{y})$ and, by Theorem 5.12, $UC(H,x)\simeq UC(H,y)$.
\end{proof}

\noindent\textbf{Example 5.16:} Let us consider the graph $H$ of Example 5.4
and Figures 6,7,8 (Section 5.1). Figure 7 shows $\mathit{Unf}(Sym(H)_{x}%
))\upharpoonright3$, and Figure 8 the result of pruning it. Figure 11 shows
the first three levels of $Sym(UC(H,x)).$ The directed paths of length 3 in
the tree $\mathit{Unf}(Sym(H)_{x})$ that start from the root $\overline{x}$
correspond bijectively to the directed walks of length 3 in $Sym(UC(H,x))$
that start from $\overline{x}.$\ In the proof of Theorem 5.15, we use a
similar observation for $ES(H)$ where $H$ is weighted. $\square$

\medskip
As a consequence of Theorem 5.15, we obtain in \cite{CouNext} a first-order
definability result for the strongly regular trees $\boldsymbol{UC}(H)$
similar to that for regular trees following from Theorem 3.20.

\bigskip
\noindent\textbf{Definition 5.17:} \emph{Regular unrooted trees}.\smallskip

We recall that the \emph{subtrees} of a (labelled) rooted tree $R$ are the
(labelled) rooted trees $R/x$ for $x\in N_{R}$. Their nodes are those of $R$
accessible from $x$.\ By Definition 3.8,\ a rooted (labelled) tree $R$ is
\emph{regular} if the set of isomorphism classes $[R/x]_{\simeq}$ for $x\in
N_{R}$\ is finite.\ In that case, its cardinality is the \emph{regularity
index} $Ind(R)$ of $R$.

A (labelled) tree $T$ without root is \emph{regular }if the rooted (labelled)
tree $T_{x}$ is regular for some $x\in N_{T}$.

\bigskip

\noindent\textbf{Proposition 5.18:} If a (labelled) tree $T$ is regular, then
the rooted (labelled) trees $T_{y}$ are regular for all $y\in N_{T}$.

\begin{proof}
Let $T_{x}$ be regular for $x\in N_{T}$. If $y$ is a neighbour of $x$, then
the subtrees of $T_{y}$ are $T_{y}$, $T_{y}/x$ and the subtrees $T_{x}/z$ for
$z\notin\{x,y\}.$ Hence there are finitely many up to isomorphism. If $y$ is
at distance $n$ of $x$, there is a path $x-z_{1}-\dots-z_{n-1}-y$ and each
rooted tree $T_{z_{1}},\dots,T_{z_{n-1}},T_{y}$ is regular by the first observation.
\end{proof}

We may have $Ind(T_{y})>Ind(T_{x})$, as shown in Example 5.21.

\bigskip
\noindent\textbf{Definition 5.19:} \emph{Strongly regular trees}.\smallskip

A possibly labelled tree $T$ is \emph{strongly regular} if it has finitely
many associated rooted trees $T_{x}$, \emph{u.t.i}, that is, if the set
$\{[T_{x}]_{\simeq}\mid x\in N_{T}\}$ is finite. $\square$

\medskip
We will prove that a strongly regular tree is regular.\ This is not an
immediate consequence of the definition as we do not require that any of the
trees $T_{x}$ is regular. However, all are.

\bigskip
\noindent\textbf{Example} \textbf{5.20:} The rooted tree $P$\ such that
$N_{P}:=\mathbb{N}$\ and $x\leq_{P}y$ if and only if $y\leq x$ is an
\emph{infinite path} $P$.\ It is regular, hence, the tree $Unr(P)$ is
regular.\ The rooted trees $Unr(P)_{x}$ are all regular but pairwise non
isomorphic. Hence, $Unr(P)$ is not strongly regular.

\bigskip
\noindent\textbf{Proposition 5.21:} Let $H$ be a finite, connected and
weighted graph.\smallskip

(1) Its universal coverings are strongly regular.

(2) For each $x\in V_{H}$ the rooted tree $UC(H,x)$ is regular.

\begin{proof}
(1)\textbf{ }If $\eta$: $T\rightarrow H$ is a universal covering, then for
each node $x$ of $T$, we have $T_{x}\simeq UC(H,\eta(x))$ by Corollary
5.8(1).\ Hence, $T$ is strongly regular.

(2) Let $x\in V_{H}$.\ The rooted tree $\mathit{Unf}(ES(H)_{x})$ from which we
get $UC(H,x)$ by pruning is regular, but this is not enough to conclude.

Let $\gamma$: $\mathit{Unf}(ES(H)_{x})\rightarrow Sym(H)$ be the homomorphism
that is the composition of the unfolding $\alpha:\mathit{Unf}(ES(H)_{x}%
)\rightarrow ES(H)_{x}$ and $\beta:ES(H)\rightarrow Sym(H)$ where $\beta$ is
the identity on vertices.

Let $u$ and $u^{\prime}$ be nodes of $UC(H,x),$ hence of $\mathit{Unf}%
(ES(H)_{x}),$ that are not the root.\ Let $e:v\rightarrow u$ and $e^{\prime
}:v^{\prime}\rightarrow u^{\prime}$ be the arcs of $\mathit{Unf}(ES(H)_{x})$
with heads $u$ and $u^{\prime}$.\ If $\gamma(e)=\gamma(e^{\prime})$ then
$\gamma(u)=\gamma(u^{\prime})$ and we have $Pr(\mathit{Unf}(ES(H)_{x}%
))/u\simeq Pr(\mathit{Unf}(ES(H)_{x}))/u^{\prime}.\ $Note that $\gamma$\ maps
$\mathit{Unf}(ES(H)_{x})$ to $Sym(H)$.\ We may have $e=(f^{+},i)$ and
$e^{\prime}=(f^{+},j)$ so that $\gamma(e)=\gamma(e^{\prime})=f^{+}$.\

It follows that $UC(H,x)/u\simeq UC(H,x)/u^{\prime}$ and that the subtrees of
$UC(H,x)$ are $UC(H,x)$ itself and those associated as above with the arcs of
$Sym(H)$.\ Hence, there are at most $1+2.\left\vert E_{H}\right\vert $
subtrees \emph{u.t.i}, and $UC(H,x)$ is regular.
\end{proof}

\noindent\textbf{Theorem 5.22:} A tree $T$ is strongly regular if and only if
it is the universal covering of a finite, connected and weighted graph if and
only if it is the universal covering of such a graph without loops.

\begin{proof}
If $T$ is the universal covering of a finite, connected and weighted graph,
then it is strongly regular by Proposition 5.21.

\medskip
Conversely, let $T$ be a strongly regular tree.\ Let $\sim$ be the equivalence
relation on $N_{T}$ such that $x\sim y$ if and only if $T_{x}\simeq T_{y}%
$.\ This equivalence relation satisfies Condition (N) of Proposition 5.13 and
has finitely many classes.\ Hence, by this proposition, $T$ is a universal
covering of the finite weighted graph $H:=T/\sim$.

Finally we show how to replace $H$ by $H^{\prime}$\ without loops.\ A loop of
weight $p$ arises in $H$\ if a node has $p$ neighbours equivalent to it. To
avoid loops, we define on $T$ a proper 2-coloring.\ We define $\sim^{\prime}$
such that $x\sim^{\prime}y$ if and only if $T_{x}\simeq T_{y}$ and $x$ and $y$
have the same color. Then $T$ is a universal covering of the finite weighted
graph $H^{\prime}:=T/\sim^{\prime}$ that has no loop\footnote{Note that
$H^{\prime}$ is a connected component of $H\times K_{2}$ defined in Definition
3.21.}.
\end{proof}

\noindent\textbf{Examples 5.23:} 1) Let $T$ consist of a biinfinite path
$B$\ (cf.\ Example 4.9(2)), where each node $x$ has, in addition, an incident
pendent edge $x-x^{\prime}$ for some new node $x^{\prime}$. The rooted trees
$T_{x}$ for $x\in N_{B}$ are all isomorphic, and so are the trees
$T_{x^{\prime}}$. The quotient graph is the edge $[x]_{\sim}-[x^{\prime
}]_{\sim}$ together with a loop at $[x]_{\sim}$ of weight 2, that yields trees
isomorphic to $T_{x}$.\ The two other half-edges have weight 1.

2) For the tree of Example 4.20(1), we get an edge with weights 3 and 4.

\bigskip
\noindent\textbf{Remark 5.24:}  1) Finite weighted graphs can be used as finite descriptions of strongly
regular trees, even of infinite degree. The construction of Theorem 5.22
defines a minimal and canonical one.

2) By Theorem 5.22,\ a strongly regular tree is the universal covering of a
finite minimal weighted graph $H$.\ It is not necessarily that of a finite
graph $G$, otherwise such a graph $G$\ would cover $H$, and Example
4.25\ shows that this may be not possible.\

\bigskip
\noindent\textbf{Corollary 5.25:} Every node-labelled strongly regular\ tree
is regular.

\begin{proof}
Immediate from Theorem 5.22\ and Proposition 5.21.
\end{proof}

\section{Common coverings of finite graphs}

Our aim is to examine the theorem by Leighton \cite{Lei} that we stated in
Theorem 4.10.\ Its proof is quite difficult. Alternative no more easier proofs
have been given that use tools from combinatorics, topology and group theory
\cite{AFS,BK,Moh,Neu,Tuc,Wo}. We will give an easy proof for particular cases,
including that of $k$-regular graphs proved in \cite{AngGar}.

\bigskip
\noindent\textbf{Theorem 6.1:} If two finite connected graphs $G$ and $H$ are
coverings of a same graph $M$, they have a common covering by a graph $K$
having at most $4\left\vert V_{G}\right\vert .\left\vert V_{H}\right\vert $
vertices. The graph $K$ has at most $\left\vert V_{G}\right\vert .\left\vert
V_{H}\right\vert $ vertices if $M$ is loop-free.

\begin{proof}
We first assume that $G,H$ and $M$ are loop-free. (They may have parallel edges).\smallskip

Let $\alpha:$ $G\rightarrow M$ and $\beta:$ $H\rightarrow M$ be coverings. It
follows that, if $x-x^{\prime}$ is an edge of $G$, then $\alpha(x)\neq
\alpha(x^{\prime}),$ and similarly for $\beta$. We construct $K$ as follows:

\medskip
$V_{K}:=\{(x,y)\in V_{G}\times V_{H}\mid\alpha(x)=\beta(y)\}$.

$E_{K}:=\{(e,f)\in E_{G}\times E_{H}\mid\alpha(e)=\beta(f)\}$.

\medskip
An edge $(e,f)$ of $K$ links $(x,y)$\ and $(x^{\prime},y^{\prime})$\ if
$e:x-x^{\prime}$ and $f:y-y^{\prime}$. We cannot not have $(e,f)$ linking also
$(x,y^{\prime})$\ and $(x^{\prime},y)$\ because this would mean that
$\beta(y)=\alpha(x)=\alpha(x^{\prime})$ contradicting a previous remark.

We define $\gamma:$ $K\rightarrow G$ as the first projection, \emph{i.e.},
$\gamma(x,y):=x$ and $\gamma(e,f):=e.$\ Similarly $\eta:$ $K\rightarrow H$ is
the second projection. It is clear that $\gamma$ and $\eta$ are homomorphisms.\

We prove that $\gamma$ is surjective. Let $x\in V_{G}.$ There is $y\in V_{H}%
$\ such that $\beta(y)=\alpha(x)$ because $\beta$\ is surjective.\ Hence
$(x,y)\in V_{K}$ and $\gamma(x,y)=x.$ Let $e:x-x^{\prime}$ be an edge of
$G$.\ Then $\alpha(e):\alpha(x)-\alpha(x^{\prime})$ is an edge of $M$. There
is in $H$ an edge $f:y-y^{\prime}$ such that $\beta(f)=\alpha(e)$. We have
$\beta(f):\beta(y)-\beta(y^{\prime})$, $\beta(y)=\alpha(x),\beta(y^{\prime
})=\alpha(x^{\prime}).$ Then, $\eta$ is surjective too.

It remains to prove that $\gamma$ and $\eta$ are coverings.\ We prove that for
$\gamma$.

\medskip
Consider $(x,y)\in V_{K}\ $and its image $x$ in $G$\ by $\gamma$. Let
$e_{1},\dots,e_{p}$ be the edges of $G$\ incident with $x$.\ Let $f_{1}%
,\dots,f_{q}$ be the edges of $H$\ incident with $y$.\ The edges of
$M$\ incident with $\alpha(x)$ are $\alpha(e_{1}),\dots,\alpha(e_{p})$ that are
pairwise distinct.\ Those incident with $\beta(y)$ are $\beta(f_{1}%
),\dots,\beta(f_{q})$, also pairwise distinct.\ But $\alpha(x)=\beta(y),$ hence,
$q=p$, and we can renumber these edges so that $\alpha(e_{i})=\beta(f_{i}%
)$\ for each $i$.\ The edges of $K$ incident with $(x,y)$ are thus
$(e_{i},f_{i})$ for $i=1,..,p$. Hence $\gamma$ is a covering as wanted. We
have $\left\vert V_{K}\right\vert \leq\left\vert V_{G}\right\vert .\left\vert
V_{H}\right\vert $.

\smallskip
If $K$ is not connected, then each of its connected components is a covering
as wanted.

\smallskip
We now consider the case where $G,H$ and $M$ may have loops.\ It follows from
Lemma 4.22\ that we have coverings $\alpha^{\prime}:$ $G\times K_{2}%
\rightarrow M\times K_{2}$ and $\beta^{\prime}:$ $H\times K_{2}\rightarrow
M\times K_{2}$. As $G\times K_{2},$ $H\times K_{2}$ and $M\times K_{2}$ have
no loops, the previous proof yields coverings $\gamma:$ $K\rightarrow G\times
K_{2}$ and $\eta:$ $K\rightarrow H\times K_{2}$.\ As $G\times K_{2}$ and
$H\times K_{2}$ cover $G$ and $H$ respectively, we have (by Proposition 4.16)
\ coverings $\gamma^{\prime}:$ $K\rightarrow G$ and $\eta^{\prime}:$
$K\rightarrow H$\ where $K$ has at most $4\left\vert V_{G}\right\vert
.\left\vert V_{H}\right\vert $ vertices.
\end{proof}

This theorem does not apply to the two graphs of Example 4.4.

\medskip
A $k$-\emph{regular} graph has all its vertices of degree $k$.\ It may have
loops.\ A loop contributes 1 to the degree of its vertex.

\bigskip
\noindent\textbf{Proposition 6.2:} Let $G$ and $H$ be finite connected graphs.
They have a common finite cover in the following cases.\smallskip

(1) They have the same degree matrix (up to a permutation of rows and
columns), that is symmetric.\

(2) They are $k$-regular.

(3) Each of them has exactly one cycle, no loops, and they have isomorphic
universal covers.

\begin{proof}
(1) The degree matrix of $G$ and $H\ $is the adjacency matrix (counting loops
and parallel edges) of a graph $M$ covered by $G$ and $H.$ Theorem 6.1 is
applicable.\ Note that $M$ has loops if and only if some values on the
diagonal of the adjacency matrix are not null.

(2) The graphs $G$ and $H\ $ cover the graph with one vertex and $k$
loops.\ Theorem 6.1 is applicable, which gives the result proved in
\cite{AngGar}. It is a special case of (1).

(3) The graph $G$ is the union of a cycle $x_{1}-x_{2}-\dots-x_{n}-x_{1}$ and
pairwise disjoint trees $T_{i}$, each of them having node $x_{i}$ and no node
$x_{j}$, for $j\neq i$.\ Its universal cover $U$ is the union of a biinfinite
path $\dots-z_{0}-z_{1}-z_{2}-\dots-z_{n}-z_{n+1}-z_{n+2}-\dots-z_{2n}-z_{2n+1}-\dots$
and, similarly, of pairwise disjoint trees $U_{i}$ containing nodes $z_{i}%
$.\ The covering homorphism $\alpha$\ maps each $z_{p+kn}$ to $x_{p}$ for
$p\in\lbrack n]$ and $k\in\mathbb{Z}$ and, isomorphically, each tree
$U_{p+kn}$ to $T_{p}$.

\medskip
Hence, $U$ can be seen, up to isomorphism, as a periodic biinfinite sequence
of at most $n$ finite trees. From $H$, we have a similar description. A
binifinite sequence of the form $X^{\mathbb{Z}}=Y^{\mathbb{Z}}$ where $X$ has
length $n$ and $Y$ has length $p$ is equal to $(X^{p}Y^{\prime n}%
)^{\mathbb{Z}}$ for a circular shift $Y^{\prime}$ of $Y$.\ From the sequence
$X^{p}Y^{\prime n}$ one can build a common cover of $G$ and $H$.\

We can alternatively apply Theorem 6.1.\ We observe that $X$ and $Y^{\prime}%
$\ are respectively $S^{q}$ and $S^{m}$ for some sequence $S$, hence, we can
define a loop-free graph $M$\ with one cycle covered by $G$ and $H$ if $S$ has
length at least 2. If $S$ has length 1, one can define such a graph $M$ with a
loop of weight 2.
\end{proof}

By Lemma 4.22, it suffices to prove Theorem 4.10\ for finite bipartite graphs,
because if two finite graphs $G,H$ have a common universal cover $T$, then
$T$\ covers also $G\times K_{2}$ and $H\times K_{2}$ that are finite and
bipartite.\ A common finite cover of $G\times K_{2}$ and $H\times K_{2}$ is
also one of $G$ and $H$. The proof of \cite{Lei} uses this observation. In
order to indicate why its proof\ is difficult, we explain informally why a
natural proof generalizing that of Theorem 6.1\ fails.

\bigskip
\noindent\textbf{Definition 6.3:} \emph{Quotients of strongly regular labelled
trees.}\smallskip

(a) Let $T$ be a tree.\ It is bipartite with bipartition ($N_{T}^{1},N_{T}%
^{2}$) of its nodes.\ Let $\alpha$ be a labelling of $N_{T}^{1}\cup N_{T}%
^{2}\cup E_{T}$. We let $\sim$\ be an equivalence relation on $N_{T}^{1}\cup
N_{T}^{2}\cup E_{T}$ such that each equivalence class is included in
$N_{T}^{1}$, or in $N_{T}^{2}$ or in $E_{T}$, and two equivalent vertices or
edges have the same label. We require that if $e$ and $e^{\prime}$ are
equivalent edges, then $e:x-y$, $e^{\prime}:x^{\prime}-y^{\prime}$ for some
$x,y,x^{\prime},y^{\prime}$ such that $x\sim x^{\prime}$ and $y\sim y^{\prime
}.\ $Furthermore, we modify as follows the condition of Definition 4.11:
\begin{quote}
If $x$ and $y$ are equivalent vertices, then, $\sim$\ defines a bijection
$E_{T}(x)\cap\lbrack e]_{\sim}\rightarrow E_{T}(y)\cap\lbrack e]_{\sim}$.
\end{quote}

We obtain a quotient graph $T/\sim$ \ and a covering $T\rightarrow T/\sim$
that preserves labels.\smallskip

(b) Let $\gamma:T\rightarrow G$ be a universal covering of a finite bipartite
graph $G$. We label $T$\ as follows.\ A node $x\in N_{T}$ is labelled by
$\gamma(x)$ and an edge $e\in E_{T}$ by $\gamma(e)$. The labelled tree
$T_{\gamma}$ is strongly regular. Assume now that $\eta:T\rightarrow H$ is a
universal covering where $H$ is also a finite and bipartite graph. We define a
labelled tree $T_{\gamma,\eta}$ that combines the labels of $T_{\gamma}$ and
of $T_{\eta}$: a node $x\in N_{T}$ is labelled by ($\gamma(x),\eta(x))$ and an
edge $e$ is labelled by ($\gamma(e),\eta(e))$.\ $\square$

\medskip
Letting $G,H,\gamma,\eta$ and $T_{\gamma,\eta}$\ be as in this definition:

\bigskip
\noindent\textbf{Proposition 6.4:} If $T_{\gamma,\eta}$ is strongly regular,
there exists a finite bipartite graph $K$ that is a covering of both $G$ and
$H$.

\begin{proof}
Let $\approx$ be the equivalence relation on $N_{T_{\gamma,\eta}}$ such that
$x\approx y$ if and only if $(T_{\gamma,\eta})_{x}\simeq(T_{\gamma,\eta})_{y}%
$. Two equivalent nodes have the same label that is a pair in $V_{G}\times
V_{H}$. (However, Example 6.5\ below shows that two nodes may have the same
label in $T_{\gamma,\eta}$ without being equivalent for $\approx$).

\medskip
Without assuming that $T_{\gamma,\eta}$ is strongly regular, we first examine
the neighbourhood of a node $x$.\ Its incident edges have labels $(e_{1}%
,f_{1}),\dots.,(e_{p},f_{p})$ and respective other ends $z_{1},\dots.,z_{p}.$ In
$G$, the vertex $\gamma(x)$ has incident edges $e_{1},\dots.,e_{p}$ and
respective other ends $\gamma(z_{1}),\dots,\gamma(z_{p}).$ In $H$, the vertex
$\eta(x)$ has incident edges $f_{1},\dots,f_{p}$ and respective other ends
$\eta(z_{1}),\dots,\eta(z_{p}).$

If $x^{\prime}\approx$ $x$, then, since $(T_{\gamma,\eta})_{x}\simeq
(T_{\gamma,\eta})_{x^{\prime}}$, the edges incident to $x^{\prime}$ have
labels $(e_{1},f_{1}),\dots,(e_{p},f_{p})$ and respective other ends
$z_{1}^{\prime},\dots.,z_{p}^{\prime}.$ Consider an isomorphism $\alpha$:
$(T_{\gamma,\eta})_{x}\rightarrow(T_{\gamma,\eta})_{x^{\prime}}$. Since the
edge labels $(e_{1},f_{1}),\dots,(e_{p},f_{p})$ are pairwise distinct, it maps
$z_{i}$ to $z_{i}^{\prime}$ for each $i$.\ Hence, it is an isomorphism
$(T_{\gamma,\eta})_{z_{i}}\rightarrow(T_{\gamma,\eta})_{z_{i}^{\prime}}$\ and
$z_{i}\approx$ $z_{i}^{\prime}$.\ It follows that we get a quotient graph
$K:=T_{\gamma,\eta}/\approx$ \ that inherits the labels of $T_{\gamma,\eta
}/\approx.$

A vertex $[x]_{\approx}$ has label $(\gamma(x),\eta(x)).$ An edge of $K$
coming from $g:x-y$ in $T_{\gamma,\eta}$\ (it links $[x]_{\approx}$\ and
$[y]_{\approx}$\ in $K$) has label $(\gamma(g),\eta(g))\in E_{G}\times E_{H}$.
This is well-defined by the above remarks about neighbourhoods in
$T_{\gamma,\eta}$.

\medskip
We claim that $K$ is a covering of both $G$ and $H.$ We let $\kappa:V_{K}\cup
E_{K}$\ $\rightarrow V_{G}\cup E_{G}$\ be defined as follows: $\kappa
([x]_{\approx}):=\gamma(x),$ the first component of the label of
$[x]_{\approx}$; if $m:[x]_{\approx}-[y]_{\approx}$ is an edge of $K$ coming
from $g:x-y$ in $T_{\gamma,\eta}$, we define $\kappa(m):=\gamma(g).$

\medskip
\emph{Claim}: $\kappa:K\rightarrow G$\ is a covering.

\smallskip
\emph{Proof}: $\kappa$ is a surjective homomorphism.\ To prove that it is a
covering, we consider a vertex $[x]_{\approx}$ of $K$ where $x$ is a node in
$T_{\gamma,\eta}$. We recapitulate the above observations.

\medskip
The edges of $T_{\gamma,\eta}$ incident with $x$ are $g_{1},\dots,g_{p}$ with
respective ends $y_{1},\dots,y_{p}$ and labels $(e_{1},f_{1})$   $,\dots,(e_{p},f_{p})$. The edges of $G$\ incident with $\gamma(x)$ are $e_{1},\dots,e_{p}$.
We get edges $[x]_{\approx}-[y_{i}]_{\approx}$ in $K$, each with label
$(e_{i},f_{i}).$ They yield by $\kappa$\ the edges $e_{1},\dots,e_{p}$. Hence,
$\kappa$\ is a bijection of $E_{K}([x]_{\approx})$ to $E_{G}(x)$. $\square$

Similarly, we have a covering $K\rightarrow H$.

Finally, if $T_{\gamma,\eta}$ is strongly regular, the equivalence $\approx$
has finitely many classes and $K$ is finite.
\end{proof}

We do not obtain a proof of Theorem 4.10\ because the tree $T_{\gamma,\eta}$
constructed from two covering homomorphisms of finite graphs $G$ and $H$ is
not necessarily strongly regular.

\bigskip
\noindent\textbf{Example 6.5:} \emph{A tree} $T_{\gamma,\eta}$ \emph{that is
not strongly regular}.\smallskip

We let $G$ be the bipartite graph such that $V_{G}^{1}=\{a\}$, $V_{G}%
^{2}=\{b\},E_{G}=\{1,2,3,4\},$ and $H$\ similarly be such that $V_{H}%
^{1}=\{c\}$, $V_{H}^{2}=\{d\},E_{H}=\{5,6,7,8\}.$ They both have two vertices
and four parallel edges. Let $\gamma:T\rightarrow G$ be a universal covering
of $G$.\

\smallskip
We choose adjacent nodes $r$ and $s$ of $T$\ such that $\gamma(r)=a,\gamma
(s)=b$ and $\gamma(e)=1$ where $e:r-s$. We get a labelled tree $T_{\gamma}$.
We will enrich its labelling so as to obtain a tree $T_{\gamma,\eta}$ for some
covering $\eta:T\rightarrow H$.

For this purpose, we replace each node label $a$ of $T_{\gamma}$ by $(a,c)$,
each label $b$ of $T_{\gamma}$ by $(b,d)$, each edge label 1 by (1,5) and each
label 2 by (2,6).\ Then for each edge in the rooted tree $T_{r}-T_{r}/s$
(obtained by deleting $T_{r}/s$ from $T_{r}$), we replace 3 by (3,7) and 4 by
(4,8); for each edge in the subtree $T_{s}-T_{s}/r$, we replace 3 by (3,8) and
4 by (4,7).

We get a labelled tree $T_{\gamma,\eta}$ related to a universal covering
$\eta:T\rightarrow H$.

It is clear that $T_{\gamma,\eta}$ is not strongly regular because the edge
labels (3,7) are present in the part $T_{r}-T_{r}/s$, but not in the other
part $T_{s}-T_{s}/r$, and these two parts are infinite.\ $\square$

\bigskip
\noindent\textbf{Questions 6.6: } Does Theorem 4.10 extend to finite weighted graphs?

\medskip
It does in a somewhat trivial way for graphs whose weights are all $\omega$
.\ Let $G$ and $H$ be two such connected weighted graphs.\ Let $K$\ be their
product with $V_{K}:=V_{G}\times V_{H}$ and $(x,y)-(x^{\prime},y^{\prime})$ in
$K$ if and only if $x-x^{\prime}$ and $y-y^{\prime}$\ in $G$ and $H$
respectively.\ Since $\omega+\omega=\omega$ the two projections $\pi_{1}:$
$V_{K}\rightarrow V_{G}\ $ and\ $\pi_{2}:V_{K}=V_{H}$ are coverings.

\medskip
The next case to consider would be when weights are 1 or $\omega$.

\section{Conclusion}

We have generalized the notions of regular trees studied in
\cite{Cou83,CouHdBk}, in \cite{Arn,Cou9,CouWal} and in
\cite{Ang,Bod,BodVL,Nor} having motivations in program semantics by attaching
weights to the arcs or edges of the digraphs or graphs of which we consider
complete unfoldings or universal coverings.\ In particular, infinite weights
yield trees with nodes of infinite degree. Our finite weighted graphs offer
effective descriptions and yield decidability results.

\smallskip
The new notion of a \emph{strongly regular tree} defined as a universal
covering of a finite weighted graph is investigated in the companion article
\cite{CouNext}.

\subsection*{Acknowledgements}
I thank Yves M\'{e}tivier for fruitful discussions and the referee for useful comments.

\bibliographystyle{fundam}

\end{document}